\documentclass[journal=aesccq,manuscript=review,layout=twocolumn]{achemso}

\pdfoutput=1

\usepackage{chemformula} 
\usepackage{graphicx,rotating}
\usepackage[T1]{fontenc} 
\usepackage{amssymb}

\author{Marta Sewi{\l}o}
\alsoaffiliation{CRESST II and Exoplanets and Stellar Astrophysics Laboratory, NASA Goddard Space Flight Center, Greenbelt, MD 20771, USA}
\affiliation{Department of Astronomy, University of Maryland, College Park, MD 20742, USA}
\email{marta.m.sewilo@nasa.gov}

\author{Steven B. Charnley}
\affiliation{Astrochemistry Laboratory, NASA Goddard Space Flight Center, Greenbelt, MD 20771, USA}

\author{Peter Schilke}
\affiliation{I. Physikalisches Institut der Universit{\"a}t zu K{\"o}ln, Z{\"u}lpicher Str. 77, 50937, K{\"o}ln, Germany}

\author{Vianney Taquet}
\affiliation{INAF, Osservatorio Astrofisico di Arcetri, Largo E. Fermi 5, 50125 Firenze, Italy}

\author{Joana M. Oliveira}
\affiliation{Lennard-Jones Laboratories, Keele University, ST5 5BG, UK}

\author{Takashi Shimonishi}
\affiliation{Frontier Research Institute for Interdisciplinary Sciences, Tohoku University, Aramakiazaaoba 6-3, Aoba-ku, Sendai, Miyagi, 980-8578, Japan}
\alsoaffiliation{Astronomical Institute, Tohoku University, Aramakiazaaoba 6-3, Aoba-ku, Sendai, Miyagi, 980-8578, Japan}

\author{Eva Wirstr{\"o}m}
\affiliation{Department of Space, Earth and Environment, Chalmers University of Technology, Onsala Space Observatory, 43992, Onsala, Sweden}

\author{Remy Indebetouw}
\affiliation{Department of Astronomy, University of Virginia, PO Box 400325, Charlottesville, VA 22904, USA}
\alsoaffiliation{National Radio Astronomy Observatory, 520 Edgemont Rd, Charlottesville, VA 22903, USA}

\author{Jacob L. Ward}
\affiliation{Astronomisches Rechen-Institut, Zentrum f{\"u}r Astronomie der Universit{\"a}t Heidelberg, M{\"o}nchhofstr. 12-14, 69120 Heidelberg Germany}

\author{Jacco Th. van Loon}
\affiliation{Lennard-Jones Laboratories, Keele University, ST5 5BG, UK}

\author{Jennifer Wiseman}
\affiliation{NASA Goddard Space Flight Center, 8800 Greenbelt Rd, Greenbelt, MD 20771, USA}

\author{Sarolta Zahorecz}
\affiliation{Department of Physical Science, Graduate School of Science, Osaka Prefecture University, 1-1 Gakuen-cho, Naka-ku, Sakai, Osaka 599-8531, Japan}
\alsoaffiliation{Chile Observatory, National Astronomical Observatory of Japan, National Institutes of Natural Sciences, 2-21-1 Osawa, Mitaka, Tokyo 181-8588, Japan}

\author{Toshikazu Onishi}
\affiliation{Department of Physical Science, Graduate School of Science, Osaka Prefecture University, 1-1 Gakuen-cho, Naka-ku, Sakai, Osaka 599-8531, Japan}

\author{Akiko Kawamura}
\affiliation{National Astronomical Observatory of Japan, 2-21-1 Osawa, Mitaka, Tokyo 181-8588, Japan}

\author{C.-H. Rosie Chen}
\affiliation{Max-Planck-Institut fu?r Radioastronomie, Auf dem H{\"u}gel 69 D-53121 Bonn, Germany}

\author{Yasuo Fukui}
\affiliation{Nagoya University, School of Science, Furo-cho, Chikusa-ku, Nagoya, JP 464-8602}

\author{Roya Hamedani Golshan}
\affiliation{I. Physikalisches Institut der Universit{\"a}t zu K{\"o}ln, Z{\"u}lpicher Str. 77, 50937, K{\"o}ln, Germany}

\title{Complex Organic Molecules in Star-Forming Regions of the Magellanic Clouds}

\abbreviations{IR,NMR,UV}


\begin{document}


\begin{abstract}

The Large and Small Magellanic Clouds (LMC and SMC), gas-rich dwarf companions of the Milky Way, are the nearest laboratories for detailed studies on the formation and survival of complex organic molecules (COMs) under metal poor conditions. To date, only methanol, methyl formate, and dimethyl ether have been detected in these galaxies -- all three toward two hot cores in the N113 star-forming region in the LMC, the only extragalactic sources exhibiting complex hot core chemistry. We describe a small and diverse sample of the LMC and SMC sources associated with COMs or hot core chemistry, and compare the observations to theoretical model predictions.  Theoretical models accounting for the physical conditions and metallicity of  hot molecular cores in the Magellanic Clouds have been able to broadly account for the existing observations, but fail to reproduce the dimethyl ether abundance by more than an order of magnitude. We discuss future prospects for research in the field of complex chemistry in the low-metallicity environment. The detection of COMs in the Magellanic Clouds has important implications for astrobiology.  The metallicity of the Magellanic Clouds is similar to galaxies in the earlier epochs of the Universe, thus the presence of COMs in the LMC and SMC indicates that a similar prebiotic chemistry leading to the emergence of life, as it happened on Earth, is possible in low-metallicity systems in the earlier Universe.
\begin{figure}
\includegraphics[width=0.47\textwidth]{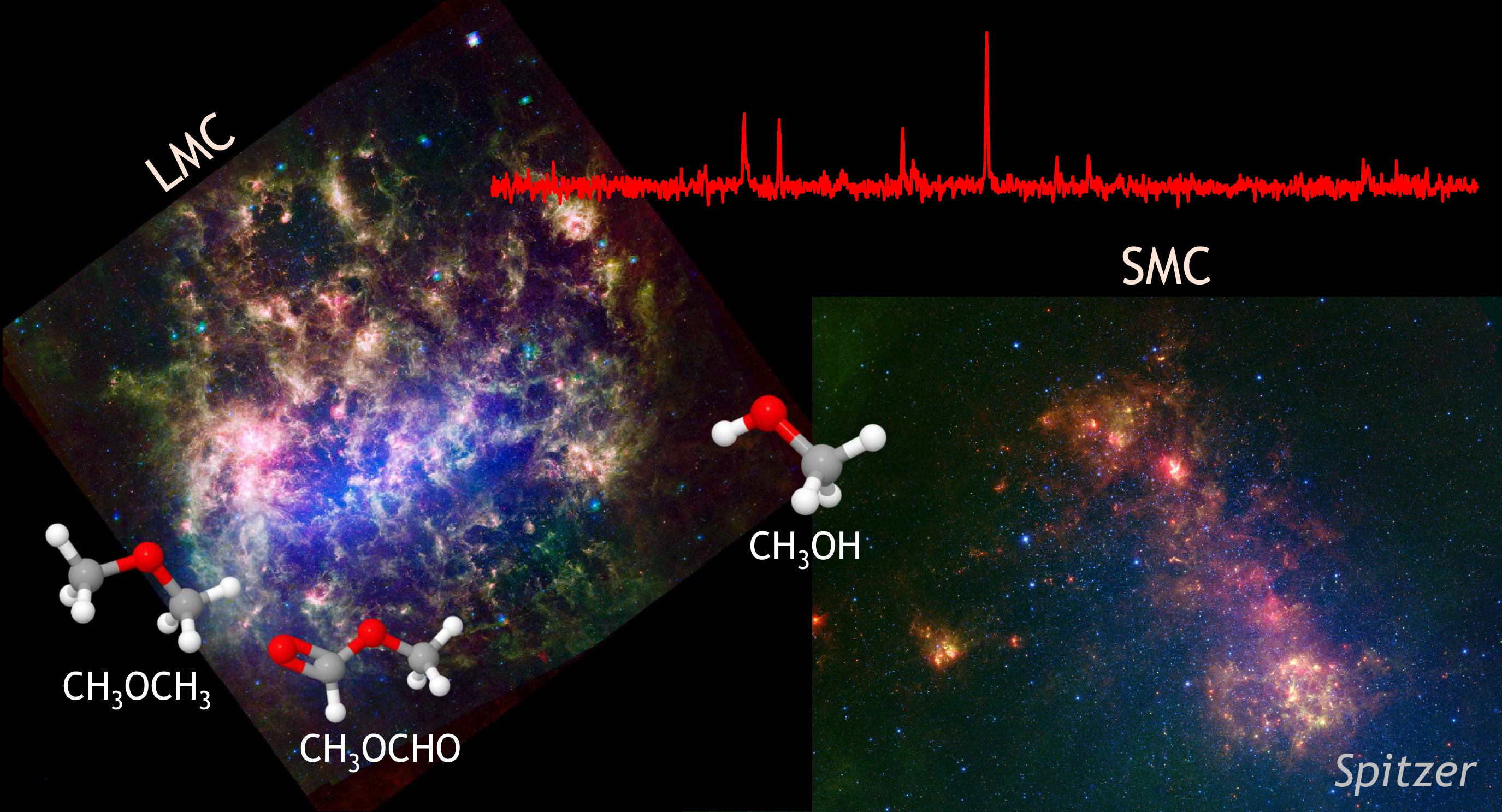}
\end{figure}

{\it Accepted for publication in the ACS Earth and Space Chemistry journal on August 27, 2019.  The article is part of the  ``Complex Organic Molecules (COMs) in Star-Forming Regions'' Special Issue.}
\end{abstract}

\section{Introduction}

An important issue for astrochemistry is to understand how the organic chemistry in low-metallicity environments (i.e., with low abundances of elements heavier than hydrogen or helium), relevant for star formation at earlier epochs of cosmic evolution, differs from that in the Galaxy.  Large aromatic organic molecules such as polycyclic aromatic hydrocarbons  (PAHs) are detected in high-redshift galaxies, but the formation efficiency of smaller complex organic molecules has been unknown.  A quantitative determination of the organic chemistry in high-redshift galaxies with different star formation histories, and lower abundances of the important biogenic elements (C, O, N, S, P), can shed light on the inventory of complex organic molecules available to planetary systems and their potential for harboring life  (e.g. \cite{ehrenfreund2000}). The nearest laboratories for detailed studies of star formation under metal poor conditions are the Large and Small Magellanic Clouds (LMC and SMC). 

\begin{figure*}
  \includegraphics[width=0.75\textwidth]{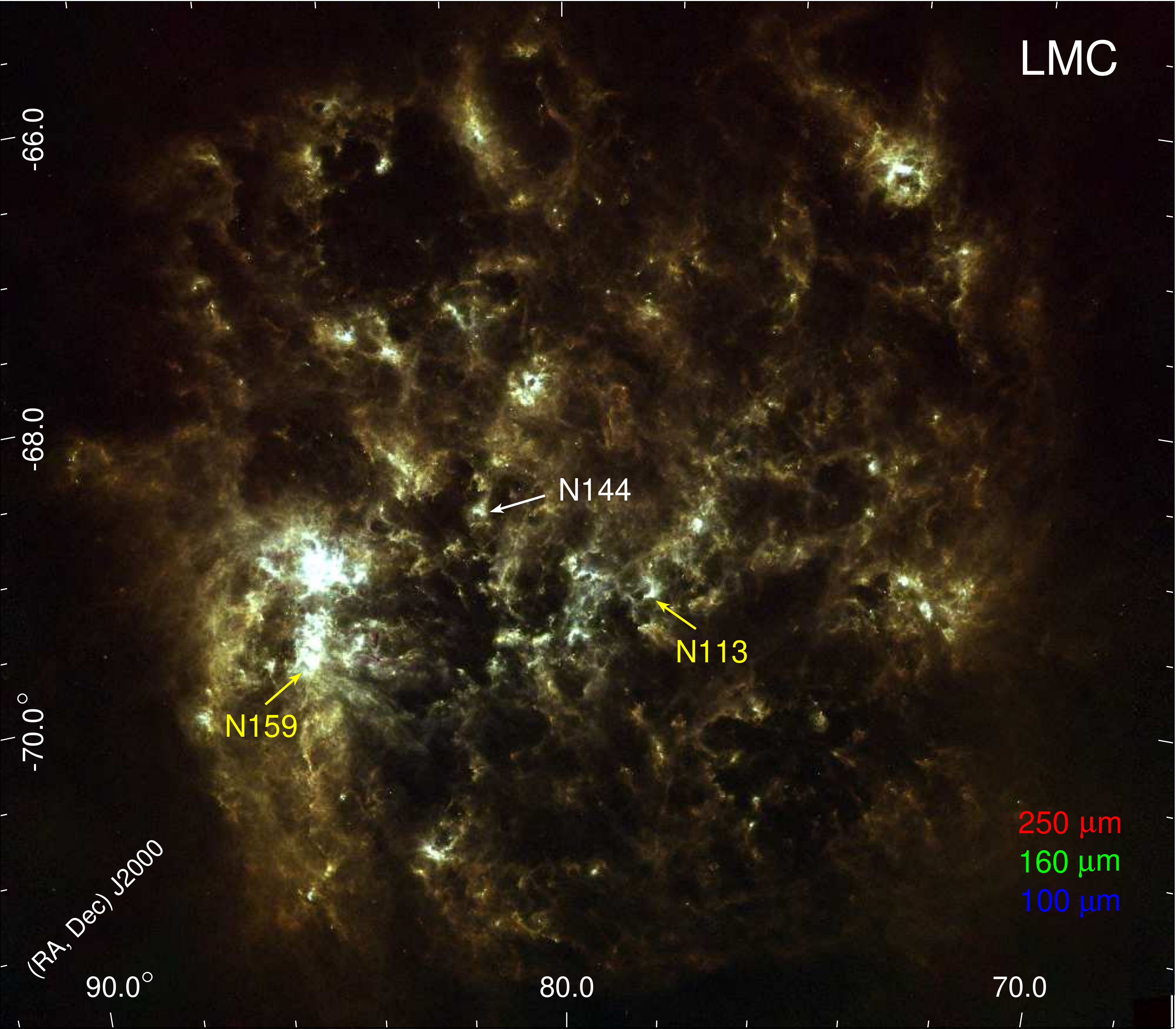}
  \includegraphics[width=0.75\textwidth]{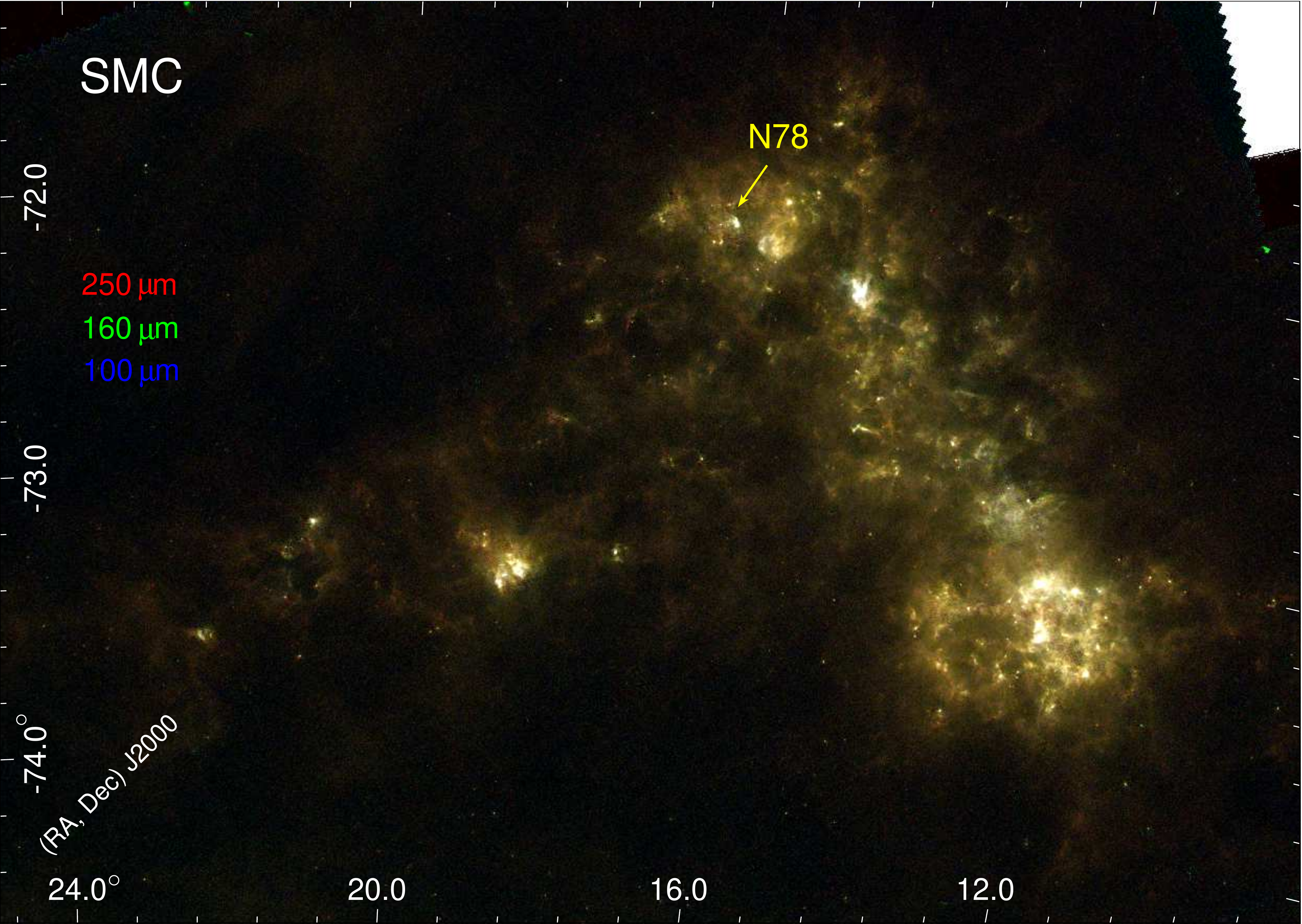}
  \caption{Three-color composite mosaics of the LMC ({\it top}) and SMC ({\it bottom}), combining the {\it Herschel}/HERITAGE 250 $\mu$m (red), 160 $\mu$m (green), and 100 $\mu$m (blue) images \cite{meixner2013}. Star-forming regions harboring sources with the detection of COMs (N\,113 and N\,159 in the LMC and N\,78 in the SMC; see Table~\ref{tbl:physical}) are indicated with yellow arrows and labeled. The location of the star-forming region N\,144 hosting ST\,11, a hot core without COMs, is indicated in white. North is up and east to the left.}
  \label{f:example}
\end{figure*}

The LMC and SMC are gas-rich dwarf companions of the Milky Way located at a distance of (50.0 $\pm$ 1.1) kpc \cite{pietrzynski2013} and (62.1 $\pm$ 2.0) kpc \cite{graczyk2014}, respectively. They are the nearest star-forming galaxies with metallicities $Z$ (the mass fractions of all the chemical elements other than hydrogen and helium) lower than that in the solar neighborhood ($Z_{\odot}$ = 0.0134)\cite{asplund2009}: $Z_{\rm LMC}$$\sim$0.3--0.5 $Z_{\odot}$ and $Z_{\rm SMC}$$\sim$0.2 $Z_{\odot}$\cite{russell1992,westerlund1997}.  Apart from the lower elemental abundances of gaseous C, O, and N atoms (e.g., \cite{dufour1984}), low metallicity leads to less shielding (due to the lower dust abundance; e.g., \cite{koornneef1984,duval2014}), greater penetration of UV photons into the interstellar medium, and consequently warmer dust grains (e.g., \cite{vanloon2010,oliveira2011}).  The interstellar ultraviolet radiation field in the LMC and SMC is 10--100 higher than typical Galactic values (e.g., \cite{browning2003}).  Gamma-ray observations indicate that the cosmic-ray density in the LMC and SMC is, respectively,  $\sim$25\% and $\sim$15\% of that measured in the solar neighborhood   (\cite{abdo2010,knodlseder2013}).  All these low metallicity effects may have direct consequences on the formation efficiency and survival of COMs, although their relative importance remains unclear. 

The range of metallicities observed in the Magellanic Clouds is similar to galaxies at redshift $z\sim1.5-2$, i.e. at the peak of the star formation in the Universe (between $\sim$2.8 and $\sim$3.5 billion years after the Big Bang; e.g., \cite{pei1999}), making them ideal templates for studying star formation and complex chemistry in low-metallicity systems in an earlier Universe where direct measurements of resolved stellar populations are not possible. Only very recently have gaseous complex organic molecules (methanol: CH$_3$OH, methyl formate: HCOOCH$_3$, dimethyl ether: CH$_3$OCH$_3$) been detected in the LMC \cite{heikkila1999,wang2009,sewilo2018} and in the SMC \cite{shimonishi2018}, using the Atacama Large Millimeter/submillimeter Array (ALMA).   These observations showed that interstellar complex organic molecules (COMs) can form in the low-metallicity environments, probably in the hot molecular cores and corinos associated with massive and low- to intermediate-mass protostars, respectively.  COMs had previously been detected outside the Milky Way, but in galaxies with metallicities comparable to or higher than solar (\cite{mcguire2018} and references therein).

In this article, we summarize the current state of knowledge concerning the organic chemistry in the Magellanic Clouds. 

\begin{table*}[t]
  \caption{\ch{CH3OH} Transitions Detected in the LMC in Single-Dish Observations\textsuperscript{\emph{a,b}}}
  \label{tbl:transitionssingle}
  \begin{tabular}{cccccc}
    \hline
    \hline
    Region & Species  & Transition & Frequency & $E_{\rm U}$  & Ref. \\
                &                &                  & (GHz)        & (K)       &    \\ 
    \hline
N\,159W & CH$_{3}$OH v$_t$=0\textsuperscript{\emph{c}}  & 2$_{-1,2}$--1$_{-1,1}$ E & 96.739362 & 12.54 & \cite{heikkila1999}\\
                                                                              && 2$_{0,2}$--1$_{0,1}$ A$^{+}$ & 96.741375 & 6.97& \cite{heikkila1999} \\
                                                                              && 2$_{0,2}$--1$_{0,1}$  E & 96.744550 & 20.09 & \cite{heikkila1999} \\
                                                                              && 2$_{1,1}$--1$_{1,0}$  E & 96.755511 & 28.01 & \cite{heikkila1999} \\
                                                                              && 3$_{-1,3}$--2$_{-1,2}$ E &  145.09747 &  19.51 & \cite{heikkila1999} \\
                                                                              && 3$_{0,3}$--2$_{0,2}$  A$^{+}$ &  145.10323 &  13.93 & \cite{heikkila1999} \\
 N\,113 & CH$_{3}$OH v$_t$=0  & 2$_{-1,2}$--1$_{-1,1}$ E & 96.739362 & 12.54 & \cite{wang2009}\\
                                                    && 2$_{0,2}$--1$_{0,1}$ A$^{+}$ & 96.741375 & 6.97& \cite{wang2009} \\
                                                    && 3$_{-1,3}$--2$_{-1,2}$ E &  145.09747 &  19.51 & \cite{wang2009} \\
                                                    && 3$_{0,3}$--2$_{0,2}$  A$^{+}$ &  145.10323 &  13.93 & \cite{wang2009} \\
 
    \hline
      \end{tabular}
        
   \textsuperscript{\emph{a}} All the observations were conducted with SEST. The SEST's half-power beam sizes are  61$''$--54$''$ and 47$''$--24$''$ in frequency ranges of 85--98 GHz and 109--219 GHz, respectively (from \cite{wang2009});      
   \textsuperscript{\emph{b}} The table uses the Cologne Database for Molecular Spectroscopy (CDMS) notation from the National Radio Astronomy Observatory (NRAO) Spectral Line Catalog (Splatalogue).  For \ch{CH3OH} ($N~K_{\rm a}~K_{\rm c}~p~{\rm v}$), a sign value  of $K_{\rm a}$ is used to differentiate $E_1$ ($+$) and $E_2$ ($-$) states (both belong to the same $E$ symmetry species);
   \textsuperscript{\emph{c}} The four transitions at $\sim$96.7 GHz are blended.

 \end{table*}
\normalsize
 
\section{Molecular Line Inventories in Star-Forming Regions in the Magellanic Clouds}
\label{s:singledish}

The proximity of the LMC/SMC enables both detailed studies of individual star-forming regions and statistical studies of molecular clouds or stellar populations on galactic scales.  The systematic studies of molecular clouds in the Magellanic Clouds started with the $^{12}$CO $J=1-0$ survey of the LMC (central 6$^{\circ}$ $\times$ 6$^{\circ}$; \cite{cohen1988}) and the SMC (2$^{\circ}$ $\times$ 2$^{\circ}$, \cite{rubio1991}) with a resolution of 8$\rlap.{'}$8 (the Columbia 1.2m Millimeter-WaveTelescope), followed by a higher resolution (2$\rlap.{'}$6) survey of both galaxies with the NANTEN 4m telescope  (e.g., \cite{fukui1999,mizuno2001}).  The more sensitive NANTEN survey that identified 272 molecular clouds in the LMC was published by \cite{fukui2008}. The most luminous CO clouds from this survey were observed with 45$''$ resolution using the Australia Telescope National Facility (ATNF) Mopra 22-m Telescope (the MAGMA survey; \cite{wong2011}). 

The molecular gas in the Magellanic Clouds was also investigated using the CO data from the ESO SEST (Swedish/ESO Submillimeter Telescope) Key-Program (CO $J = 1-0$ and $2-1$ lines with 50$''$ and 25$''$ half-power beam width, HPBW; \cite{israel1993}) toward 92 and 42 positions in the LMC and SMC, respectively, mostly coincident with IRAS sources; the results were reported in multiple papers, including detailed studies of major star-forming giant molecular clouds (GMCs) and less active cloud complexes in the LMC and small molecular cloud complexes in the SMC (see \cite{israel2003} and references therein).  Many other CO studies of individual star-forming regions with SEST in both the LMC and SMC followed  (e.g., LMC regions N159W, N113, N44BC, and N214DE \cite{heikkila1998}).  

The first unbiased spectral line survey of an individual star-forming region in the LMC was performed by \cite{johansson1994} using SEST and centered on the N\,159W molecular peak in N\,159 -- the brightest NANTEN GMC with H\,{\sc ii} regions.  The observations covered a 215--245 GHz frequency range in several bands and selected bright lines at lower frequencies ($\sim$100 GHz; HPBW$\sim$23$''$ at 220 GHz and 50$''$ at 100 GHz). No COM transitions covered by these observations (e.g., propyne: \ch{CH3CCH} $J=5-4$ and $J=6-5$ and \ch{CH3OCH3} $J=6-5$ lines) were detected. \cite{johansson1994} found that fractional abundances with respect to \ch{H2} for molecules detected in N\,159W ($^{13}$CO, C$^{18}$O, CS, SO, \ch{HCO+}, HCN, HNC, CN, and \ch{H2CO} -- formaldehyde) are typically a factor of 10 lower than those observed in the Galaxy. 

The multi-region, multi-line SEST 3 mm (85--115 GHz) survey by \cite{chin1997} focused on one SMC (LIRS 36 or N\,12) and four LMC (N\,113, N\,44BC, N\,159HW, and N\,214DE) molecular cloud cores.  They detected the \ch{C2H}, HCN, \ch{HCO+}, HNC, CS, $^{13}$CO, CN, $^{12}$CO transitions, but the \ch{CH3OH} $J=2-1$ line at $\sim$96.74 GHz covered by their observations was not detected toward any of the regions.  The follow-up study on N\,12 in the SMC included the 2, 1.3, and 0.85 mm SEST bands, but also failed to detect  \ch{CH3OH} or other COMs \cite{chin1998}. 
  
The first detection of a COM in the Magellanic Clouds was reported by \cite{heikkila1999} in their SEST 0.85, 1.3, 2, and 3 mm survey of the SMC LIRS\,49 (or N\,27) and LMC 30\,Dor--10, 30\,Dor--27, N\,159W, N\,159S,  and N\,160 regions, covering a range of metallicities and UV radiation strengths.  N\,159W was the only region where \ch{CH3OH} was detected:  $J=2-1$ and two $3-2$ lines at $\sim$96.7 and $\sim$145.1 GHz, respectively (see Table~\ref{tbl:transitionssingle}). The  \cite{heikkila1999}'s observations covered the $J=6-5$ and $J=8-7$ lines of \ch{CH3CCH}, but only marginal ($\sim$2.5$\sigma$) detections are reported toward one region -- N\,159W.  \cite{heikkila1999} estimated the \ch{CH3OH} fractional abundance with respect to \ch{H2} of 4.3 $\times$ 10$^{-10}$ in N\,159W.  The authors indicated that the derived physical and chemical quantities are averages over spatial scales probed by their observations (10--15 pc in the 3 mm, and 5--10 pc in the 2/1.3 mm bands). Overall, the molecular abundances in N\,159W where \ch{CH3OH} was detected are typically five to twenty times lower than those measured in the Galactic molecular clouds, assuming that hydrogen is mainly in molecular form within the clouds.   \cite{heikkila1999} concluded that reduced abundances are likely to be a combined effect of lower abundances of carbon and oxygen and higher photodissociation rates of molecules, both due to the lower metallicity in the LMC.

An extensive spectral line study on N\,113, previously observed by \cite{chin1997} at 3 mm (see above), provided new detections of \ch{CH3OH}  \cite{wang2009}.  Their SEST observations covered the 0.85, 1.3, 2, and 3 mm bands. The rich spectral line inventory included two $J=2-1$ and two $J=3-2$ lines of \ch{CH3OH} (see Table~\ref{tbl:transitionssingle}). The \ch{CH3OH} abundance with respect to \ch{H2} was estimated to be $\sim$5 $\times$ 10$^{-10}$, similar to that observed in N\,159W by \cite{heikkila1999}, and an order of magnitude lower than that observed in the Galaxy. 
 
N\,113 and N\,159W remain the best-studied regions in the LMC in terms of chemical inventory. The initial single-dish surveys were followed by interferometric observations with the Australia Telescope Compact Array (ATCA) with 4$''$--6$''$ angular resolutions by \cite{wong2006} (N\,113 only) and \cite{seale2012} (N\,113, N\,159, N\,105, and N\,44) that focused on dense gas tracers such as \ch{HCO+} and \ch{HCN}. \cite{ott2010} searched for \ch{NH3} (1,\,1) and (2,\,2) in seven star-forming regions in the LMC (N\,113, N\,159W, N\,159S, 30\,Dor\,10, N\,44\,BC, N\,105\,A, and N\,83\,A) with ATCA (HPBW$\sim$9$''$--$17''$) and only detected it toward N\,159W with an abundance of $\sim$4 $\times$ 10$^{-10}$ with respect to \ch{H2} -- 1.5--5 orders of magnitude lower than observed in Galactic star-forming regions. 

The most recent multi-region, multi-line single-dish survey was conducted by \cite{nishimura2016a} using the Mopra 22-m telescope's 3 mm band (HPBW$\sim$38$''$ at 90 GHz and 30$''$ at 115 GHz).  The sample included seven H\,{\sc ii} regions:  NQC2, CO Peak 1, N\,79, N\,44C, N\,11B, N\,113, and N\,159W.  That work confirmed lower molecular abundances in the LMC star-forming regions. Those observations failed to detect \ch{CH3OH} and other COMs. 

 The results of the single dish molecular line surveys of individual H\,{\sc ii} regions in the LMC and SMC indicate a deficiency of \ch{CH3OH} in these low-metallicity galaxies. This deficiency is also supported by the low detection rate of \ch{CH3OH} masers in the Magellanic Clouds. \cite{green2008} conducted a complete systematic survey for 6668-MHz \ch{CH3OH} and 6035-MHz excited-state OH masers in both the LMC and SMC using the Methanol Multibeam (MMB) survey receiver on the 64-m Parkes telescope (HPBW$\sim$3$\rlap.{'}$3), supplemented by higher sensitivity targeted observations of known star-forming regions. The survey resulted in a detection of only one maser of each kind, increasing the number of known 6668-MHz \ch{CH3OH} masers in the LMC to four 
  \cite{sinclair1992,ellingsen1994,beasley1996}.  \cite{ellingsen2010} later detected a single 12.2 GHz \ch{CH3OH} maser in the LMC.  To date, no \ch{CH3OH} masers have been detected in the SMC.   \cite{green2008} estimated that  \ch{CH3OH} masers in the LMC are underabundant by a factor of $\sim$45 (or $\sim$4--5 after correcting for differences in the star formation rates between the galaxies) and interstellar OH and H$_2$O masers by a factor of $\sim$10 compared to the Galaxy. 

The {\it Spitzer Space Telescope} ({\it Spitzer}; \cite{werner2004}) and {\it Herschel Space Observatory} ({\it Herschel}; \cite{pilbratt2010}) enabled studies of young stellar object (YSO) populations in the LMC and SMC, both galaxy-wide and in individual star-forming regions. Using the data from the ``{\it Spitzer} Surveying the Agents of Galaxy Evolution'' (SAGE; \cite{meixner2006}) and ``Surveying the Agents of Galaxy Evolution in the Tidally Stripped, Low Metallicity Small Magellanic Cloud'' (SAGE-SMC;  \cite{gordon2011}), thousands of YSOs were identified in the Magellanic Clouds, dramatically increasing the number of previously known YSOs (from 20/1 in the LMC/SMC; e.g., \cite{whitney2008,gruendl2009,chen2010,romita2010,carlson2012,sewilo2013}). YSOs detected by {\it Spitzer} include mostly Stage I (protostars with accreting envelopes and disks) and also Stage II (disk-only) YSOs; the youngest Stage 0/I YSOs (protostars in the main accretion stage) were identified with {\it Herschel} using the data from the ``HERschel Inventory  of  The  Agents  of  Galaxy  Evolution''  (HERITAGE, \cite{meixner2013}; e.g., \cite{sewilo2010,seale2014}) survey.   The \textit{AKARI} Large-area Survey of the Large Magellanic Cloud (LSLMC) also carried out near- to mid-infrared photometric survey and near-infrared slitless spectroscopic survey towards the LMC, whose datasets also are used to search for YSOs in the LMC \cite{ita2008,kato2012,shimonishi2013}. On a galaxy-wide scale, YSO lists can be utilized to determine a star formation rate in each galaxy.  On scales of individual star-forming clouds, YSO catalogs can also be used to study star formation rate, star formation efficiency, the evolutionary stage of the GMCs, and provide targets for follow-up detailed observations.  

Spectroscopic follow-up observations of {\it Spitzer}, {\it Herschel}, and  \textit{AKARI} YSOs in the LMC and SMC revealed chemical differences in the YSO envelopes between these galaxies and compared to Galactic YSOs, which can be explained as a consequence of differences in metallicity (see Section~\ref{s:spec}).


\section{Near- to Mid-Infrared Spectroscopy: More Evidence for Chemical Differences between the LMC, SMC, and the Milky Way} 
\label{s:spec}

The mid-IR studies on the LMC YSOs with {\it Spitzer} Infrared Spectrograph (IRS) \cite{oliveira2009,oliveira2011,oliveira2013,seale2011} and near-IR studies with the AKARI satellite \cite{shimonishi2008,shimonishi2010} and ground-based instrumentation \cite{oliveira2011,oliveira2013} found differences in ice chemistry between the LMC and the Galaxy.  In cold molecular clouds, layers of ice form on the surface of dust grains. The main constituent of ice in envelopes of YSOs is \ch{H2O}, mixed primarily with CO and \ch{CO2} \cite{vandishoeck2014}.  Since the \ch{H2O} ice abundances ($\sim$10$^{-4}$; e.g., \cite{whittet1988,pontoppidan2014}) exceed by a few orders of magnitude the gas-phase abundances, understanding gas--grain processes is crucial to fully understand the chemistry in the YSO envelopes.
 
In the LMC and by comparison to Galactic samples, \ch{CO2} ice column densities are enhanced with respect to \ch{H2O} ice \cite{oliveira2009,shimonishi2010,seale2011}, while relative CO-to-\ch{CO2} abundances are unchanged \cite{oliveira2011}.  \cite{oliveira2011} proposed the scenario where the high \ch{CO2}/\ch{H2O} ratio is due to the low abundance of \ch{H2O} in the LMC.  If there is a difference in the optical extinction $A_{\rm V}$ threshold for \ch{H2O} and \ch{CO2} ices (as the observations suggest), there could be an envelope of less shielded material that shows \ch{H2O} ice but no \ch{CO2} ice.  The strong interstellar radiation in the LMC penetrates deeper into the YSO envelopes as compared with Galactic YSOs, possibly destroying \ch{H2O} ice in less-shielded outer layers without affecting \ch{CO2} and \ch{H2O} ice mixtures that exist deeper in the envelope.

\begin{figure*}[t]
  \includegraphics[width=\textwidth]{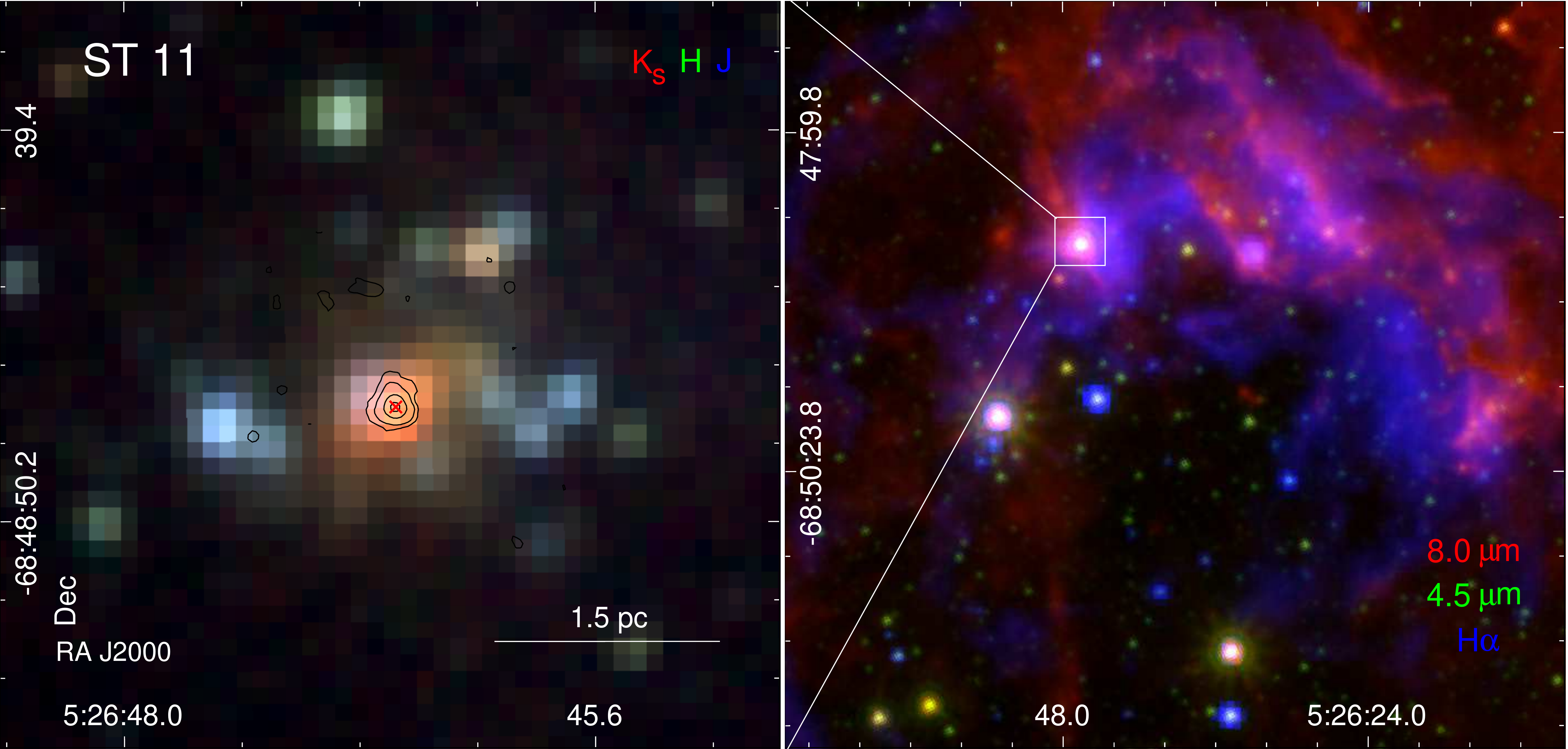}
  \caption{{\it Left:} A three-color mosaic combining the InfraRed Survey Facility (IRSF) $K_s$ (red), $H$ (green), and $J$ (blue; \cite{kato2007}) images  for the LMC source ST\,11 -- the source with a hot core chemistry, but no COMs detection \cite{shimonishi2016b}.    {\it Right:} a three-color mosaic combining the {\it Spitzer}/SAGE 8.0 $\mu$m (red), 4.5 $\mu$m (green; \cite{meixner2006}), and the MCELS H$\alpha$ (blue; \cite{smith1998}) images. The ALMA 840 $\mu$m continuum contours are overlaid on the image in the left panel; the contour levels are (3, 6, 20, 40) $\times$ 0.94 mJy beam$^{-1}$, the 840 $\mu$m image rms noise level. The ALMA beam size is 0$\rlap.{''}$5 $\times$ 0$\rlap.{''}$5. The red `$\times$' symbol in the left panel indicates the position of the {\it Spitzer} YSO -- the bright source in the mosaic shown in the right panel.} 
  \label{f:OtherMosLMCST11}
\end{figure*}

 In their Very Large Telescope (VLT)'s Infrared Spectrometer And Array Camera (ISAAC) near-IR spectroscopic observations, \cite{shimonishi2016a} marginally detected  \ch{CH3OH} ice absorption band toward two embedded massive YSOs in the LMC with the abundances suggesting that solid \ch{CH3OH} is less abundant for high-mass YSOs in the LMC than those in the Galaxy. They proposed a model which explains both the low abundance of \ch{CH3OH} and the higher abundance of solid \ch{CO2} found in previous studies. In this ``warm ice chemistry'' model, the grain surface reactions at a relatively high temperature ($T \gtrsim 20$ K) are responsible for the observed characteristics of ice chemical composition in the LMC. The high dust temperature in the low-metallicity environment caused by the strong interstellar radiation field leads to inefficient H atom sticking and CO hydrogenation to \ch{CH3OH} on grain surfaces. At the same time, the production of \ch{CO2} is enhanced as a result of the increased mobility of parent species (CO and OH; see Section~\ref{sec:magmodels} for a discussion on theoretical models).

The {\it Spitzer}/IRS spectroscopy also revealed several COMs in the {\it circumstellar} environment of SMP\,LMC\,11 (or LHA\,120--N\,78) - an object in the LMC classified as a carbon-rich planetary nebula (PN) in the optical, but with infrared properties consistent with it being in the transition from the post-asymptotic giant branch (AGB) to the PN stage (e.g., \cite{bernard2006} and references therein).  \cite{bernard2006} identified the diacetylene (\ch{C4H2}), ethylene (\ch{C2H4}), triacetylene (\ch{C6H2}), benzene (\ch{C6H6}), and possibly methylacetynele (\ch{CH3C2H}; commonly known as propyne which is found also in Titan's atmosphere) absorption bands in the {\it Spitzer}/IRS spectrum of SMP\,LMC\,11. \ch{C6H6} is a simplest building block of polycyclic aromatic hydrocarbons (PAHs), which are  abundant and ubiquitous in the interstellar medium (e.g., \cite{tielens2008}). In the subsequent analysis and modeling of the same {\it Spitzer}/IRS  spectrum,  \citep{malek2012} questioned the detection of \ch{C6H2}, but confirmed the detection of \ch{CH3C2H}.  SMP\,LMC\,11 shows a peculiar chemistry, not typical for PN or post-AGB objects in general, e.g., it is one of only two evolved stars in which hydrocarbons up to benzene in absorption are detected (e.g., \cite{cernicharo2001,malek2012,jones2017}).

\begin{figure*}[t]
  \includegraphics[width=\textwidth]{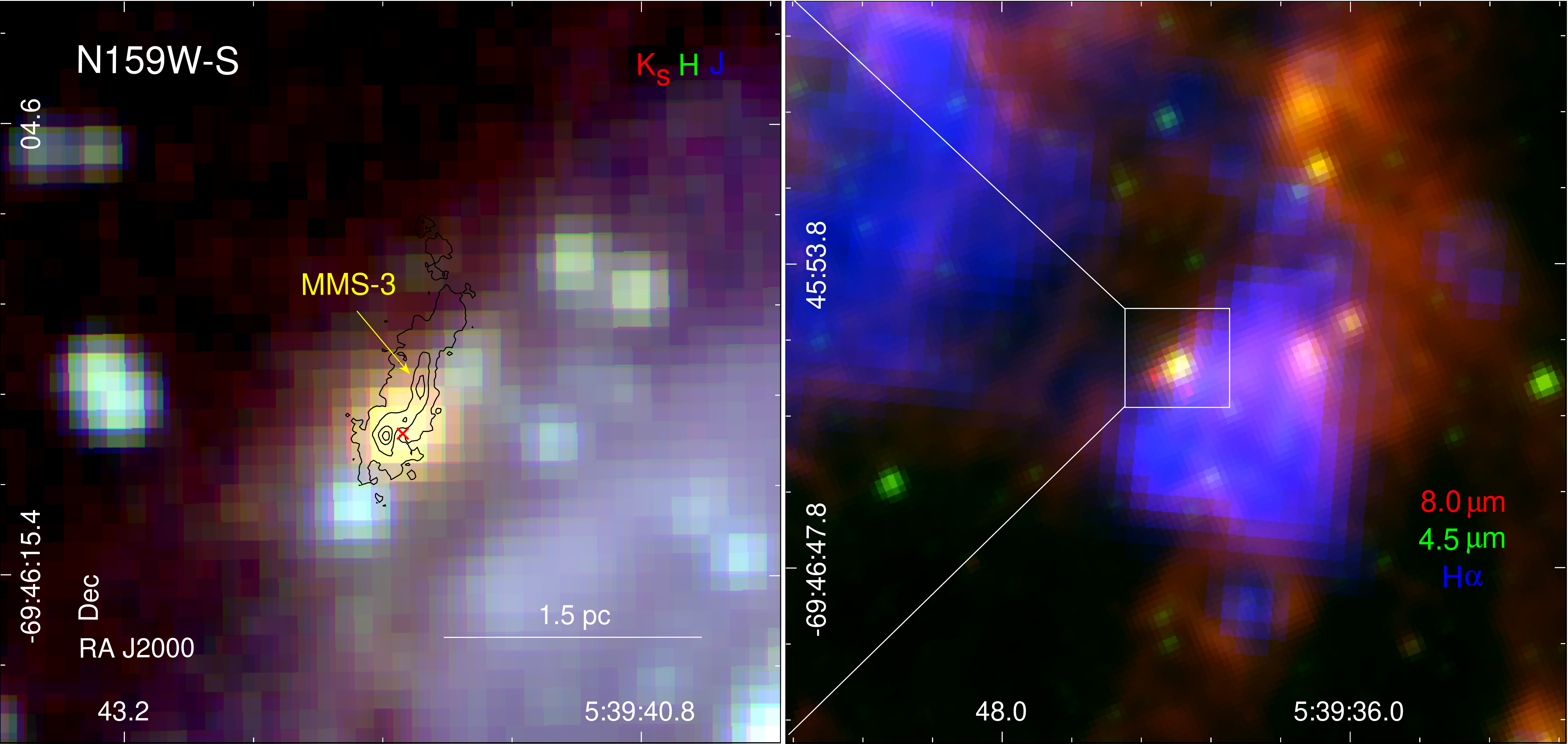}   
  \caption{The same as Fig.\ref{f:OtherMosLMCST11}, but for N159W--S -- the source with a cold methanol detection (PI Peter Schilke, see Sections~\ref{s:N159WS} and ~\ref{s:newalma}). The ALMA 1.3 mm continuum contours are overlaid on the image in the left panel; the contour levels are (0.1, 0.3, 0.5, 0.7) mJy beam$^{-1}$, the image rms noise level is 0.027 mJy beam$^{-1}$  \cite{tokuda2018}. The ALMA beam size is  0$\rlap.{''}$26 $\times$ 0$\rlap.{''}$23.} 
  \label{f:OtherMosLMCN159}
\end{figure*}

\section{The Quest for Detecting COMs in the Magellanic Clouds with ALMA}
\label{sec:questforcoms}


The detection of \ch{CH3OH} and more complex molecules in the Magellanic Clouds has been accelerated by the advent of ALMA. ALMA provides high spatial resolution and sensitivity which enabled studies in the LMC/SMC that in the pre-ALMA era were only possible in our Galaxy. The wide frequency coverage of ALMA observations (up to four $\sim$2 GHz spectral windows in the submm/mm wavelength range) allows for simultaneous observations of multiple molecular species and enables serendipitous discoveries.

\subsection{Hot Core without COMs: LMC ST\,11} 
\label{s:ST11}

ST\,11 is a high-mass YSO ($\sim$5$\times$10$^{^5}$ L$_{\odot}$, {\it Spitzer} YSO 052646.61$-$684847.2; \cite{whitney2008,gruendl2009,seale2009,shimonishi2010,seale2014}, or IRAS\,05270--6851) located in the N\,144 star-forming region in the LMC (Fig.~\ref{f:example} and \ref{OtherMosLMCST11}). In the pre-ALMA era, the spectral properties of the source were well studied with mid-infrared and near-infrared spectroscopy, which have detected absorption bands due to \ch{H2O} ice, \ch{CO2} ice, and silicate dust, as well as emission due to Polycyclic Aromatic Hydrocarbon (PAH) and hydrogen recombination lines \cite{seale2009,shimonishi2010}. 

The high-resolution ($\sim$0$\rlap.{''}$5 or $\sim$0.12 pc) submillimeter ($\sim$0.86 mm) observations toward ST\,11 with ALMA revealed the presence of a hot molecular core associated with the source \cite{shimonishi2016b}. According to the rotational diagram analysis of multiple $^{32}$SO$_2$ and $^{34}$SO$_2$ lines, the gas temperature of the hot core region is estimated to be $\sim$100--200 K. The total gas density of the source is estimated to be higher than 2$\times$10$^6$ cm$^{-3}$ based on the analysis of the submillimeter dust continuum. Emission lines of CO, C$^{17}$O, \ch{HCO+}, H$^{13}$CO$^{+}$, \ch{H2CO}, NO, \ch{H2CS}, $^{33}$SO,  $^{33}$SO$_2$, and SiO are also detected from the compact region associated with the source. High-velocity components are detected in the line profile of CO (3--2), which suggests the presence of molecular outflow in this source. 

\ch{CH3OH} and larger COMs are not detected toward ST\,11, despite the high gas temperature that is sufficient for the ice sublimation and despite the detection of  \ch{H2CO}, a small organic molecule. The estimated upper limit on the \ch{CH3OH} gas fractional abundance of $8\times10^{-10}$ is significantly lower than those of Galactic hot cores by a few orders of magnitude (\cite{shimonishi2016b}, see Table~\ref{tbl:galactic}).  \cite{shimonishi2016b} hypothesized that the inhibited formation of \ch{CH3OH} on the dust surface before the hot core stage could be responsible for this deficiency.   \cite{shimonishi2016b} also estimated upper limits on fractional abundances with respect to \ch{H2} for (\ch{CH3OCH3}, \ch{HCOOCH3}, \ch{C2H5OH}) of ($<$1.5$\times$10$^{-8}$, $<$1.8$\times$10$^{-8}$, $<$3.7$\times$10$^{-9}$). These upper limits indicate that in ST\,11 the fractional abundance of \ch{CH3OCH3} is at least an order of magnitude lower compared to Galactic hot cores, while the  \ch{HCOOCH3} and \ch{C2H5OH} fractional abundances are comparable to the average abundances observed in the Galactic counterparts.

\begin{figure*}
  \includegraphics[width=\textwidth]{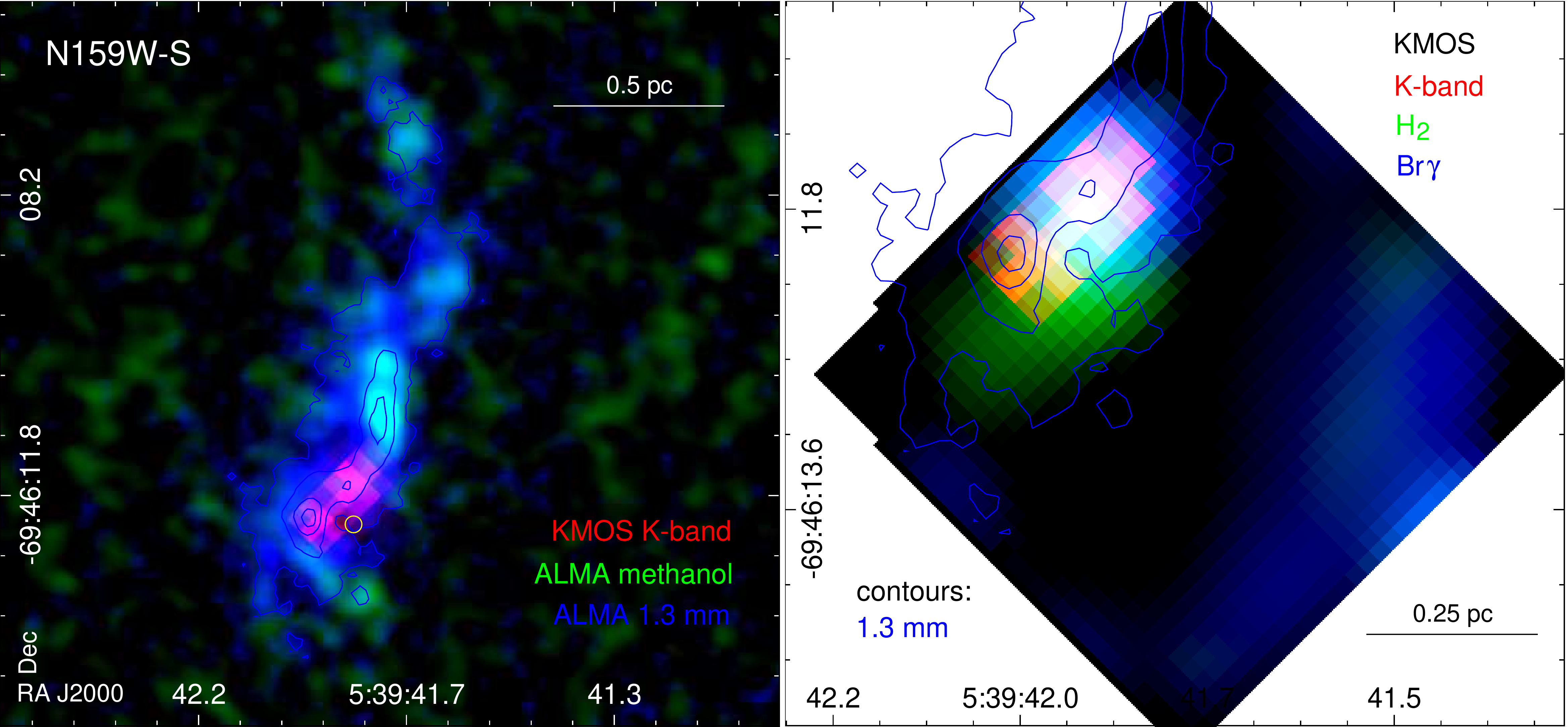} 
  \caption{{\it Left}:  Three-color mosaic combining the {\it VLT}/KMOS {\it K}-band (red), ALMA \ch{CH3OH} (green), and ALMA 1.3 mm (blue) images of N\,159W--S. The contours correspond to the 1.3 mm continuum emission with contour levels the same as in Fig.~\ref{f:OtherMosLMCN159}. The yellow circle shows the catalog position of the {\it Spitzer} YSO \cite{gruendl2009}. {\it Right}: Three-color mosaic combining KMOS images: {\it K}-band (red), \ch{H2} (green), and Br$\gamma$ (blue).  The 1.3 continuum contours (the same as in the left panel) are overlaid for reference.}
  \label{f:N159WSKMOS}
 \end{figure*}

\begin{figure*}
   \includegraphics[width=0.7\textwidth]{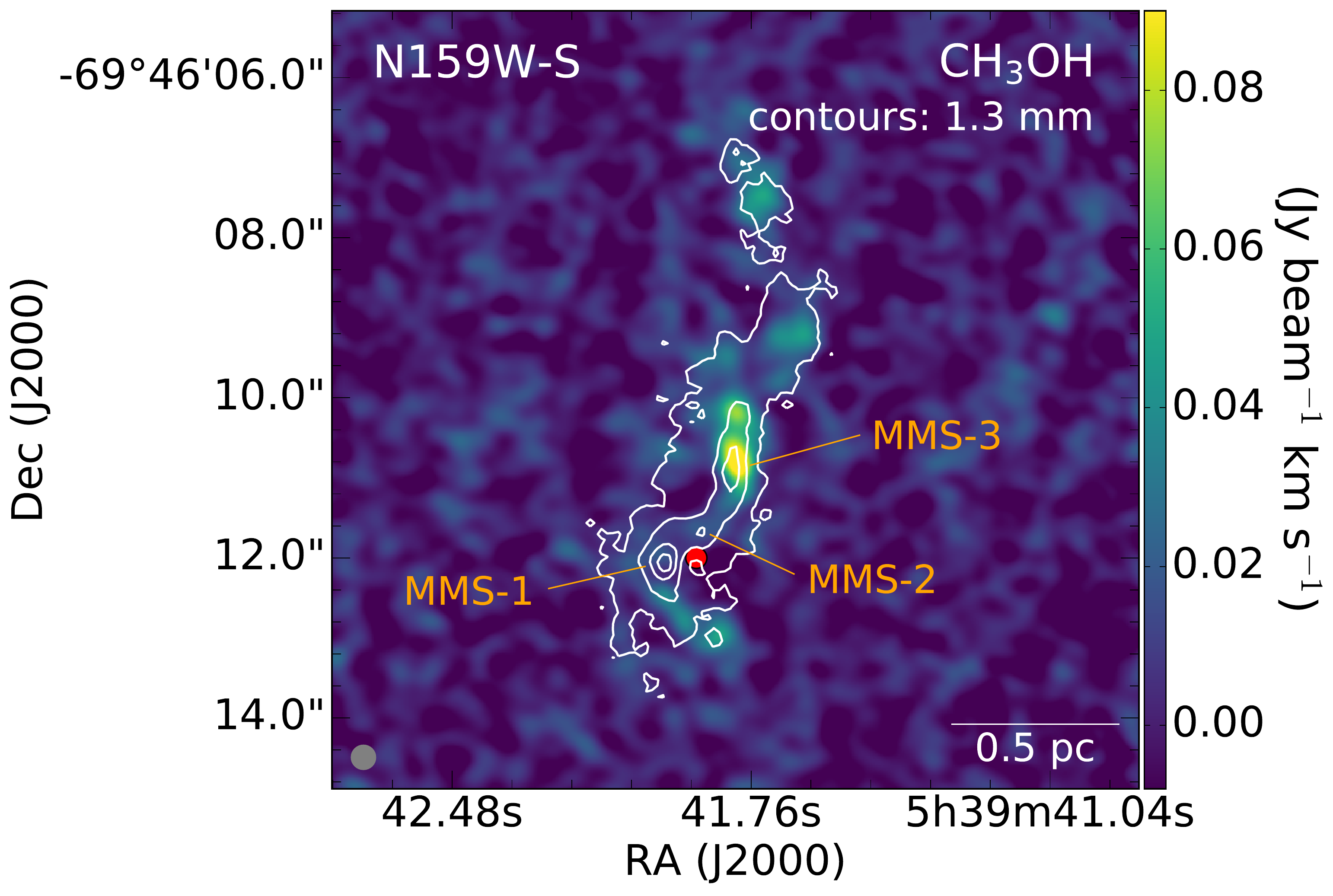}
  \caption{The integrated intensity image of \ch{CH3OH} for N\,159W--S with the 1.3 mm continuum contours and the positions of the continuum peaks (MMS--1, MMS--2, and MMS--3; \cite{tokuda2018}) overlaid.  The position of the {\it Spitzer} YSO is indicated with the red filled circle (e.g., \cite{chen2010}). The 1.3 mm continuum contour levels and the ALMA beam size are the same as in Fig.~\ref{f:OtherMosLMCN159}. The ALMA synthesized beam size of the \ch{CH3OH} observations (shown in the lower left corner) is 0$\rlap.{''}$3 $\times$ 0$\rlap.{''}$3.}
  \label{f:N159WSmeth}
 \end{figure*}

\scriptsize
\begin{sidewaystable*}
  \caption{Positions, Physical Properties, and Fractional Abundances for Sources Described in Section~\ref{sec:questforcoms} with COMs Detection or Measured Upper Limits with ALMA}
  \label{tbl:physical}
  \begin{tabular}{ccccccccc}
    \hline
    \hline
    Source  & R.A. (J2000) & Decl. (J2000) & Molecule& $T_{\rm rot}$ & $N$ & $N/N(\ch{H2}$) & Type\textsuperscript{\emph{m}} & Ref. \\
            &  ($^{\rm h}$ $^{\rm m}$ $^{\rm s}$) & ($^{\rm \circ}$ $'$ $''$) & & (K) & (cm$^{-2}$) & & & \\ 
    \hline 
    \multicolumn{9}{c}{LMC} \\ 
    \hline
    N\,113 A1\textsuperscript{\emph{a}} & 05:13:25.17	& $-$69:22:45.5 &  \ch{CH3OH}\textsuperscript{\emph{g}} & 134$\pm$6 & (1.6$\pm$0.1)$\times$10$^{16}$ & (2.0$\pm$0.3)$\times$10$^{-8}$ & HC  & \cite{sewilo2018}\\
              & && \ch{CH3OCH3}& \dotfill \textsuperscript{\emph{h}}& (1.8$\pm$0.5)$\times$10$^{15}$ & (2.2$\pm$0.7)$\times$10$^{-9}$ & & \\
              & && \ch{HCOOCH3}& \dotfill \textsuperscript{\emph{h}} & (1.1$\pm$0.2)$\times$10$^{15}$ & (1.4$\pm$0.4)$\times$10$^{-9}$ & & \\
    N\,113 B3\textsuperscript{\emph{a}} & 05:13:17.18 & $-$69:22:21.5 & \ch{CH3OH}\textsuperscript{\emph{g}} & 131$\pm$15 & (6.4$\pm$0.8)$\times$10$^{15}$ & (9.1$\pm$1.7)$\times$10$^{-9}$ & HC& \cite{sewilo2018} \\
              & && \ch{CH3OCH3}& \dotfill \textsuperscript{\emph{h}} & (1.2$\pm$0.4)$\times$10$^{15}$ & (1.7$\pm$0.7)$\times$10$^{-9}$ & & \\
              &&& \ch{HCOOCH3}\textsuperscript{\emph{f}}& \dotfill \textsuperscript{\emph{h}} & $<3.4$$\times$10$^{15}$ & $<0.5$$\times$10$^{-9}$  &  & \\ 
    N159W--South\textsuperscript{\emph{b,j}} & 05:39:41 & $-$69:46:06 & \ch{CH3OH} &\dotfill&\dotfill&\dotfill & CM  & \textsuperscript{\emph{k}} \\
    ST\,11\textsuperscript{\emph{c,i}} &  05:26:46.60 & $-$68:48:47.0 &  \ch{CH3OH} & 100\textsuperscript{\emph{l}}  & $<3.5$$\times$10$^{14}$ & $<8$$\times$10$^{-10}$  & HC$^{\star}$  & \cite{shimonishi2016b} \\
               & &&                                          \ch{CH3OCH3}& 100\textsuperscript{\emph{l}}  & $<1.3$$\times$10$^{15}$ & $<3$$\times$10$^{-9}$  & & \\
               & &&                                         \ch{HCOOCH3}& 100\textsuperscript{\emph{l}}  & $<7.1$$\times$10$^{15}$ &  $<2$$\times$10$^{-8}$ & & \\
                 & &&                                       \ch{C2H5OH}   & 100\textsuperscript{\emph{l}}  & $<2.2$$\times$10$^{15}$ &  $<5$$\times$10$^{-9}$ & & \\             
    \hline 
    \multicolumn{9}{c}{SMC} \\ 
    \hline
    IRAS\,01042$-$7215 P2\textsuperscript{\emph{d,i}}  & 01:05:49.54 &	$-$71:59:48.9 & \ch{CH3OH} & 18$\pm$5 & (1.4$^{+1.1}_{-0.6})\times10^{14}$  & (5.0$^{+4.2}_{-2.3})\times10^{-10}$  & CM & \cite{shimonishi2018} \\ 
    IRAS\,01042$-$7215 P3\textsuperscript{\emph{e,i}}  & 01:05:49.50 &	$-$71:59:49.4 &\ch{CH3OH} & 22$\pm$6 & (1.6$^{+1.1}_{-0.6})\times10^{14}$ & (1.5$^{+1.0}_{-0.6})\times10^{-9}$ & CM & \cite{shimonishi2018}\\
    \hline
      \end{tabular}
      
\textsuperscript{\emph{a}} The positions of N\,113 A1 and B3 correspond to the $\sim$224.3 GHz continuum peaks; 
\textsuperscript{\emph{b}} Multiple \ch{CH3OH} peaks are detected across the N159W--South molecular clump (see Section~\ref{s:N159WS} and Fig.~\ref{f:N159WSmeth});  
\textsuperscript{\emph{c}} The position of ST\,11 corresponds to the 359 GHz continuum peak; 
\textsuperscript{\emph{d}} The position of IRAS\,01042$-$7215 P2 corresponds to the C$^{33}$S, \ch{H2CS}, and \ch{SiO} peak; 
\textsuperscript{\emph{e}} The position of IRAS\,01042$-$7215 P3 corresponds to the \ch{CH3OH} peak;
\textsuperscript{\emph{f}} a tentative detection;
     \textsuperscript{\emph{g}} $T_{\rm rot}$ and $N$ were determined using the {\sc madcubaij} software with the initial estimates  based on the rotational diagram analysis of \ch{CH3OH};
     \textsuperscript{\emph{h}} $T_{\rm rot}$ for \ch{CH3OCH3} and \ch{HCOOCH3} was assumed to be equal to $T_{\rm rot}$ determined for \ch{CH3OH} \cite{sewilo2018};
     \textsuperscript{\emph{i}} $T_{\rm rot}$ and $N$ were determined based on the rotational diagram of \ch{CH3OH}; the values presented in the table are the revised values from \cite{shimonishi2018} based on new ALMA observations (PI T. Shimonishi);  
     \textsuperscript{\emph{j}} The analysis of the N159W--South data is preliminary, thus we do not provide physical parameters for this region; 
     \textsuperscript{\emph{k}} PI P. Schilke;
      \textsuperscript{\emph{l}} The average rotation temperature of \ch{SO2}, $^{34}$SO$_2$, and $^{33}$SO$_2$;  $N$ and $N/N(\ch{H2}$) upper limits are at the 2$\sigma$ level;
           \textsuperscript{\emph{m}} Type: HC -- a hot core; HC$^{\star}$ -- a hot core with no COMs; CM -- a source with cold methanol.
\end{sidewaystable*}
\normalsize

\subsection{Cold Methanol: LMC N159W-South and SMC IRAS\,01042 $-$7215} 
\label{almaobs}

\subsubsection{N159W--South}
\label{s:N159WS}

\begin{figure*}
  \includegraphics[width=\textwidth]{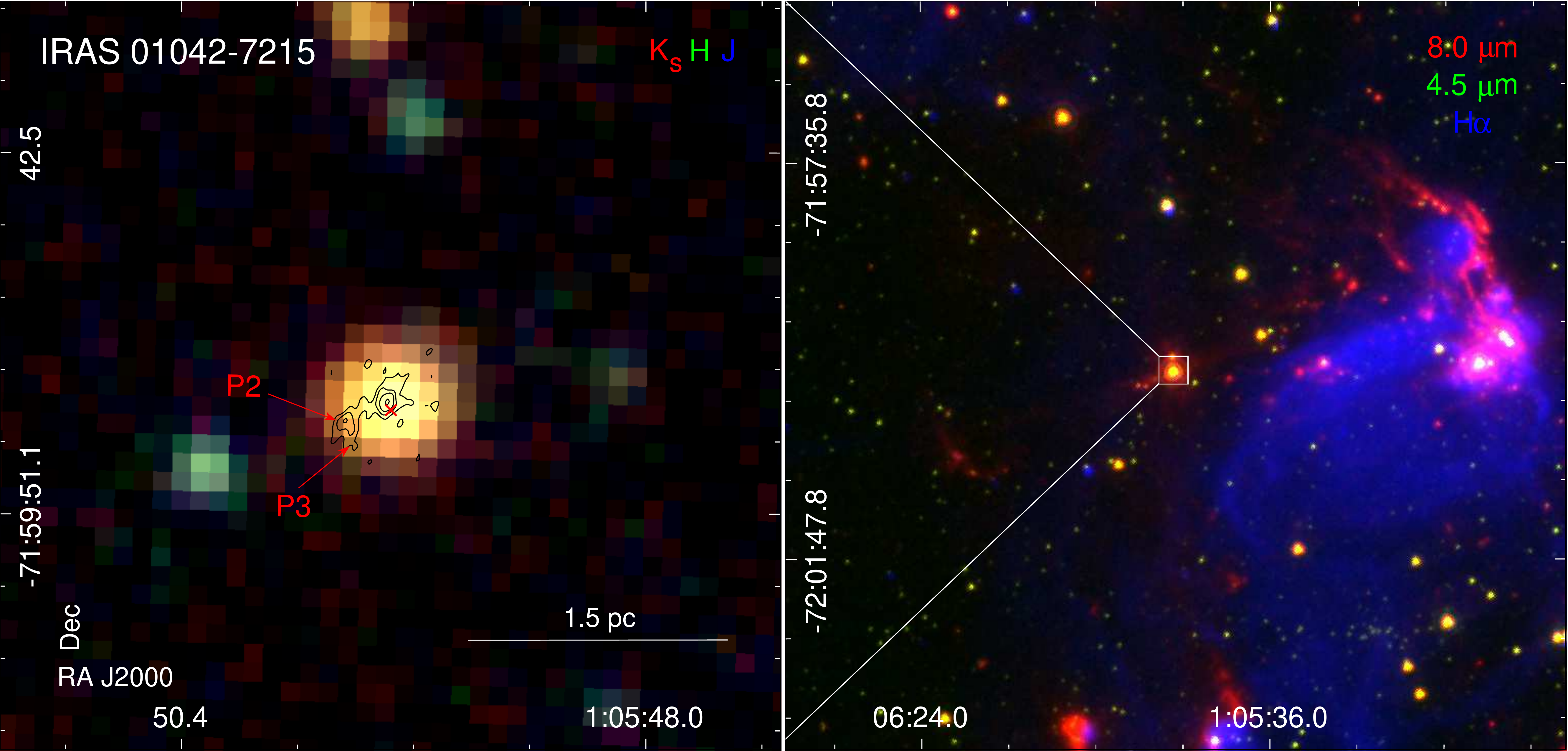} 
  \caption{The same as Fig.\ref{f:OtherMosLMCST11}, but for the field around the SMC source IRAS\,01042$-$7215 where \ch{CH3OH} was detected toward sources P2 and P3 (indicated in the left panel) corresponding to the \ch{CH3OH} peaks \cite{shimonishi2018}. The ALMA 1.2 mm continuum contours are overlaid. The continuum contour levels are (3, 6, 10, 20) $\times$ 21 mJy beam$^{-1}$, the 1.2 mm image rms noise level. The ALMA synthesized beam size is 0$\rlap.{''}$35 $\times$ 0$\rlap.{''}$22.} 
  \label{f:OtherMosSMC}
\end{figure*}

The N\,159 star-forming region located south from 30 Doradus is one of the best studied areas in the LMC hosting H\,{\sc ii} regions, young stellar clusters, {\it Spitzer} and {\it Herschel} massive YSOs, and a water maser (e.g., \cite{chen2010} and references therein; \cite{ellingsen2010}).  As described in Section~\ref{s:singledish}, N\,159 was the target of detailed spectral line observations with single-dish telescopes.  It is the brightest region in the LMC in $^{12}$CO at SEST/Mopra resolution. One of the three giant molecular clouds in N\,159, N\,159W, is one of only two regions with a \ch{CH3OH} detection in the pre-ALMA era.  

The first high-resolution ($\sim$1$''$ or $\sim$0.25 pc)  $^{13}$CO and $^{12}$CO (2--1) ALMA observations of N159W resolved the molecular gas emission into filaments and detected two sources with outflows \cite{fukui2015}. It was the first detection of both  filamentary structure and outflows outside the Galaxy.  These observations identified two main regions of high-mass star formation which were dubbed N\,159W--North and N\,159W--South (or N\,159W--N and N\,159W--S, respectively).  Both N\,159W--N and N\,159W--S are associated with {\it Spitzer}/{\it Herschel} YSOs \cite{whitney2008,gruendl2009,chen2010,carlson2012,seale2014}.  Two Stage 0/I YSOs with outflows have masses $>$30 M$_{\odot}$ and are associated with 1.3 mm continuum peaks, one in each N\,159W--N (YSO--N or {\it Spitzer} 053937.56$-$694525.4) and N\,159W--S (YSO--S or {\it Spitzer} 053941.89$-$694612.0; e.g., \cite{chen2010}).  Based on the kinematics of the molecular gas and the location of YSO--S at the intersection of two $^{13}$CO filaments, \cite{fukui2015} proposed that star-formation in N\,159W--S was triggered by the collision of two filaments $\sim$10$^{5}$ years ago. 

This model was revised by \cite{tokuda2018} based on the four times higher resolution observations of CO isotopes ($\sim$0$\rlap.{''}$25 or 0.06 pc), which resolved the filaments in N\,159W--S into a complex, hub--filament structure.   Multiple protostellar sources with outflows separated by 0.2--2 pc were detected along the main massive filament. They correspond to the three major 1.3 mm continuum peaks dubbed MMS--1, MMS--2, and MMS--3 (see Fig.~\ref{f:N159WSmeth}). The \cite{tokuda2018}'s observations are consistent with a scenario that a large-scale ($>$100 pc), rather than local ($\sim$10 pc; \cite{fukui2015}), collision triggered the formation of both filaments and massive protostars in N\,159W--S. 

\begin{figure*}
  \includegraphics[width=0.5\textwidth]{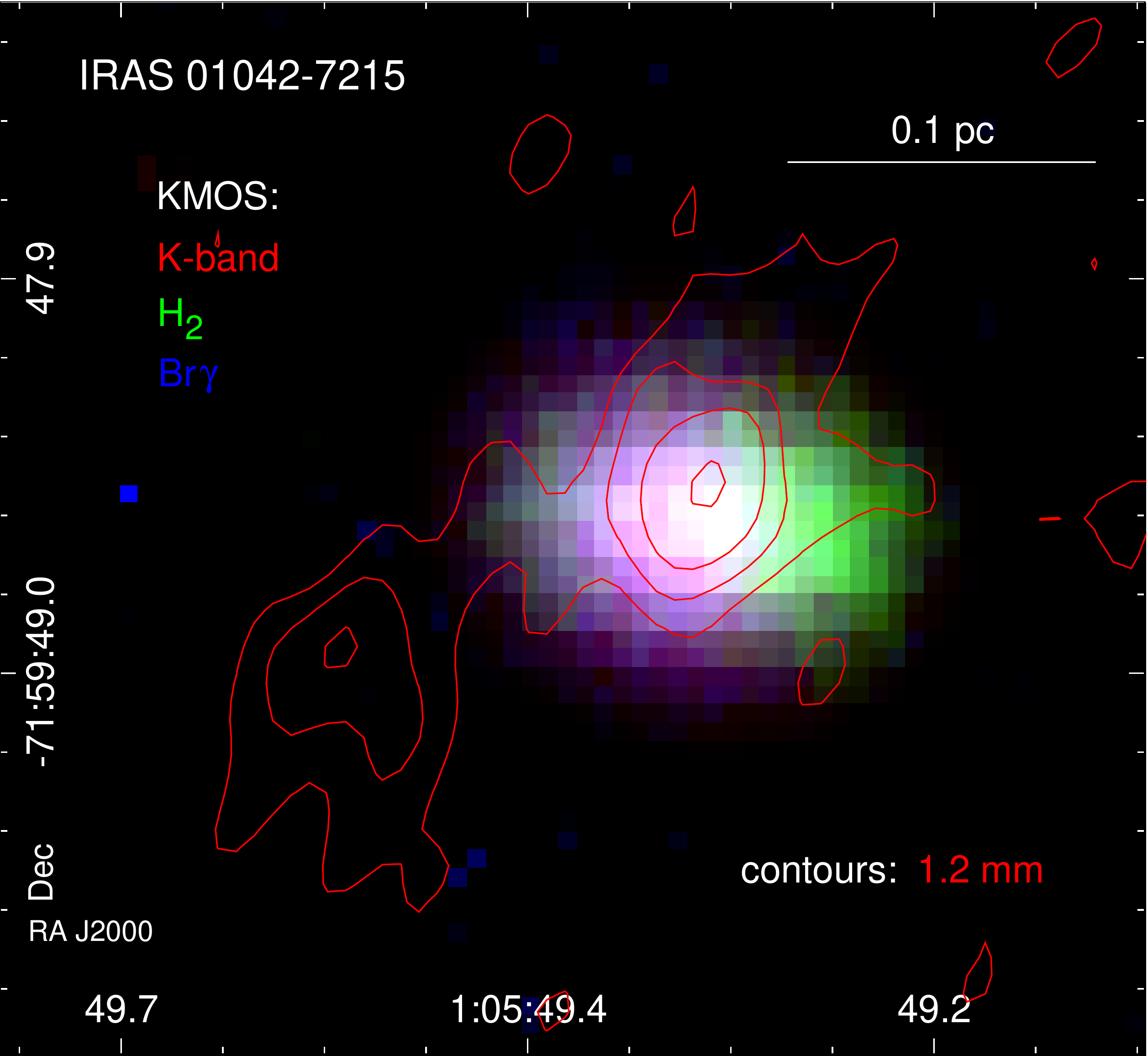} 
  \caption{Three-color mosaic of the IRAS\,01042--7215 field combining the {\it VLT}/SINFONI {\it K}-band (red), \ch{H2} (green), and Br$\gamma$ (blue) images \cite{ward2017}. Red contours correspond to the ALMA 1.2 mm continuum emission with contour levels the same as in Fig.~\ref{f:OtherMosSMC}.}
  \label{f:IRASSINFONI}
\end{figure*}  

 MMS--1 and MMS--2 continuum sources are associated with two near-infrared sources N\,159\,A7 `121' and `123', respectively, detected by \cite{testor2006} using high-resolution ($\sim$0$\rlap.{''}$2) {\it VLT}/NACO observations.  The position of the {\it Spitzer} YSO (YSO--S in \cite{fukui2015}; see Fig.~\ref{f:OtherMosLMCN159}) is offset to the west hinting at the possibility that a cluster is being formed.  No infrared source is detected toward MMS--3, indicating that it is the youngest of the three protostars.  Recent {\it VLT}/KMOS observations (a pixel scale 0$\rlap.{''}$2, a full width at half maximum seeing $\sim$0$\rlap.{''}$4; PI Jacob Ward, see Section~\ref{s:newkmos} for technical details) targeting YSO--S (see Fig.~\ref{f:OtherMosLMCN159}) detected the {\it K}-band continuum, the  Brackett-gamma emission line of hydrogen (Br$\gamma$), and the \ch{H2} emission line toward N\,159\,A7 121 (MMS--1) and 123 (MMS--2).  As the \ch{H2} and Br$\gamma$ lines trace shocks and accretion, respectively, they can shed light on the nature of the sources and thus possible formation scenarios of COMs (e.g., non-thermal origin  in shocks traced by \ch{H2}). The right panel in Fig.~\ref{f:N159WSKMOS} shows a combination of the three KMOS images with the ALMA continuum contours overlaid for reference.  The extended \ch{H2} emission is associated with outflows from N\,159\,A7 121 (MMS--1) and 123 (MMS--2) detected by \cite{tokuda2018} with ALMA. The Br$\gamma$ emission is only detected toward N\,159\,A7 123 (MMS--2).

The most recent ALMA observations of N\,159W--S detected multiple \ch{CH3OH} peaks throughout the region (see Fig.~\ref{f:N159WSmeth}; PI P. Schilke) with the brightest emission associated with the continuum source MMS--3. The left panel in Fig.~\ref{f:N159WSKMOS} shows a three-color mosaic combining the KMOS {\it K}-band, ALMA \ch{CH3OH}  and 1.3 mm continuum emission. No \ch{CH3OH} emission is detected toward MMS--1 and MMS--2 that have infrared counterparts.  Although these observations were designed to detect multiple COMs (e.g, \ch{CH3OH}, \ch{CH3CN} -- methyl cyanide, \ch{HCOOCH3}, \ch{CH3OCH3}), only \ch{CH3OH} was found. The detection of multiple peaks hints at the possibility that there is an underlying more extended distribution of \ch{CH3OH} that is resolved out. 

\begin{figure*}
  \includegraphics[width=0.65\textwidth]{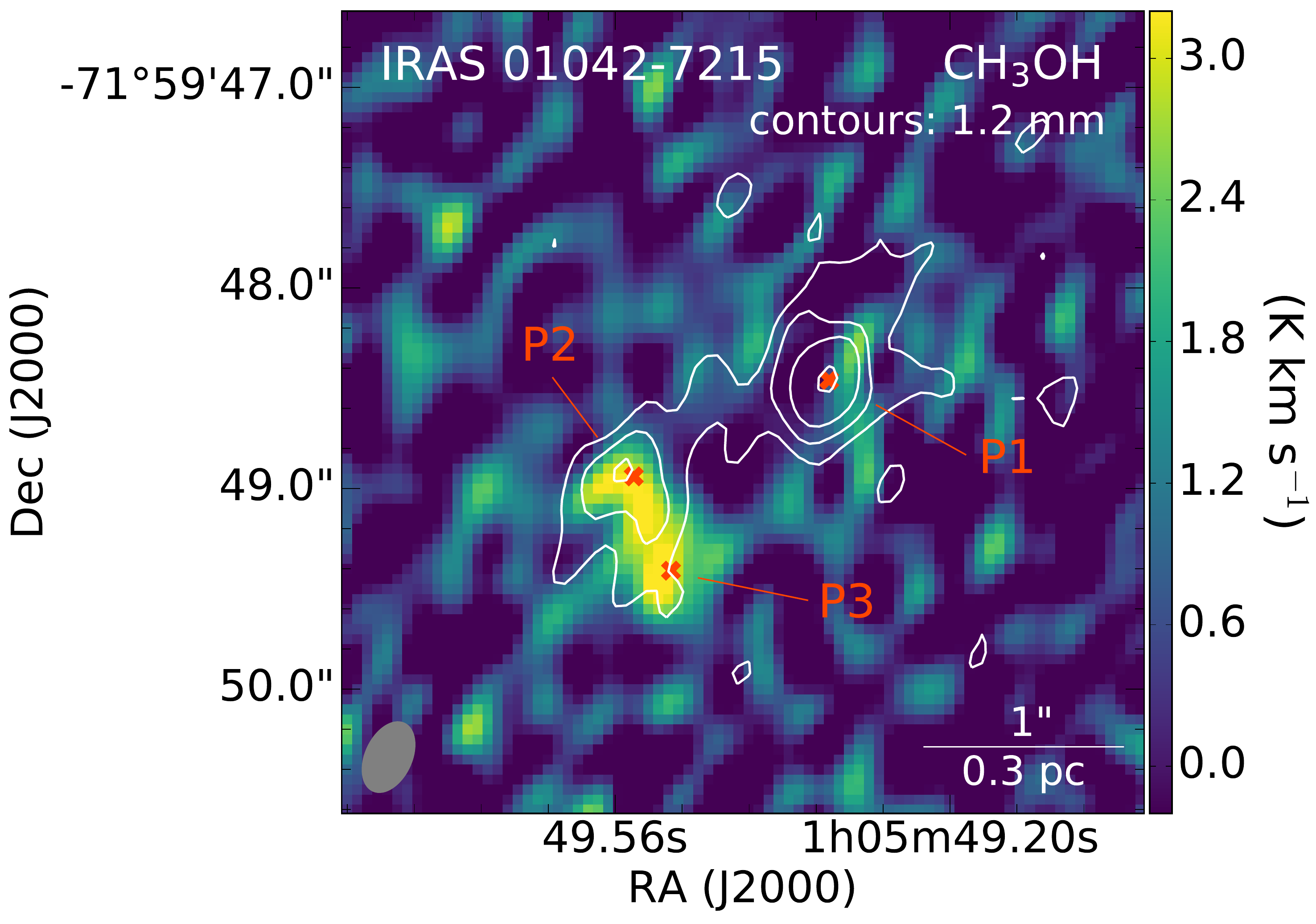} 
  \caption{The \ch{CH3OH} integrated intensity image (the average of the \ch{CH3OH} 5$_{-1,5}-4_{-1,4}$ E and 5$_{0,5}-4_{0,4}$ A$^{+}$ lines) of IRAS  01042$-$7215 in the SMC. The 1.2 mm continuum contours are overlaid; the contour levels and the size of the ALMA synthesized beam shown in the lower left corner are the same as in Fig.~\ref{f:OtherMosSMC}.}
  \label{f:IRASP2P3methanol}
\end{figure*}

Preliminary Local Thermodynamic Equilibrium (LTE) fitting using multiple \ch{CH3OH}  lines ($\sim$241.7--241.9 GHz; see Table~\ref{tbl:transitions}) detected toward the two brightest peaks indicate that the gas is cold ($\sim$14 K).  Methanol forms on grain surfaces and has to be released to the gas phase by energetic events to be detectable. The desorption mechanisms include sublimation by infrared heating by a forming star, photodesorption by UV photons, sputtering by shocks, and chemical desorption (see a discussion in Section~\ref{sec:theory}).  
None of these mechanisms is directly supported by the observations of N\,159W--S: the gas is cold (desorption by the infrared and UV heating is less likely) and the line widths are narrow (broad lines are expected in the shock sputtering scenario). Moreover, the average cosmic ray flux in the LMC is low, making the cosmic ray induced UV radiation less effective than in the Galaxy.   Since the analysis of the ALMA \ch{CH3OH} data is preliminary, we do not provide physical parameters for N159W--S in Table~\ref{tbl:physical}.

\subsubsection{IRAS\,01042$-$7215}
\label{s:IRAS01042}

IRAS\,01042--7215 is a high-mass YSO ($\sim$2$\times$10$^4$ L$_{\odot}$; e.g., \cite{vanloon2008,oliveira2011}; {\it Spitzer} YSO SSTISAGEMA J010549.30--715948.5, \cite{sewilo2013}) located in the northeast region of the SMC bar at the outskirts of the N\,78 star-forming region (see Figs.~\ref{f:example} and \ref{f:OtherMosSMC}). The source is well studied through near-infrared and mid-infrared spectroscopy; absorption bands due to \ch{H2O} ice, \ch{CO2} ice, and silicate dust are detected, while prominent PAH bands and ionized metal lines are not detected \cite{vanloon2008,vanloon2010,oliveira2011,oliveira2013}. No \ch{CH3OH} ice has been detected toward IRAS\,01042--7215.  

Recent {\it VLT}/SINFONI {\it K}-band observations (a pixel scale of 0$\rlap.{''}$1, a full width at half maximum seeing $\sim$0$\rlap.{''}$6; \cite{ward2017}) detected the Br$\gamma$  and \ch{H2} emission lines towards  IRAS\,01042$-$7215 \cite{ward2017}.  Figure~\ref{f:IRASSINFONI} shows a three-color mosaic combining the SINFONI {\it K}-band continuum, \ch{H2}, and Br$\gamma$ images. The SINFONI image reveals a single compact embedded source ($A_{\rm v} = 16\pm8$ mag) with little or no extended line emission. The measured \ch{H2} emission line ratios are consistent with expectations for shocked emission \cite{shull1978}.

The high-resolution ($\sim$0$\rlap.{''}$22--0$\rlap.{''}$37 or $\sim$0.07--0.11 pc at the distance of the SMC) 1.2 mm ALMA observations detected two continuum peaks -- `P1' and `P2/P3' toward IRAS\,01042--7215. P2/P3 is further resolved into two sources (P2 and P3) corresponding to the \ch{CH3OH} emission peaks where multiple \ch{CH3OH} transitions are detected (Fig.~$\ref{f:IRASP2P3methanol}$; \cite{shimonishi2018}). 

The 1.2 mm continuum source P1 is associated with the high-mass YSO corresponding to the IRAS source, while no infrared sources are detected at the positions of P2 and P3 (see Fig.~\ref{f:IRASSINFONI}  and \cite{shimonishi2018}). P2 and P3 correspond to the two \ch{CH3OH} emission peaks in the region; no \ch{CH3OH} is detected toward P1 (Fig.~\ref{f:IRASP2P3methanol} and \cite{shimonishi2018}).  Besides \ch{CH3OH}, CS, C$^{33}$S, SO, \ch{SO2}, H$^{13}$CO$^{+}$, H$^{13}$CN, SiO are detected toward P2/P3, but no COMs larger than \ch{CH3OH}. 

\begin{figure*}[ht!]
  \includegraphics[width=\textwidth]{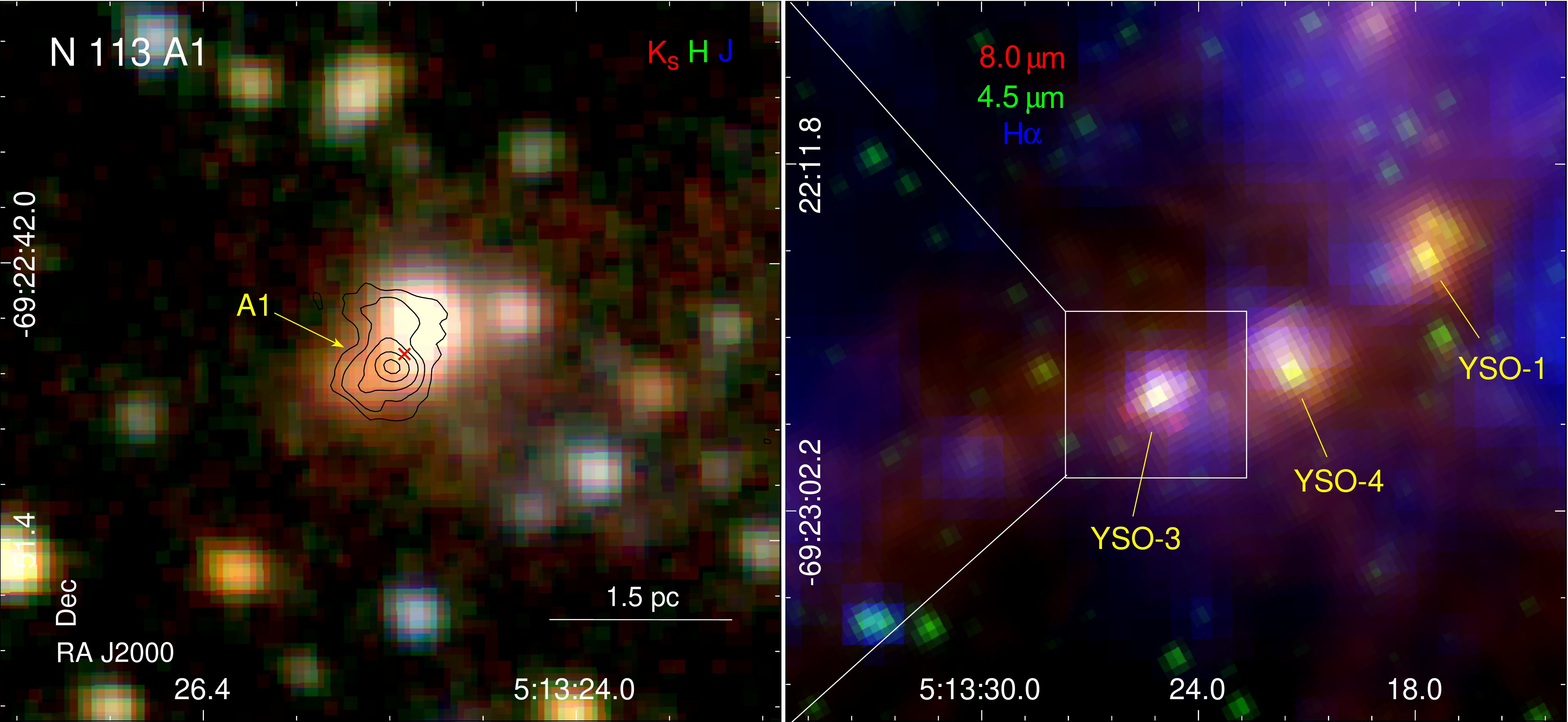}  
   \includegraphics[width=\textwidth]{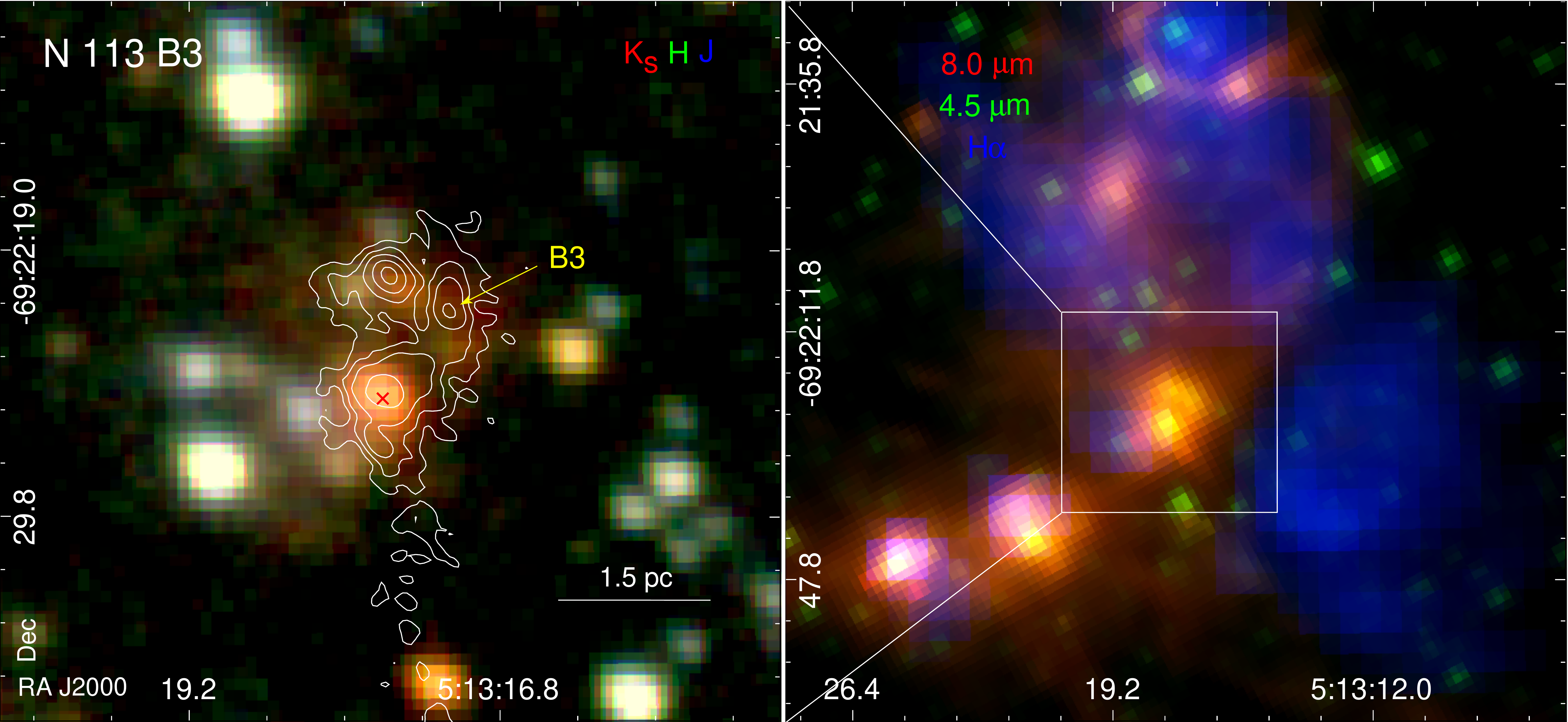} 
  \caption{{\it Left:} The same as in Fig.~\ref{f:OtherMosLMCST11}, but for N\,113 A1 ({\it top}) and B3 ({\it bottom}) hot cores. The contours in the left panel correspond to the 1.3 mm continuum emission \cite{sewilo2018}. The contour levels are (5, 10, 20, 50) $\times$ the image rms noise level of 0.1 mJy beam$^{-1}$.} 
  \label{f:N113mos}
\end{figure*}

The rotation diagram analysis of the \ch{CH3OH} lines for IRAS\,01042--7215 P2 and P3 suggests that the temperature of the \ch{CH3OH} gas is low ($\sim$20~K),  well below the sublimation temperature of the \ch{CH3OH} ice ($\sim$80~K; \cite{tielens2005}). This is another example of a source in the Magellanic Clouds with the detection of `cold methanol'.  The \ch{CH3OH} abundances relative to \ch{H2} for P2 and P3 are an order of magnitude lower than those for hot cores A1 and B3 in N\,113 (see Table~\ref{tbl:physical} and Section~\ref{s:N113}). The origin of \ch{CH3OH} gas in P2/P3 is still under debate. \cite{shimonishi2018} suggest that a possible origin could be sputtering of the ice mantles by shocks triggered by the outflow from nearby YSOs, or chemical desorption of \ch{CH3OH} from dust surfaces.  The presence of shocked but cold \ch{CH3OH} gas has been suggested in Galactic infrared dark clouds, indicating that \ch{CH3OH} may be tracing relatively old shocks (e.g., \cite{sakai2010}).

\subsection{Hot Cores with COMs More Complex than Methanol: LMC N\,113 A1 and B3}
\label{s:N113}

To date, COMs with more than six atoms have only been reliably detected in the Magellanic Clouds toward two sources, both in the LMC N\,113 star-forming region. The physical and chemical characteristics of N\,113 make it one of the most interesting star-forming regions in the LMC.   Like N\,159, it was the target of detailed spectral line studies with single-dish telescopes (see Section~\ref{s:singledish}) that resulted in the detection of \ch{CH3OH} \cite{wang2009}.

N\,113 contains:  (1) one of the most massive ($\sim$10$^{5}$ M$_{\odot}$) and richest GMCs in the LMC \cite{wong2011}; (2) signatures of both recent (H$\alpha$ emission) and ongoing (e.g., maser and bright IR emission) star formation \cite{wong2006,oliveira2006}; (3) signatures of star formation triggered by winds from massive stars; (4) the largest number of H$_2$O and OH masers and the brightest H$_2$O maser in the entire LMC \cite{whiteoak1986,lazendic2002,oliveira2006,green2008,ellingsen2010}; (5) and it has among the highest concentrations of {\it Spitzer}/{\it Herschel} Stage 0--II  YSO candidates in the LMC \cite{whitney2008,gruendl2009,sewilo2010,carlson2012,seale2014}.

\begin{figure*}[ht!]
  \includegraphics[width=0.46\textwidth]{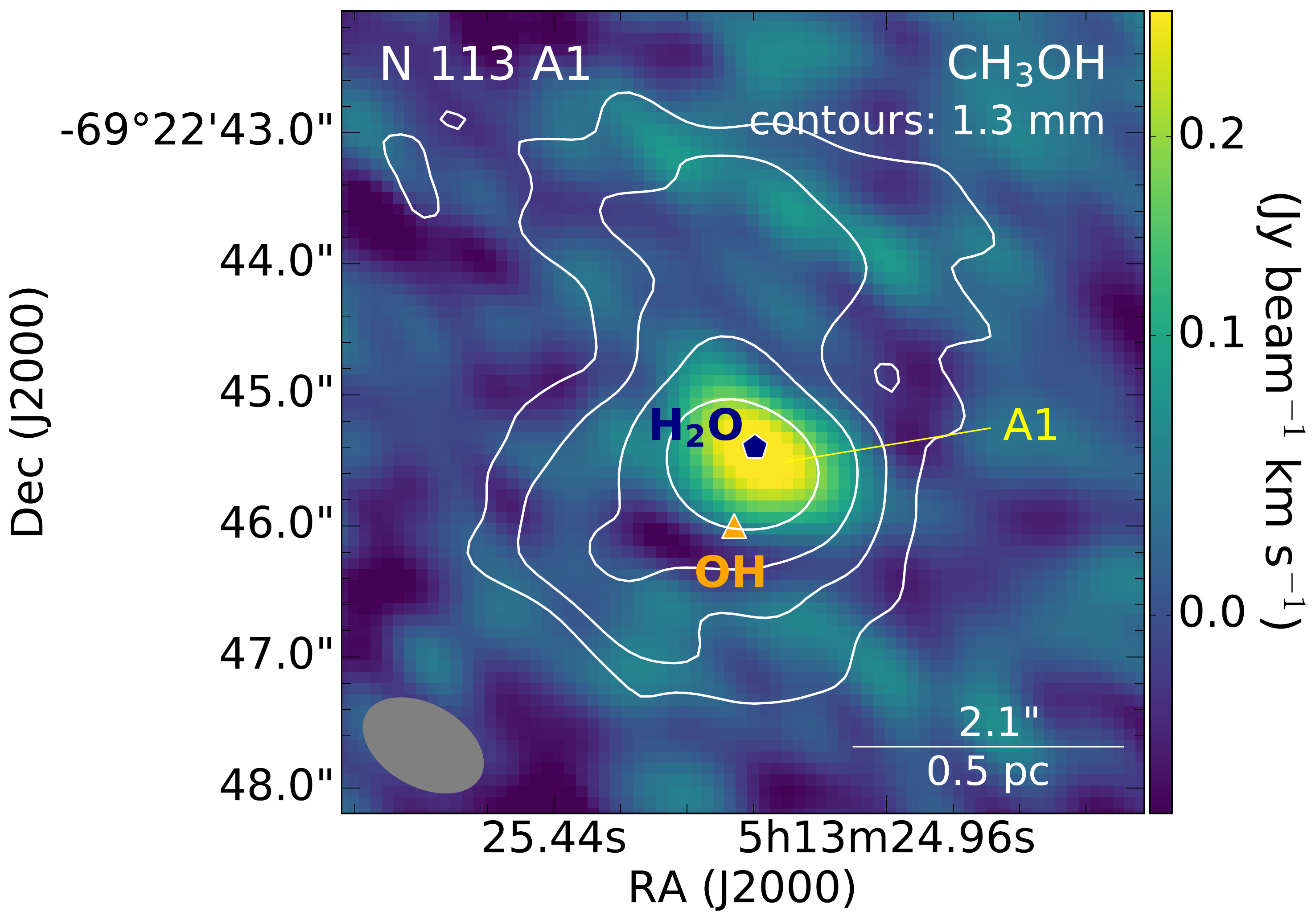} 
  \includegraphics[width=0.46\textwidth]{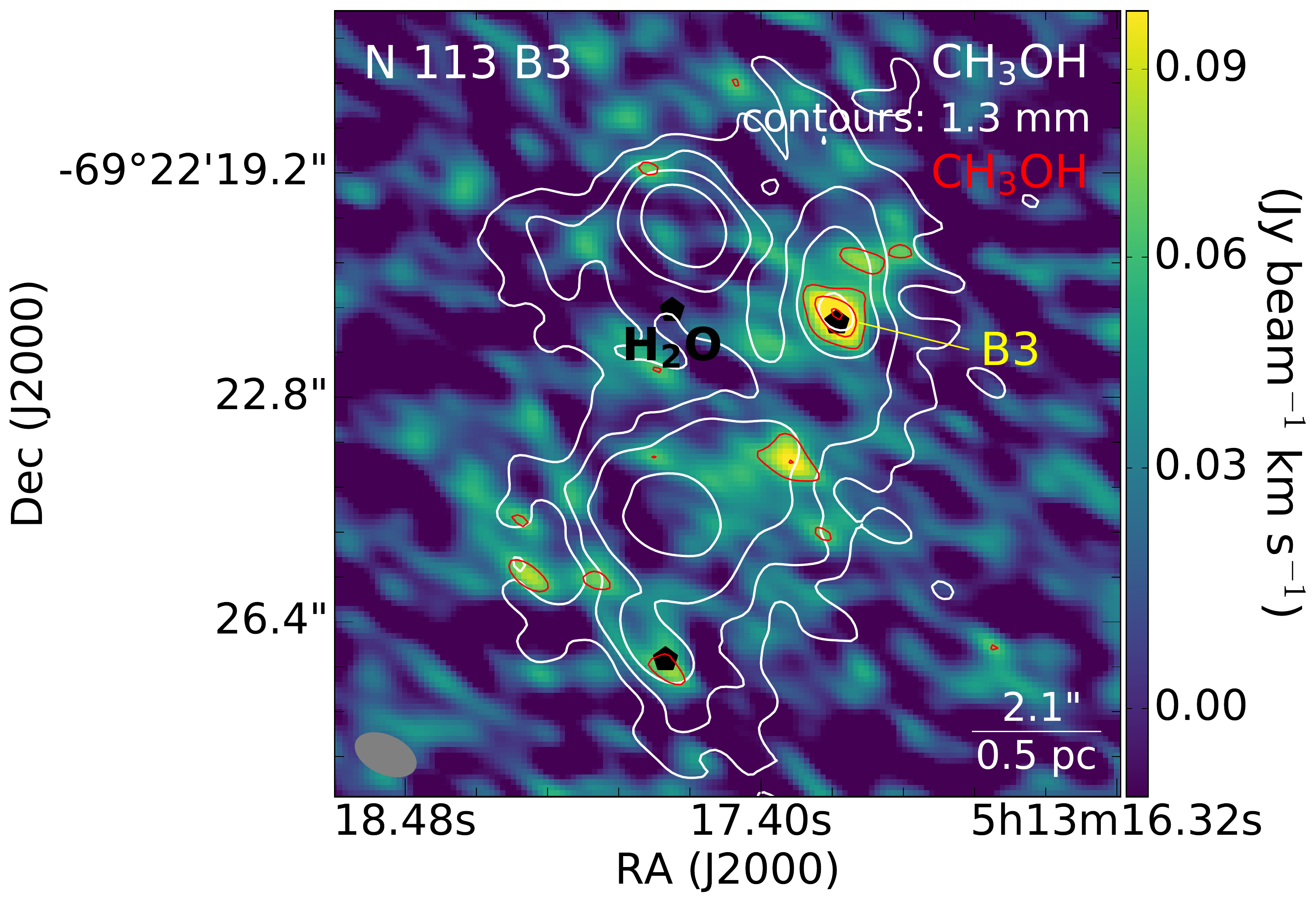} 
  \includegraphics[width=0.46\textwidth]{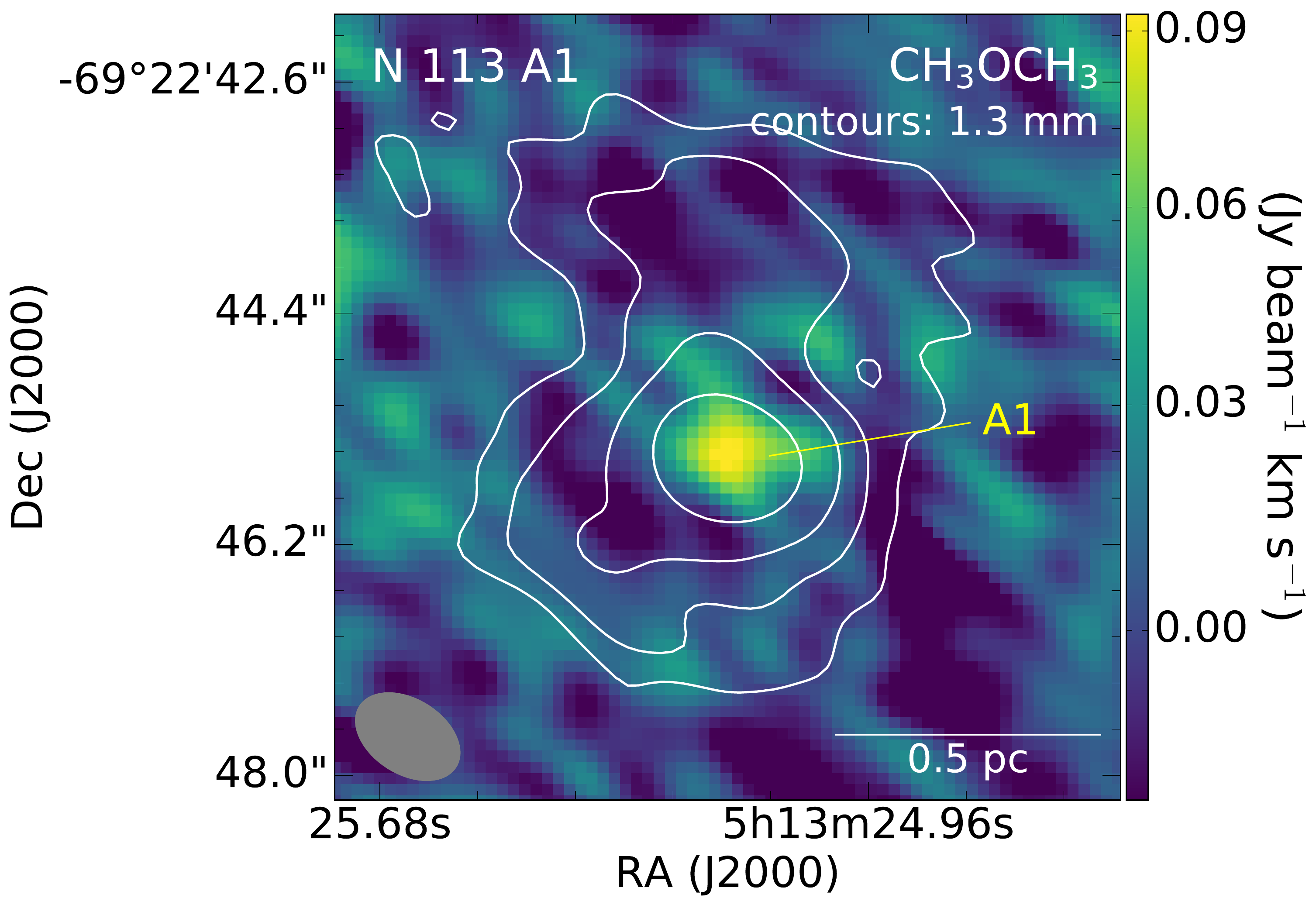} 
  \includegraphics[width=0.46\textwidth]{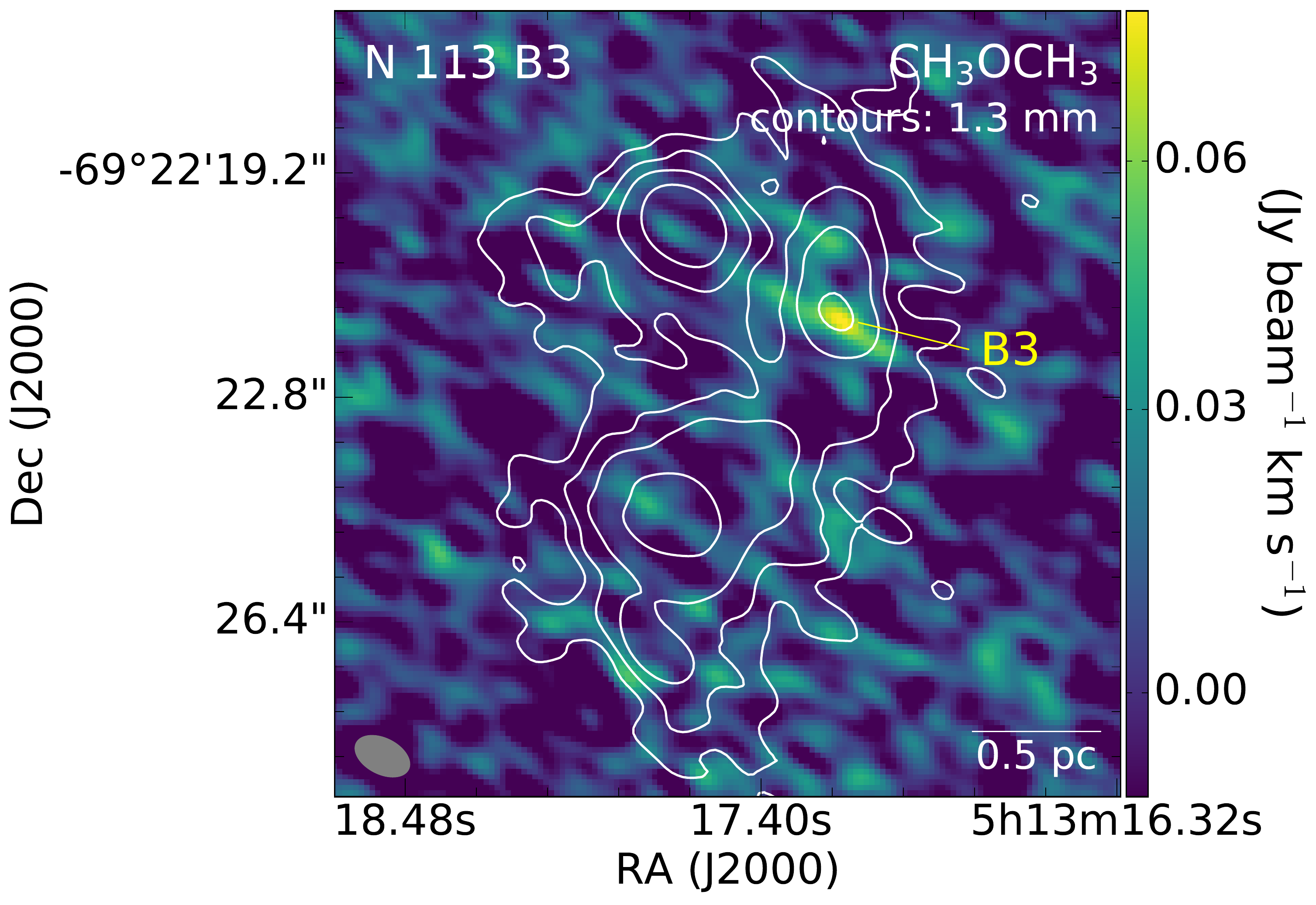} 
  \includegraphics[width=0.46\textwidth]{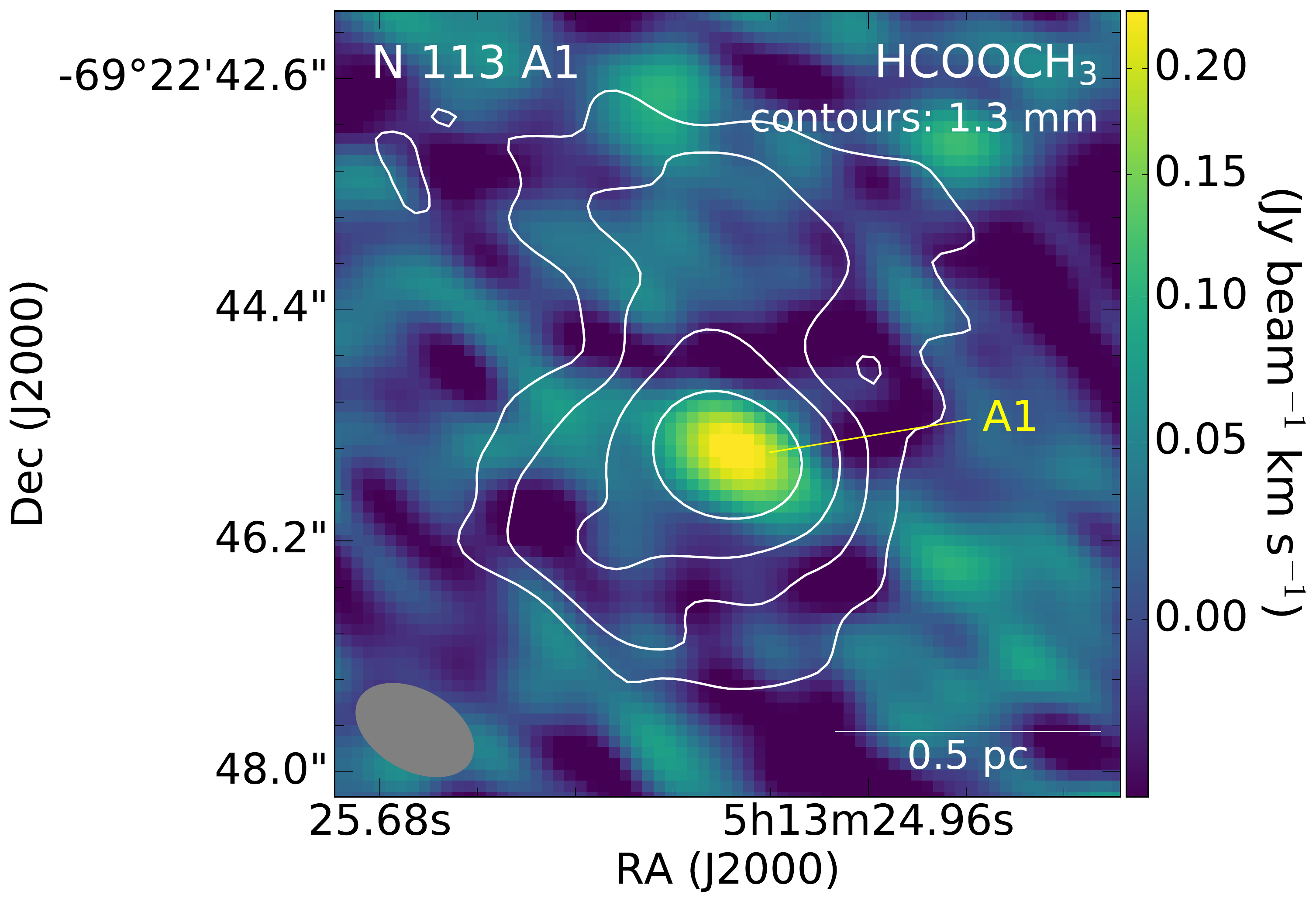} 
  \includegraphics[width=0.46\textwidth]{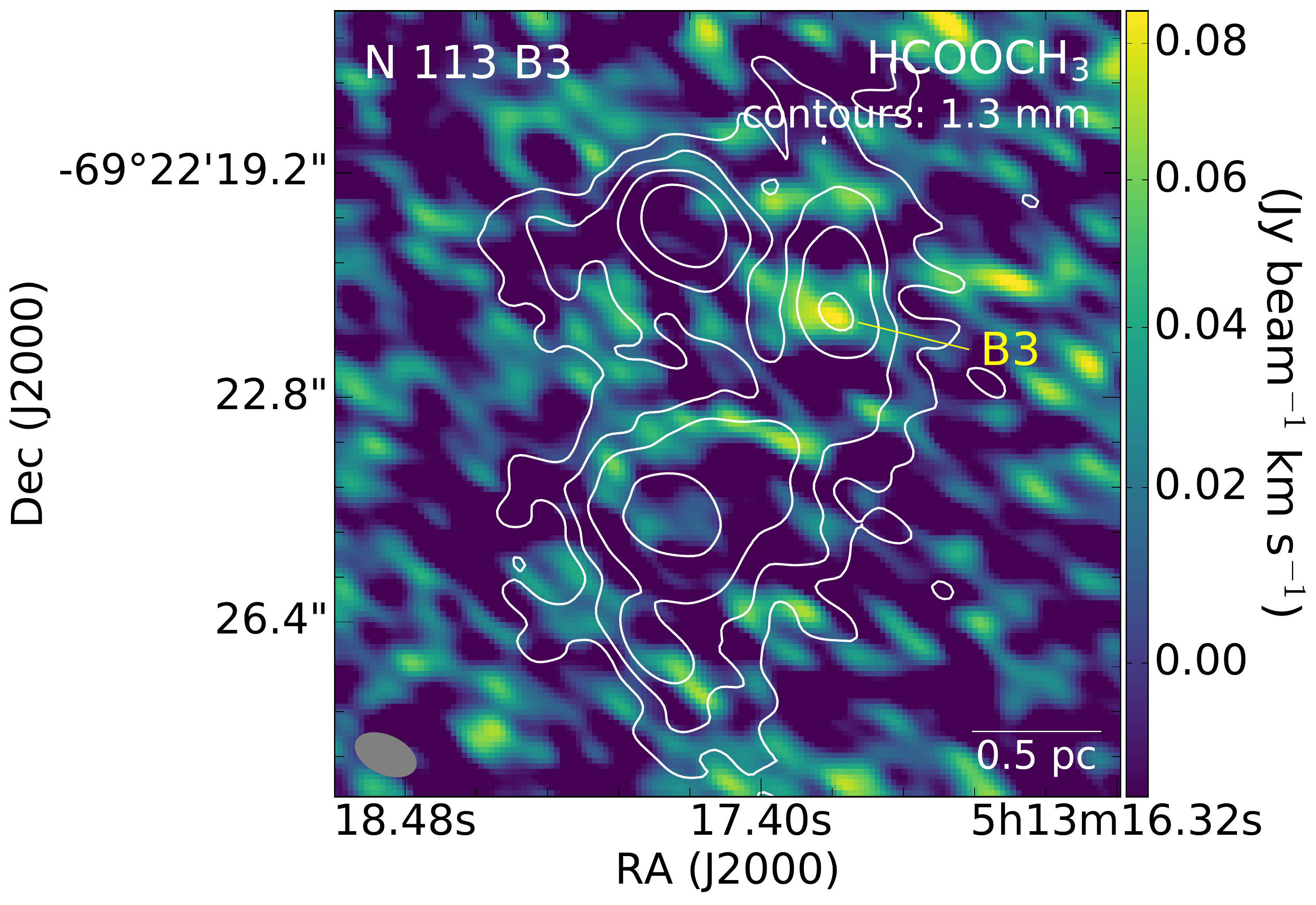} 
  \caption{Integrated intensity images for \ch{CH3OH} ({\it top panel}; made using the channels corresponding to all \ch{CH3OH} transitions in the 216.9 GHz spectral window; see Table~\ref{tbl:transitions}), \ch{CH3OCH3} ({\it middle panel}; the 13$_{0,13}$--12$_{1,12}$ transition); and \ch{HCOOCH3} ({\it bottom panel}; integrated over all detected transitions) for the A1 ({\it left column}) and B3 ({\it right column}) hot cores in the N\,113 star-forming region. The \ch{H2O} and \ch{OH} masers are indicated. The white contours correspond to the 1.3 mm continuum emission; the contour levels are the same as in Fig.~\ref{f:N113mos}. The red contours in the top right panel correspond to the \ch{CH3OH} emission with contour levels of (3, 5, 7) $\times$ 21 mJy beam$^{-1}$, the rms noise level of the \ch{CH3OH} integrated intensity image. The image shows the \ch{CH3OH} detection in other spots in the region (`Region B' in \cite{sewilo2018}) where no other COMs are detected. The results presented in this figure were reported in \cite{sewilo2018}. The ALMA beam size is shown in the lower left corner in each image.} 
  \label{f:N113COMs}
\end{figure*}

\begin{figure*}[ht!]
  \includegraphics[width=0.49\textwidth]{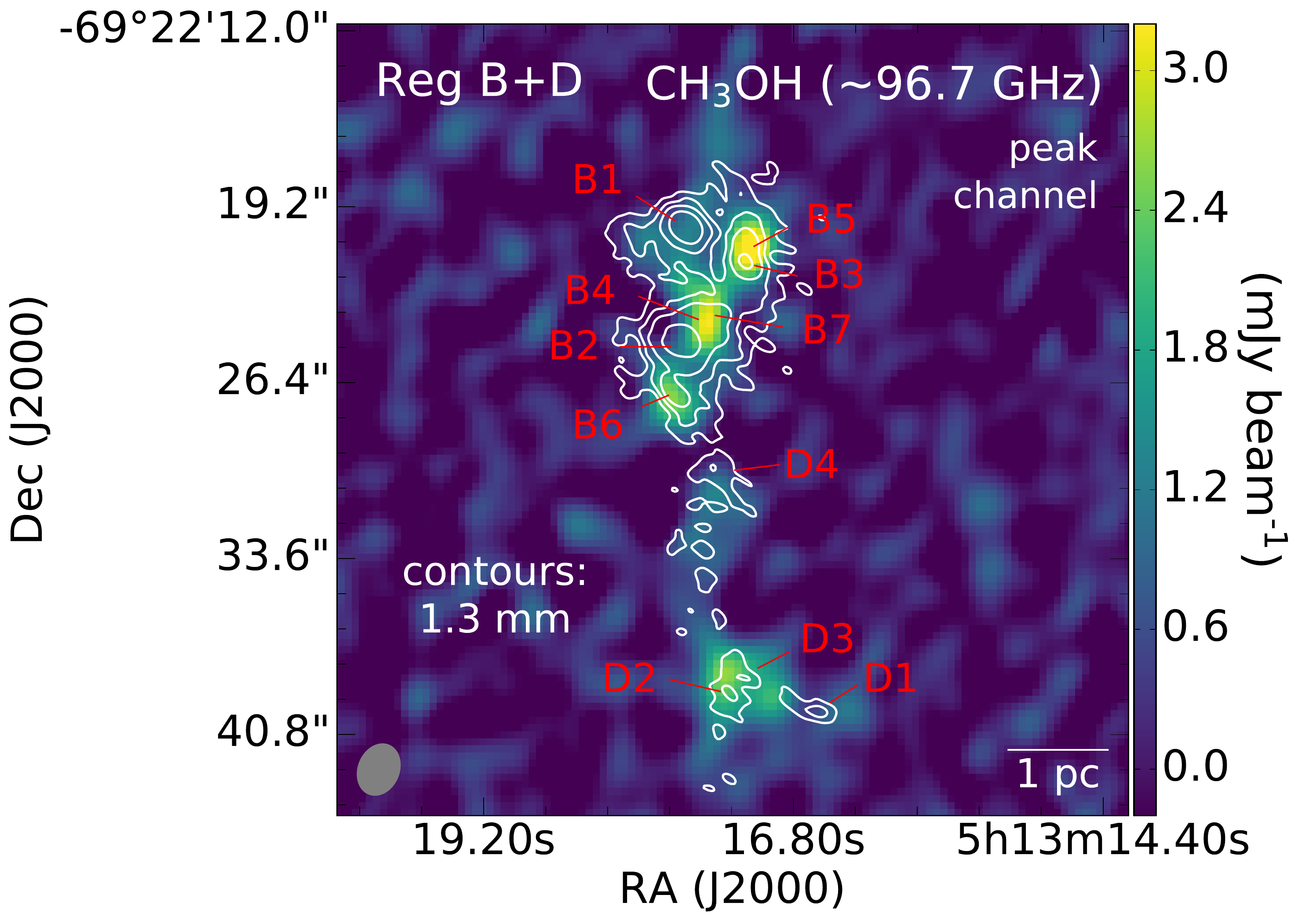}  
  \includegraphics[width=0.49\textwidth]{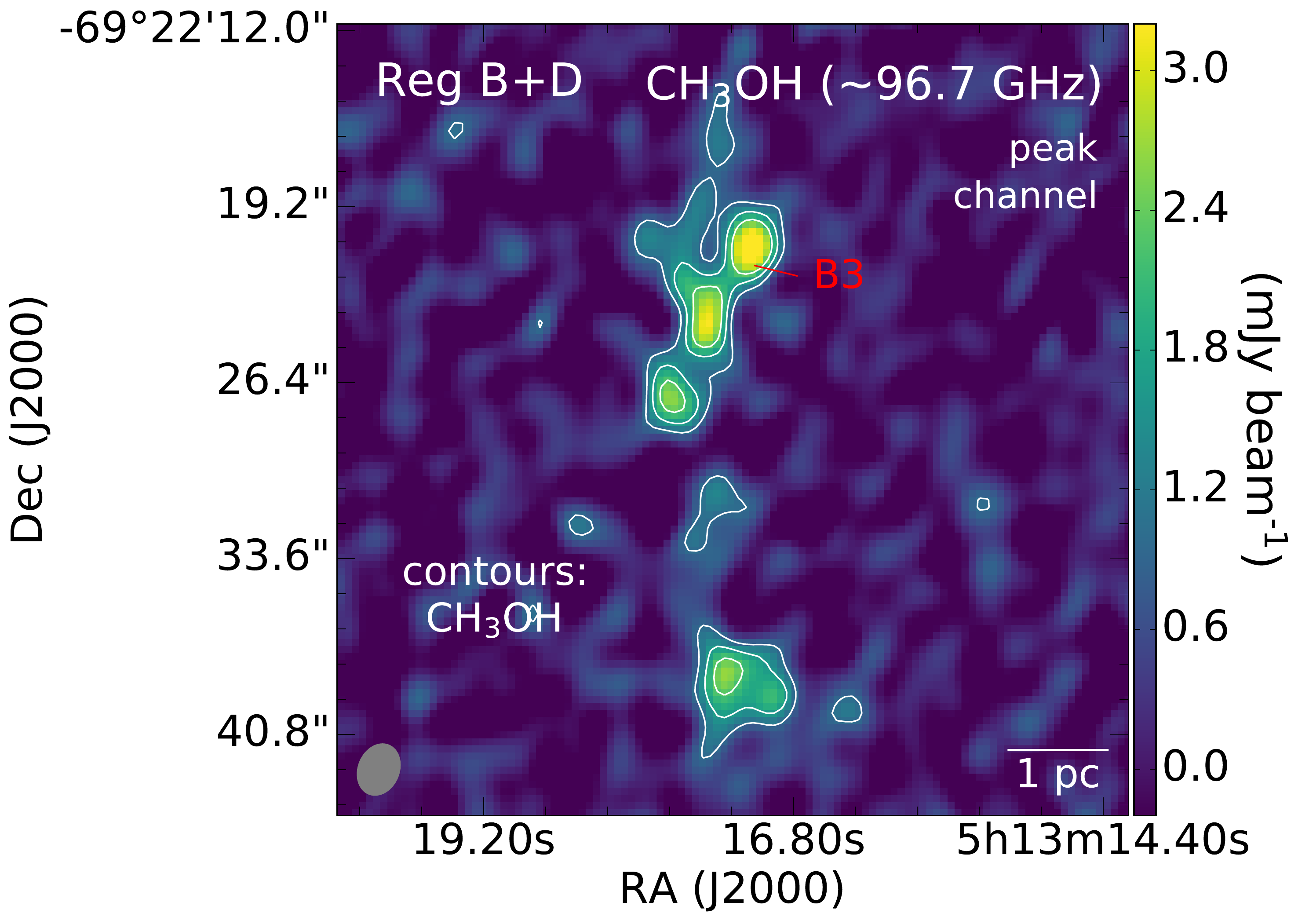}   
  \includegraphics[width=0.49\textwidth]{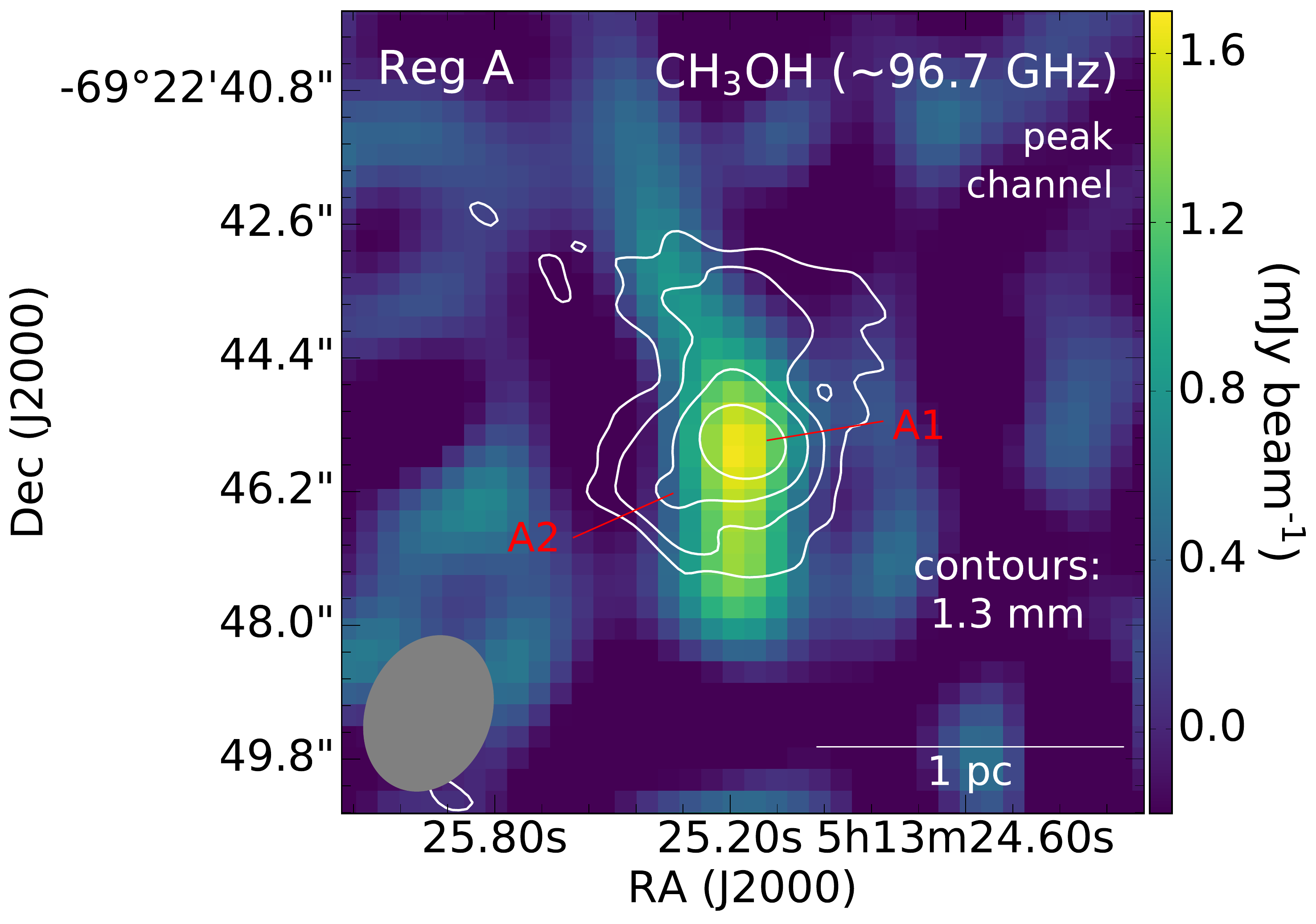} 
   \includegraphics[width=0.49\textwidth]{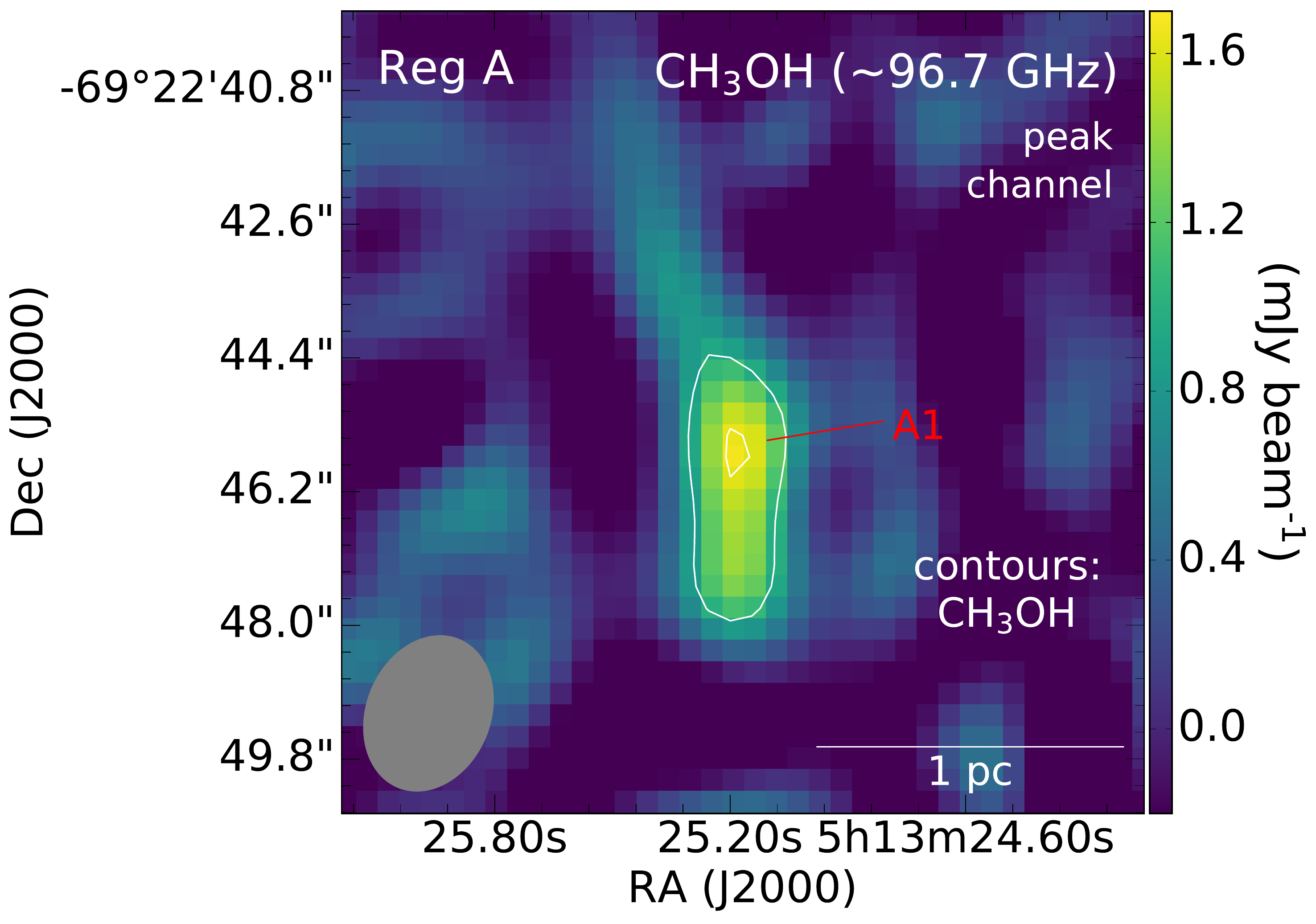} 
  \caption{The single channel maps corresponding to the peak of the unresolved rotational transition quartet $J=2_K-1_K$ of \ch{CH3OH} at 96.7 GHz for regions in N\,113 around and south of B3 ({\it top}) and toward A1 ({\it bottom}) hot cores (Sewi{\l}o et al., in prep.; ALMA 12m--Array data).  The white 1.3 mm continuum contours in the left panel are the same as in Fig.~\ref{f:N113mos}. The white \ch{CH3OH} contours in the right panel correspond to (3, 5, 7) $\times$ rms noise level of the \ch{CH3OH}  single-channel map of 0.32 mJy beam$^{-1}$.  The ALMA beam size (2$\rlap.{''}$15 $\times$ 1$\rlap.{''}$66) for the \ch{CH3OH} observations is shown in the lower left corner in all images.}
  \label{f:N113extendedmeth}
\end{figure*}

Mid-IR sources, maser emission, and compact H\,{\sc ii} regions in the central part of the GMC associated with the extended gas and dust emission, reveal sites of the current star formation in N\,113 (see Fig.~\ref{f:N113mos}).  The {\it Spitzer} (3.6--160 $\mu$m) and {\it Herschel} (160--500 $\mu$m) images with resolutions ranging from  $\sim$2$''$ to 38$''$ show that this region is dominated by three bright sources, which were identified as 30--40 M$_{\odot}$ Stage I YSOs (YSO--1, YSO--3,  and YSO--4 in Fig.~\ref{f:N113mos}: {\it Spitzer} sources 051317.69$-$692225.0, 051325.09$-$692245.1,  and 051321.43$-$692241.5, respectively; \cite{gruendl2009}) and are characterized by distinct physical conditions  (\cite{gruendl2009,ward2016}; J. M. Oliveira, in prep.).

Detailed {\it Spitzer} and {\it Herschel} studies on the YSO population in N\,113 were followed by ALMA Band 6 ($\sim$1.3 mm) observations at $\sim$0.18 pc (or $\sim$0$\rlap.{''}$8) resolution in the molecular transitions that probe a wide density range (10$^2$--10$^7$ cm$^{-3}$) to study both the dense clumps/cores and the lower density gas in the inter-clump regions. The ionized gas and outflow tracers were also included in the program.  These observations resulted in a serendipitous discovery of COMs methanol (CH$_3$OH), methyl formate (HCOOCH$_3$), and dimethyl ether (CH$_3$OCH$_3$) toward two locations in the region (\cite{sewilo2018}; see Figs.~\ref{f:N113mos}--\ref{f:N113COMs}).  The detected transitions are listed in Table~\ref{tbl:transitions}. This was the first conclusive detection of COMs more complex than methanol outside the Galaxy and in a low-metallicity environment.  COMs were detected toward two 1.3 mm continuum sources `A1' and `B3' in the field around {\it Spitzer} YSO--3 (`Region A' in \cite{sewilo2018}) and YSO--1 (`Region B').  Figure~\ref{f:N113mos} shows three-color mosaics covering A1 and B3 combining near-infrared bands with 1.3 mm contours overlaid, as well as a larger scale environment in N\,113 traced by the {\it Spitzer} 4.5 and 8.0 $\mu$m, and H$\alpha$ emission.  Figure~\ref{f:N113COMs} shows the integrated intensity \ch{CH3OH}, \ch{CH3OCH3}, and \ch{HCOOCH3} images of Regions A and B.   

\begin{figure*}[ht!]
  \includegraphics[width=0.48\textwidth]{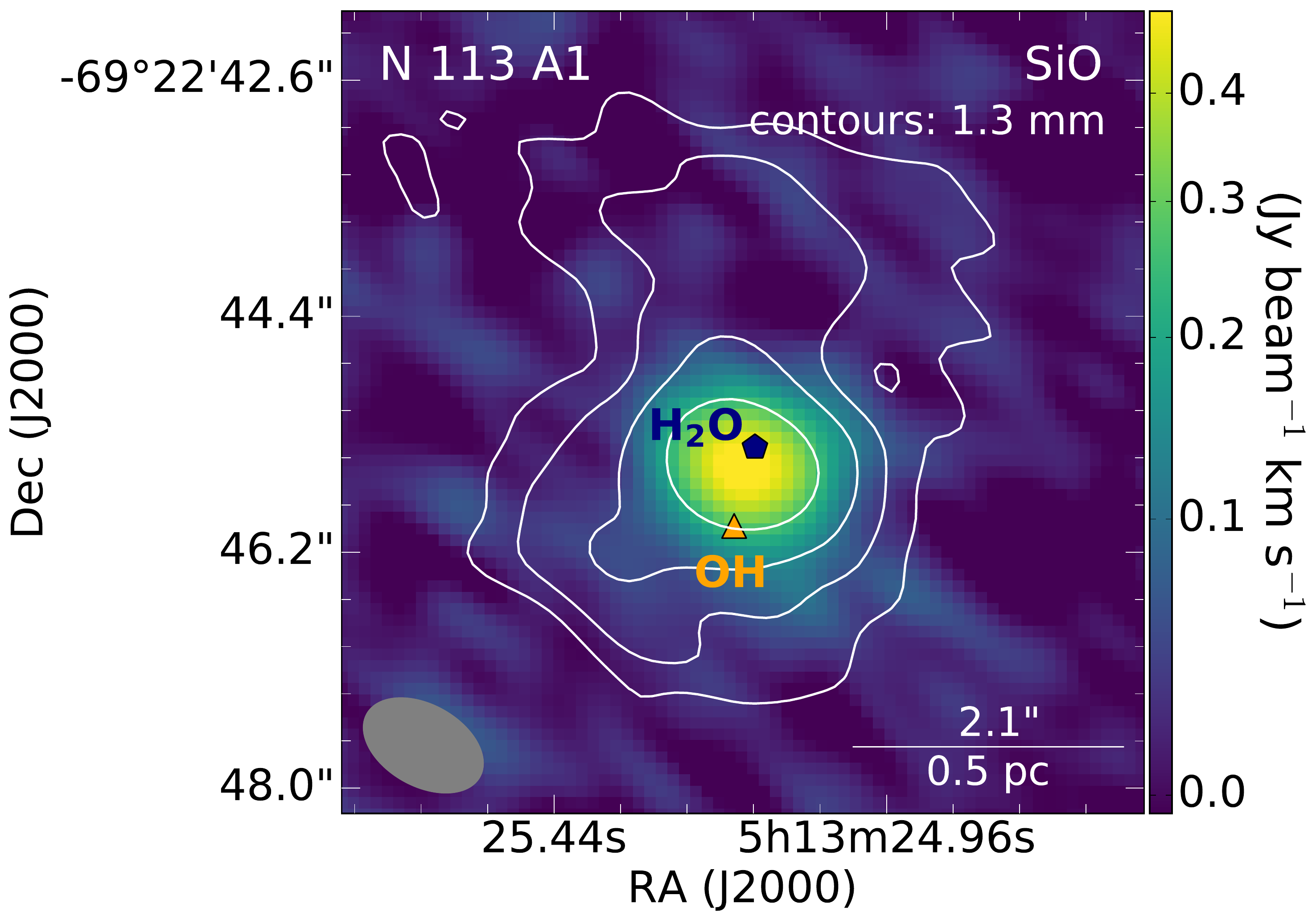} 
  \includegraphics[width=0.48\textwidth]{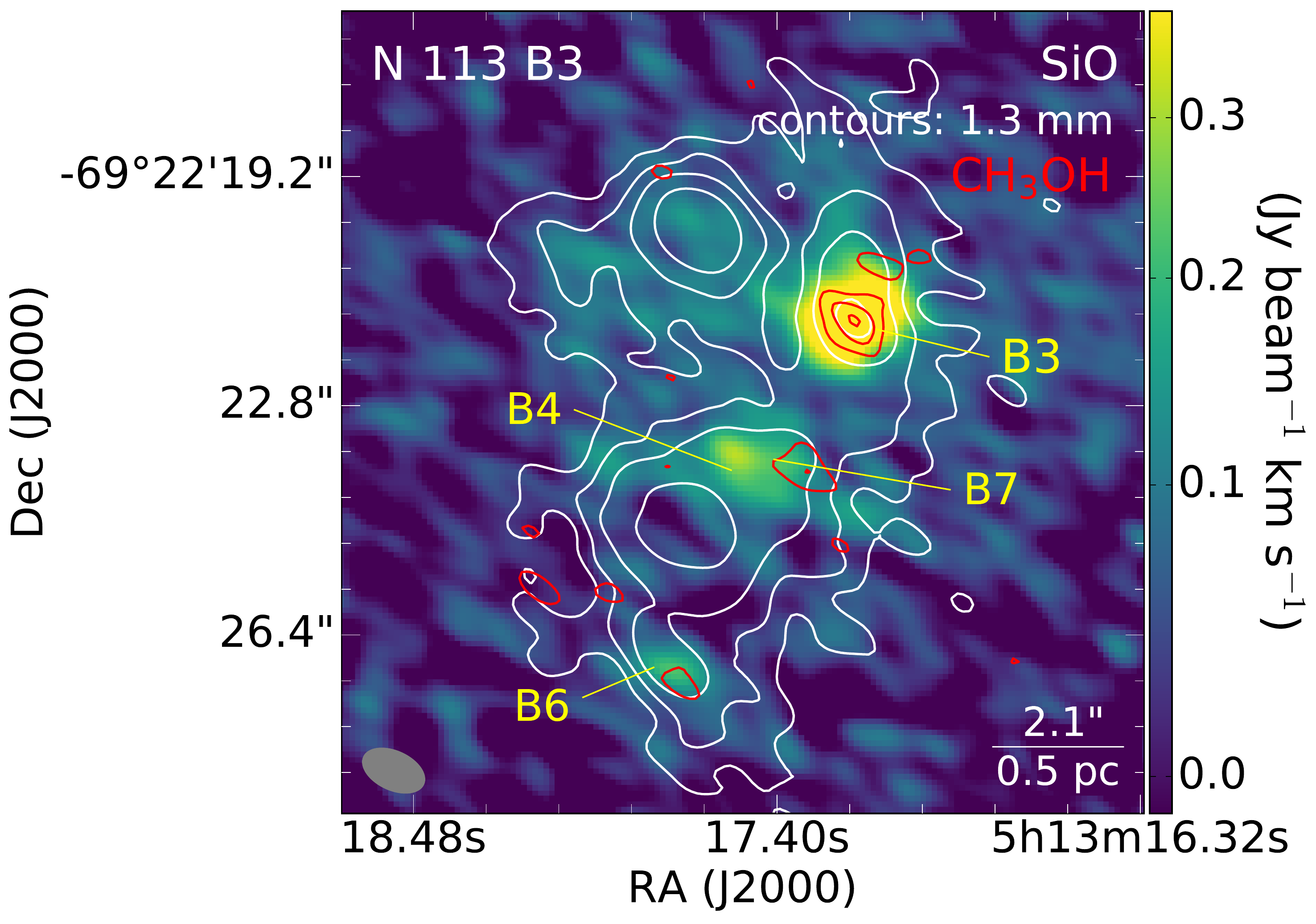} 
  \caption{The SiO (5--4) integrated intensity images of N\,113 A1 ({\it left}) and B3 ({\it right}) hot cores (Sewi{\l}o et al., in prep.). The contours are the same as in Fig.~\ref{f:N113mos} (1.3 mm; {\it white}) and Fig.~\ref{f:N113COMs} (\ch{CH3OH}; {\it red}). The positions of \ch{H2O} and \ch{OH} masers, and 1.3 mm continuum sources associated with the \ch{CH3OH} emission are indicated. The ALMA beam size (0$\rlap.{''}$98 $\times$ 0$\rlap.{''}$61) is shown in the lower left corner in both images.}
  \label{f:N113SiO}
\end{figure*}

A1 and B3 are compact (diameter $D\sim0.17$ pc) and are associated with H$_2$O (A1 and B3) and OH (A1) masers.   H$_2$ column densities of  $(8.0\pm1.2)\times10^{23}$~cm$^{-2}$  and $(7.0\pm0.9)\times10^{23}$~cm$^{-2}$ and number densities of $\sim$1.6$\times$10$^6$~cm$^{-3}$ and $\sim$1.4$\times$10$^6$~cm$^{-3}$  for A1 and B3, respectively, were estimated using the 1.3 mm continuum data.  The LTE analysis of six \ch{CH3OH} transitions resulted in  rotational temperatures and total column densities ($T_{\rm rot}$, $N_{\rm rot}$) of 
($134\pm6~{\rm K}, 1.6\pm0.1\times10^{16}~{\rm cm}^{-2}$) and 
($131\pm15~{\rm K}, 6.4\pm0.8\times10^{15}~{\rm cm}^{-2}$) 
for A1 and B3, respectively \cite{sewilo2018}.  
These sizes, number and column densities, and temperatures of A1/B3 are consistent with classic hot cores observed in the Galaxy.  COMs emission and association with masers are also among the main characteristics of hot cores.  These are the first bona fide detections of complex hot core chemistry outside the Galaxy.  

\begin{table*}[ht!]
\caption{Chemical Abundances Relative to H$_2$ for COMs Detected in the LMC/SMC for Selected Galactic Hot Cores}
\centering 
\label{tbl:galactic}
\begin{tabular}{lcccc}
\hline\hline
\multicolumn{1}{c}{Source} & \multicolumn{3}{c}{$N(X)$/$N({\rm H_2})$} \\
                           & \ch{CH3OH} & \ch{HCOOCH3} & \ch{CH3OCH3} & Ref.\textsuperscript{\emph{a}} \\
\\
\hline
Orion Mol. Cloud:  & & & &  \\
\hspace{0.4cm} Hot core \hfill & 1.4 $\times$ 10$^{-7}$ & 1.4 $\times$ 10$^{-8}$ & 8.0 $\times$ 10$^{-9}$ & 1\\
\hspace{0.4cm} Compact Ridge \hfill & 4.0 $\times$ 10$^{-7}$ & 3.0 $\times$ 10$^{-8}$ & 1.9 $\times$ 10$^{-8}$ &  1\\
G34.26$-$0.15:  & & & &  \\
\hspace{0.4cm} NE \hfill & 8.5 $\times$ 10$^{-7}$ & 7.3 $\times$ 10$^{-8}$ & 1.4 $\times$ 10$^{-7}$ & 1 \\
\hspace{0.4cm} SE \hfill & 6.4 $\times$ 10$^{-7}$ & 6.8 $\times$ 10$^{-8}$ & 8.5 $\times$ 10$^{-8}$ & 1\\
Sgr B2N & 2.0 $\times$ 10$^{-7}$ & 1.0 $\times$ 10$^{-9}$ & 3.0 $\times$ 10$^{-9}$ & 1 \\
G327.3$-$0.6  & 2.0 $\times$ 10$^{-5}$ & 2.0 $\times$ 10$^{-6}$ & 3.4 $\times$ 10$^{-7}$ & 1 \\
AFGL\,2591 & 7.0 $\times$ 10$^{-7}$ & $<$3.2 $\times$ 10$^{-7}$ & $<$1.0 $\times$ 10$^{-7}$ & 2 \\
G24.78$+$0.08 & 6.5 $\times$ 10$^{-7}$ & 7.3 $\times$ 10$^{-8}$ & 3.0 $\times$ 10$^{-7}$ & 2 \\
G75.78$+$0.34 & 9.2 $\times$ 10$^{-7}$ & 5.9 $\times$ 10$^{-8}$ & 1.9 $\times$ 10$^{-7}$ & 2 \\
NGC\,6334 IRS\,1 & 4.0 $\times$ 10$^{-6}$ & 4.6 $\times$ 10$^{-7}$ & 2.4 $\times$ 10$^{-6}$ & 2 \\
NGC\,7538 IRS\,1 & 5.7 $\times$ 10$^{-7}$ & 6.7 $\times$ 10$^{-8}$ & $<$7.6 $\times$ 10$^{-8}$ & 2 \\
W3(H$_2$O) & 5.4 $\times$ 10$^{-6}$ & 2.9 $\times$ 10$^{-7}$ & 8.3 $\times$ 10$^{-7}$ & 2 \\
W33A & 7.3 $\times$ 10$^{-7}$ & 9.6 $\times$ 10$^{-8}$ & 1.0 $\times$ 10$^{-7}$ & 2 \\
\hline
\end{tabular}

\textsuperscript{\emph{a}} References: (1) \cite{mookerjea2007} and references therein; (2) \cite{bisschop2007}. For more information on Galactic hot cores, see also, e.g., \cite{kurtz2000,herbst2009,widicus2017} and references therein. 
\end{table*}

The (CH$_3$OH, CH$_3$OCH$_3$, HCOOCH$_3$) column densities are  $(16\pm1, 1.8\pm0.5, 1.1\pm0.2)\times10^{15}$ cm$^{-2}$ for A1 and $(6.4\pm0.8, 1.2\pm0.4, <0.34)\times10^{15}$ cm$^{-2}$ for B3.  Column densities for  \ch{HCOOCH3} and \ch{CH3OCH3} were estimated  using the same $T_{\rm rot}$ as for CH$_3$OH,  assuming that these molecular species are located in the same region as \ch{CH3OH} in A1 and B3. 

The (CH$_3$OH, CH$_3$OCH$_3$, HCOOCH$_3$) fractional abundances with respect to \ch{H2} are 
$(20\pm3, 2.2\pm0.7,1.4\pm0.4)\times10^{-9}$ for A1 and 
$(9.1\pm1.7,1.7\pm0.7,<0.5)\times10^{-9}$ for B3 (see Table~\ref{tbl:physical}).  
The \ch{CH$_3$OH} fractional abundances in A1/B3 are over an order of magnitude larger than an upper limit estimated for ST\,11 by \cite{shimonishi2016b} (see Section~\ref{s:ST11}).  Table~\ref{tbl:galactic} lists CH$_3$OH, CH$_3$OCH$_3$, HCOOCH$_3$ fractional abundances for a set of Galactic hot cores. When scaled by a factor of 2.5 to account for the lower metallicity in the LMC (the ratio between the solar metallicity and the metallicity of the LMC, assuming $Z_{\rm LMC}$ = 0.4 $Z_{\odot}$),  the abundances of COMs detected in N\,113 are comparable to those found at the lower end of the range in Galactic hot cores.  This was a surprising result because previous observational and theoretical studies indicated that the abundance of \ch{CH3OH} in the LMC is very low.  \cite{sewilo2018} concluded that COMs observed in the N113 hot cores could either originate from grain surface chemistry or in post-desorption gas chemistry.

The quartet  $J=2_K -1_K$ of \ch{CH3OH} at $\sim$96.7 GHz was later detected serendipitously toward N\,113 with ALMA in the project targeting two transitions (1--0 and 2--1) of three CO isotopes ($^{12}$CO, $^{13}$CO, and C$^{18}$O) toward three LMC and three SMC star-forming regions with a resolution of $\sim$2$''$ (or $\sim$0.48 pc).  N\,113 is the only region in this sample with \ch{CH3OH} detection.  The spectral window with the \ch{CH3OH} line was dedicated to the continuum observations and has a spectral resolution of $\sim$50 km s$^{-1}$. Such low spectral resolution does not allow for any quantitative analysis, but surprisingly the data revealed several additional well-defined \ch{CH3OH} peaks and extended emission in Region B (see Fig.~\ref{f:N113extendedmeth}). The \ch{CH3OH} emission reported in \cite{sewilo2018} was limited to two compact sources (hot cores A1 and B3) with some fainter emission detected in other locations in Region B (see Fig.~\ref{f:N113COMs}); however, it was unclear whether the latter traces physically distinct sources or shocked lobes of the outflows.  

In addition to hot cores, the $\sim$96.7 GHz \ch{CH3OH} emission is detected toward continuum sources  B4, B7, and B6 in Region B and D1--D4 located toward the south (Fig.~\ref{f:N113extendedmeth}; Sewi{\l}o et al., in prep.).   In Region A, the $\sim$96.7 GHz \ch{CH3OH} emission peak coincides with hot core A1; however, the emission extends toward the south.  
 
The $\sim$96.7 GHz \ch{CH3OH} emission peaks in Regions A and B overlap with the SiO (5--4) peaks (with some faint extended emission visible throughout Region B, see Fig.~\ref{f:N113SiO}; \cite{sewilo2018}, Sewi{\l}o et al., in prep.), tracing shocks. The same regions are associated with the DCN (3--2) emission indicating their youth and are distributed along the dense gas filament.  The observational and theoretical studies indicate that the abundance of all deuterated species strongly depends on temperature: it drops rapidly with increasing gas temperature as an object evolves (see, e.g., \cite{fontani2011,caselli2012,albertsson2013,gerner2015}) .

\section{Theory} 
\label{sec:theory}

We can gain some insight as to the origin of the COMs detected in the Magellanic Clouds by comparing  observations with current astrochemical theories of Galactic COM production in star-forming regions. Here we briefly review COM formation mechanisms and  how they have been applied to low-metallicity environments. 

\subsection{Complex Molecule Formation in Star-forming Regions of the Milky Way} \label{sec:mwcoms}     
 
 In the Galaxy, the largest COMs have until now been found primarily in the hot molecular cores associated with massive protostars, as well as with the so-called hot corinos found in regions of low-mass star formation (e.g.,  \cite{herbst2009}).  In cold, dense molecular clouds some of these organics have been known to be present for quite some time: \ch{HNCO}, \ch{HCOOH}, \ch{CH2CO}, \ch{CH3CHO}, \ch{CH3OH} \cite{ohishi1992}. More recently, the larger `hot core' COMs, including \ch{CH3OCH3} and \ch{HCOOCH3}, have also been detected in several cold clouds (\cite{bacmann2012,cernicharo2012,vastel2014,jimenez2016,taquet2017}).  These detections have raised several challenges for existing theories of COM formation. These include the long-standing problem of identifying the mechanism responsible for returning ice-mantle molecules to the gas in cold clouds, and the fact that radical reactions on 10~K grains will not occur. In the case of the cold methanol detected in the LMC, it would appear that reactive desorption (i.e., using part of the energy released upon formation)  is a plausible explanation \cite{minissale2016,chuang2018}.

Hot cores are regions of high gas density and extinction where elevated dust temperatures have led to the  sublimation  of molecular ice mantles that have previously grown on cold dust grains.  The first model of hot core chemical evolution was that of ~ \cite{brown1988} . This involves a long period of cold ($\sim$10~K) chemistry during which molecules form by gas--phase reactions and on dust grains, as the gas freezes out; this chemistry is similar to that found in cold dark clouds (e.g., \cite{bergin2007}). At some point gravitational collapse ensues, a protostar is formed, and the resulting large increase in luminosity heats the surrounding envelope of gas and dust to temperatures above about 100~K, leading to the  sublimation  of ice mantles and the formation of the hot core/corino.  Further refinements to this picture have been (a) the addition of a slow `warm-up' phase that can allow UV photolysis and radical--radical reactions to occur in the ice mantles, prior to formation of the  hot core proper (\cite{garrod2006}), and (b) gas--phase synthesis in the hot core, driven by the  sublimated  molecules (\cite{charnley1992}).  Further refinements have included the spatio-temporal evolution of the gas--grain chemistry (\cite{rodgers2003,aikawa2008}).  

Thus, within this simple picture, the complex molecules observed in hot cores can have four chemical origins: (1) formation in cold gas followed by freeze out on dust; (2) formation on the surfaces of cold (10~K) dust grains; (3) formation in  UV-photolyzed ice mantles at $\sim$30~K; (4) {\it in situ} formation in hot core gas following ice sublimation.
  
\subsubsection{Cold Gas--Phase Reactions in Dark Clouds}     
 \label{sec:gas}

Cosmic-ray ionization of molecular hydrogen drives dense cloud chemistry by initiating sequences of ion-neutral and neutral--neutral reactions. This chemistry leads to the production of CO and many of the simple molecules detected in  dark interstellar  clouds.  The complex molecules produced tend to be long hydrocarbon chains and associated radicals,  such as the cyanopolyynes and various carbenes.  The dark cloud TMC-1 in Taurus is recognized as the best example of this chemistry (e.g., \cite{gratier2016}) and chemical modeling demonstrates that gas-phase chemistry can account quite well for its composition (e.g., \cite{maffucci2018}). It is now appreciated that methanol is formed entirely on dust grains (see Section~\ref{sec:coldsurf}) and so some (non-thermal) desorption mechanism is required to explain its presence in dark clouds like TMC-1 since dust temperatures are never sufficiently high for methanol and water to desorb.  However, in dense cores where low-mass protostars are forming, it appears that some dust heating  (to $\sim$30--35 K) has occurred  and this can lead to the  sublimation  of more volatile molecules.  In this Warm Carbon Chain Chemistry (WCCC),  sublimation  of large abundances of  methane from the ice mantles greatly increases the efficiency of carbon-chain growth \cite{sakai2013}.  Although we do not discuss CH$_4$ injection further, the similarly warm  dust temperatures found in the Magellanic Clouds may make this a viable route to widespread carbon chain formation there.  

\subsubsection{Cold Grain--Surface Reactions}     
\label{sec:coldsurf}
  
\begin{figure*} 
\begin{center} 
 \includegraphics[angle=0,width=0.65\textwidth]{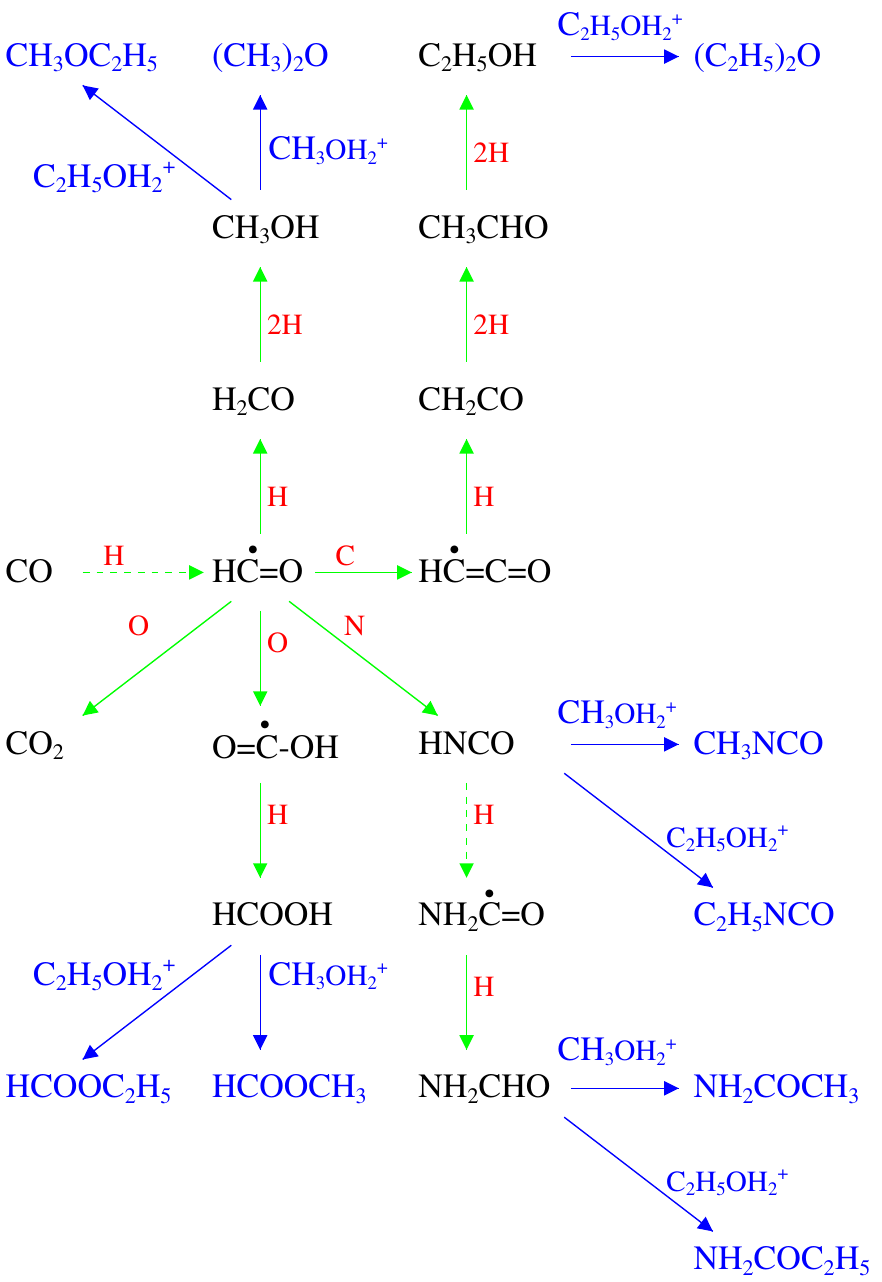}  
 \caption{Interstellar gas--grain surface chemistry.  Molecules involved in surface reactions are in black.
Hydrogen atom addition to unsaturated molecules creates reactive radicals and additions of C, O and N atoms, allows  a rich organic
chemistry seeded by carbon monoxide to develop.  Broken arrows indicate reactions with activation energy barriers; where 2H is shown, a barrier penetration reaction followed by an exothermic addition is implicitly indicated.   Once these ice mantles have been  sublimated  into the hot core gas,  methanol and ethanol can become protonated and take part in  ion--molecule alkyl cation transfer reactions with themselves and with other molecules  that were formed in the ices.  These processes are denoted in blue and each arrow indicates the initial ion--molecule reaction and the production of the neutral COM, either through electron dissociative recombination or  proton transfer to ammonia  (adapted from \cite{charnley2008} and references therein).}   
\label{fig:toe}
\end{center} \end{figure*}

Ice mantles grow on cold dust grains through the sticking of atoms of O, C and N followed by reactions with H atoms to form H$_2$O, CH$_4$ and NH$_3$. Molecules formed in the gas, such as CO and N$_2$,  can also accrete and become incorporated in the mantle. Unsaturated molecules can undergo (tunneling) addition reactions with hydrogen atoms to produce molecular radicals   that can further react and form new molecules  (e.g., \cite{tielens1982}). Figure \ref{fig:toe} shows that, in the case of CO,  these processes can lead to a rich organic chemistry  (e.g., \cite{charnley2008}) through  the growth of linear chains by single atom additions and their subsequent saturation.   This surface scheme can explain the presence of \ch{CH3OH} and many of the COMs known in hot cores and originally made predictions \cite{charnley2001} for new surface molecules (not shown)  that were subsequently detected:   glycolaldehyde \cite{hollis2000, hollis2004b}, ethylene glycol ($\rm (CH_2OH)_2 $, \cite{hollis2002}),  acrolein and propionaldehyde \cite{hollis2004a}.   Numerous surface chemistry experiments have validated many of these addition processes \cite{watanabe2008,Ioppolo2011,linnartz2015}.

\subsubsection{COM Chemistry Driven by Sublimated Ice Mantles} 
\label{sec:comsburst}
             
\cite{charnley1992} demonstrated that differences  in mantle composition could, following  sublimation ,  account for the chemical differentiation seen in the Orion--KL star-forming region. Rather than calculate the mantle composition {\it ab initio}, they assumed a composition based on observations and showed that the relative presence, or not, of methanol and ammonia could drive different  chemistries  in the hot gas (see also \cite{caselli1993}; \cite{rodgers2001}).  \cite{blake1987}  first suggested that self-methylation of methanol could be the source of dimethyl ether in hot cores, although they attributed its origin to be protostellar outflows rather than grain-surface chemistry.  The self-methylation reaction is 
\begin{equation}
{\rm CH_3OH_2^+ + CH_3OH} \rightarrow {\rm CH_3OCH_4^+ + H_2O}\label{eq:dme}
\end{equation}
where the neutral CH$_3$OCH$_3$ molecule is produced in  reactions with  either an electron  or a base,  such as ammonia.  Based on laboratory experiments, \cite{charnley1995} showed that, apart from CH$_3$OCH$_3$, many more large ethers and esters could be formed in this manner, in reactions involving ethanol, propanol and butanol \cite{karpas1989}.  Figure~\ref{fig:toe} includes alkylation reactions between protonated methanol and ethanol, the neutral alcohols, and formic acid.  Methyl formate formation by methylation of \ch{HCOOH} has also been studied in the laboratory \cite{cole2012} and shown to be a viable mechanism. In these experiments, the {\it trans}--\ch{HCOOH} conformer of formic acid produced the protonated form of {\it trans}--\ch{HCOOCH3}; electron dissociative recombination, or proton transfer (see below), would then lead to the neutral ester. However, apart from one detection of  {\it trans}--\ch{HCOOCH3} \cite{neill2011}, it is the more stable {\it cis}--\ch{HCOOCH3} conformer that is found in hot cores. It has therefore been suggested that interconversion between  {\it trans}--\ch{HCOOCH3} and {\it cis}--\ch{HCOOCH3} structures could occur in the proton loss process \cite{neill2012}. Chemical models have demonstrated that this is a very competitive  pathway to \ch{HCOOCH3} around protostars \cite{taquet2016}, assuming that interconversion  proceeds with high efficiency \cite{laas2011}. Of the COM products shown in Fig.~\ref{fig:toe}, apart from dimethyl ether and methyl formate, both ethyl formate and ethyl methyl ether have now been detected \cite{belloche2009,tercero2015}.  

A problem with this scenario has been that experiments  on electron dissociative recombination of protonated COMs show that recovery of the neutral molecule occurs with a probability much lower than previously assumed, typically $<5 \%$ as compared to 100\% (e.g., \cite{hamberg2010}), and subsequently  to COM abundances much lower than observed. However, if NH$_3$ is injected from grain mantles  at moderate abundances, comparable with those measured in Galactic ices \cite{boogert2015},  then proton transfer, e.g. 
\begin{equation}
{\rm CH_3OH_2^+ + NH_3} \rightarrow {\rm CH_3OCH_3 + NH_4^+}\label{eq:nh3}
\end{equation}
can dominate COM formation  and mitigate against the destructive effects of electron recombinations \cite{taquet2016,skouteris2019}.

As also shown in Fig.~\ref{fig:toe}, gas phase COM formation driven by alkyl cation transfer reactions involving surface-formed \ch{HNCO} and  \ch{NH2CHO} could plausibly  be a source of new molecules, perhaps explaining the hot core detections of methyl isocyanide (\ch{CH3NCO}) and acetamide (\ch{NH2COCH3}) \cite{cernicharo2016,hollis2006}. However, the associated gas phase reactions have not yet been studied experimentally or theoretically and so should be regarded as speculative.

\subsubsection{Radical Reactions in Icy Grain Mantles}      
\label{sec:radical}

As well as chemical reactions on grain surfaces driven by accretion of reactive species, it is also known that the ices can undergo bulk processing due to radiation damage caused by photons and energetic particles. Cosmic rays  interact  directly with icy mantles and recently chemical models  have been developed to explore their effects on COM  formation (see \cite{shingledecker2018} and references therein). The environment of young protostars receives  large doses of UV and EUV radiation \cite{benz2010}.  Even in dark clouds, a weak ambient flux of internal UV photons will exist \cite{prasad1983} as energetic electrons produced by cosmic-ray ionization of molecular hydrogen can subsequently collide with, and excite, other H$_2$ molecules which undergo de-excitation by emission of Lyman and Werner band photons. The resulting UV flux is  $\sim10^3$  photons s$^{-1}$ cm$^{-2}$  which can be significant for  chemistry in dense clouds.  Laboratory experiments on UV photolysis of interstellar ice analogs show that an extensive organic chemistry  can occur, primarily through reactions involving various radials produced from the major (parent)  ice species  (e.g. \cite{bernstein1995}) .    Cold H additions to CO  (see Section~\ref{sec:coldsurf}) have also to be considered since  interstellar organic synthesis by photolysis requires that the methanol  be present {\it ab initio } \cite{bernstein1995,oberg2009}. 
   
 Early theoretical models  for organic synthesis from the recombination of radicals  on cold grains   appealed to physically unrealistic  processes to attain the high mobilities needed  for heavy particle migration and reaction \cite{allen1977,dhendecourt1986,brown1990,hollis2001,sorrell2001}.   Detailed modeling has shown that sustained UV photolysis and warming during the early stages of hot core evolution can  produce many new organics \cite{garrod2006}.  In this picture, the ices are subjected to the Prasad--Tarafdar UV photons during the initial cold phase of core formation, producing a population of radicals. This irradiation can continue as  the core is being gradually (as opposed to instantaneously)  heated -- the so called `warm-up' phase \cite{viti1999,garrod2006}. When the dust grain  temperature reaches, and can be maintained close to,  about 30 K, the associated  increase in radical  mobility allows them to migrate and react. Radical--radical recombinations  between HCO, CH$_3$O, CH$_2$OH, COOH  and other simple  radicals   (e.g., CH$_3$,  CH$_2$, NH$_2$, CN, OH) then produce many of the COMs observed in hot cores \cite{garrod2008,belloche2009,ligterink2017}.

\begin{table*}[ht!]
\caption{Model Ice Mantle Compositions\textsuperscript{\emph{a}}}
\centering 
\label{tab:ices}
\begin{tabular}{lcc}
\hline\hline
\multicolumn{1}{c}{Molecule} & \multicolumn{1}{c}{LMC} & \multicolumn{1}{c}{SMC} \\
\hline
H$_2$O  & $4.5 \times 10^{-5}$  & $2.2 \times 10^{-5}$      \\	
 CO  & $1.7 \times 10^{-5}$  & $7.6 \times 10^{-6}$    \\	
 CO$_2$  & $1.3 \times 10^{-5}$  & $6.0 \times 10^{-6}$         \\	
CH$_4$  & $2.1 \times 10^{-6}$  & $1.0 \times 10^{-6}$         \\
 NH$_3$ & $1.0  \times 10^{-6}$  & $3.2 \times 10^{-7}$        \\ 
  N$_2$    &$3.2 \times 10^{-6}$   &  $1.0 \times 10^{-6}$         \\	 	
 H$_2$CO &  $1.1 \times 10^{-6}$   &  $5.0 \times 10^{-7}$         \\		 	
 CH$_3$OH &  $3.2 \times 10^{-6}$   &  $1.4 \times 10^{-6}$         \\	 	
 HCOOH  &  $6.4 \times 10^{-7}$    &  $3.2 \times 10^{-7}$         \\	 
\hline
\end{tabular}

\textsuperscript{\emph{a}} Abundances are given relative to total density of hydrogen nuclei.
\end{table*}

\subsection{Models of COM Formation in the Magellanic Clouds}      
\label{sec:magcoms}

Here we summarise and discuss  theoretical models of COM formation at reduced metallicity in the light of recent observations.
 
\subsubsection{Chemical Models of the LMC and SMC}      
\label{sec:magmodels}
  
\cite{millar1990} considered the chemical evolution of putative dark clouds in the LMC and SMC at the appropriate elemental depletions.   They considered the purely gas-phase chemistry of fairly simple molecules, with only CH$_3$OH and CH$_2$CO being the only COMs.  These models could produce CH$_3$OH abundances of $\sim$10$^{-8}$; however, the main formation mechanism was through a radiative association process that is no longer considered viable. 
 
 More recently,  \cite{acharyya2015,acharyya2016} developed gas--grain models of the LMC and SMC. Again, cold dark clouds were modeled for a wider range of physical parameters, including appropriately warmer gas and dust temperatures. In these models CH$_3$OH  was formed by CO hydrogenation on dust grains.  For the LMC, they found that molecular abundances similar to Milky Way dark clouds could be obtained, whereas for the SMC the abundances were lower for most molecules.  The model predictions were compared to several star-forming regions in the LMC (as no data on Magellanic dark clouds then existed):  N\,159W, N\,159S, N\,160, and 30 Dor--10.  
When compared to the Magellanic source N\,159W with a thermal methanol detection,  the calculated abundances were found to be $\sim$10--$10^{3}$ times lower than determined by \cite{heikkila1999}.  The reason for this discrepancy was explained as being due to the warm dust temperatures, appropriate for Magellanic dark clouds, that were considered in the models. Dust temperature affects grain-surface chemistry and can inhibit methanol formation since, for temperatures above about 15~K, hydrogen atoms will desorb before they can react with CO. Above about 25~K,  CO molecules residing in CO-rich outer layers of the ice mantle will also  sublimate , whereas those trapped in the \ch{H2O} ice matrix require higher dust temperatures \cite{collings2004}.  In both the LMC and SMC models, gaseous  NH$_3$ abundances  of $\sim$10$^{-8}$ were found.  For models with dust temperatures more similar to the Milky Way ($\sim$10--15~K), the dust ice mantles  are predicted to contain large fractions  of  CH$_3$OH ($\sim$10--40\%) and NH$_3$ ($\sim$1--7\%).  Hence, the presence of \ch{CH3OH} in the Magellanic Clouds points to formation in a cold pre-hot core phase.

\cite{pauly2018} studied the formation of ice mantles in the Magellanic Clouds but instead employed a correct stochastic calculation of the grain chemistry.  The relative composition of ices comprising of H$_2$O,  CO, CO$_2$, CH$_3$OH, NH$_3$ and CH$_4$ were calculated as a  function of the interstellar UV radiation field.  The CH$_3$OH and NH$_3$ ice fractions were found to be  lower  ($\sim \rm few \%$)  than those found by \cite{acharyya2015,acharyya2016} and more compatible with those of Galactic ices \cite{boogert2015}.
  
\begin{figure*}[ht!] 
\begin{center}
\hbox{  
\includegraphics[width=0.48\textwidth]{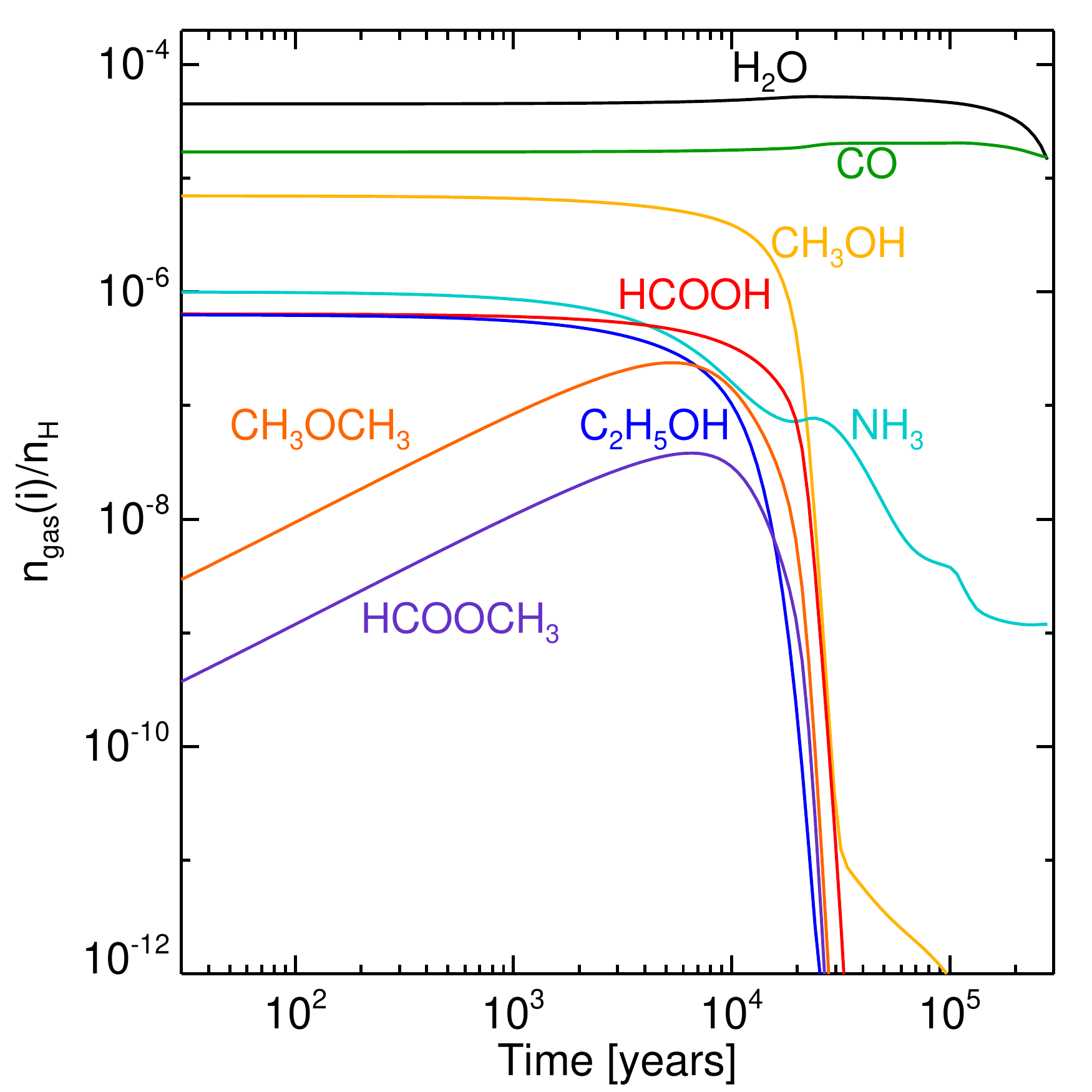}     
\includegraphics[width=0.48\textwidth]{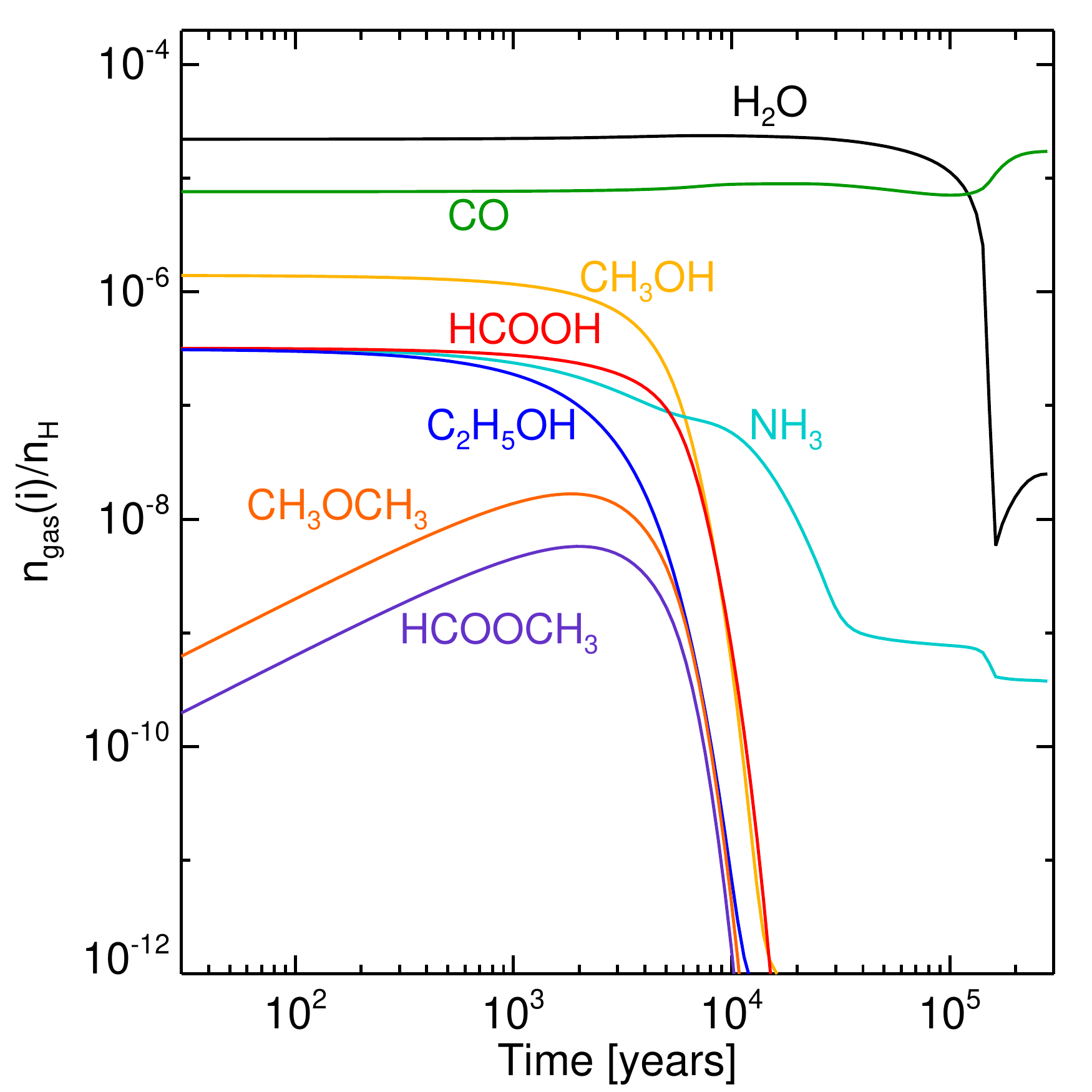}  
}
\caption{Chemical evolution in hot core models of the LMC ({\it left panel}) and SMC ({\it right panel}).  Ice mantles with the 
 compositions of Table~\ref{tab:ices} are instantaneously injected into the gas at $t=0$.
 }  \label{fig:VT}
\end{center}
\end{figure*}

\cite{acharyya2018} reported the first {\it bona fide} hot core models of  the  Magellanic Clouds.   Their model was a standard three-phase model  (cold quasistatic phase, collapse phase with out without `warm-up', and hot core phase; see Section~\ref{sec:mwcoms}) and included the radical chemistry induced during the `warm-up' phase  as well as  many of the  COMs known in Galactic hot cores (\cite{garrod2006, garrod2008}; see Section~\ref{sec:radical}).  When compared to the N\,113 observations, only  one of the models  (a 100~K core in which  the initial phase of core formation proceeded at 10~K, Model 1) reproduced  the observed CH$_3$OH abundances in both A1 and B3 at times (post- sublimation ) when CH$_3$OCH$_3$ and HCOOCH$_3$ had non-trivial abundances.  As CH$_3$OH is the proposed parent of CH$_3$OCH$_3$ and HCOOCH$_3$ in both grain-surface reactions  (see Section~\ref{sec:radical}) and in ion--molecule reactions (see Fig.~\ref{fig:toe}) the comparison of predicted  CH$_3$OCH$_3$/CH$_3$OH and HCOOCH$_3$/CH$_3$OH ratios with observations provides the most stringent tests of their origin (see Section~\ref{sec:icegas}). The relevant  \cite{acharyya2018} model yields HCOOCH$_3$/CH$_3$OH ratios of $\sim$0.04 in good agreement with that of both N\,113 cores;  whereas the CH$_3$OCH$_3$/CH$_3$OH ratios are about 5 times greater than observed.  
 
Overall, these  models indicate that the hot cores in the Magellanic Clouds must have undergone a period of chemical evolution when the dust was cold, as in theories of Galactic hot core formation.

\subsubsection{Models for COM Ice--Gas Synthesis}     
 \label{sec:icegas}
 
Given the importance of NH$_3$ for gaseous COM formation, the Magellanic Clouds offer a potential opportunity to test this scenario in a low-metallicity environment.  \cite{acharyya2018} emphasized the sublimation of COM-containing ice mantles into the hot gas, not active formation {\it in situ}, as the possible source of Magellanic COMs. Post-sublimation gas-phase synthesis (see Section~\ref{sec:comsburst}) has not yet been studied in the context of the Magellanic Clouds and here we report relevant model calculations as a first approximation.

\begin{figure*}[ht!] 
\begin{center} 
\includegraphics[angle=0,width=0.6\textwidth]{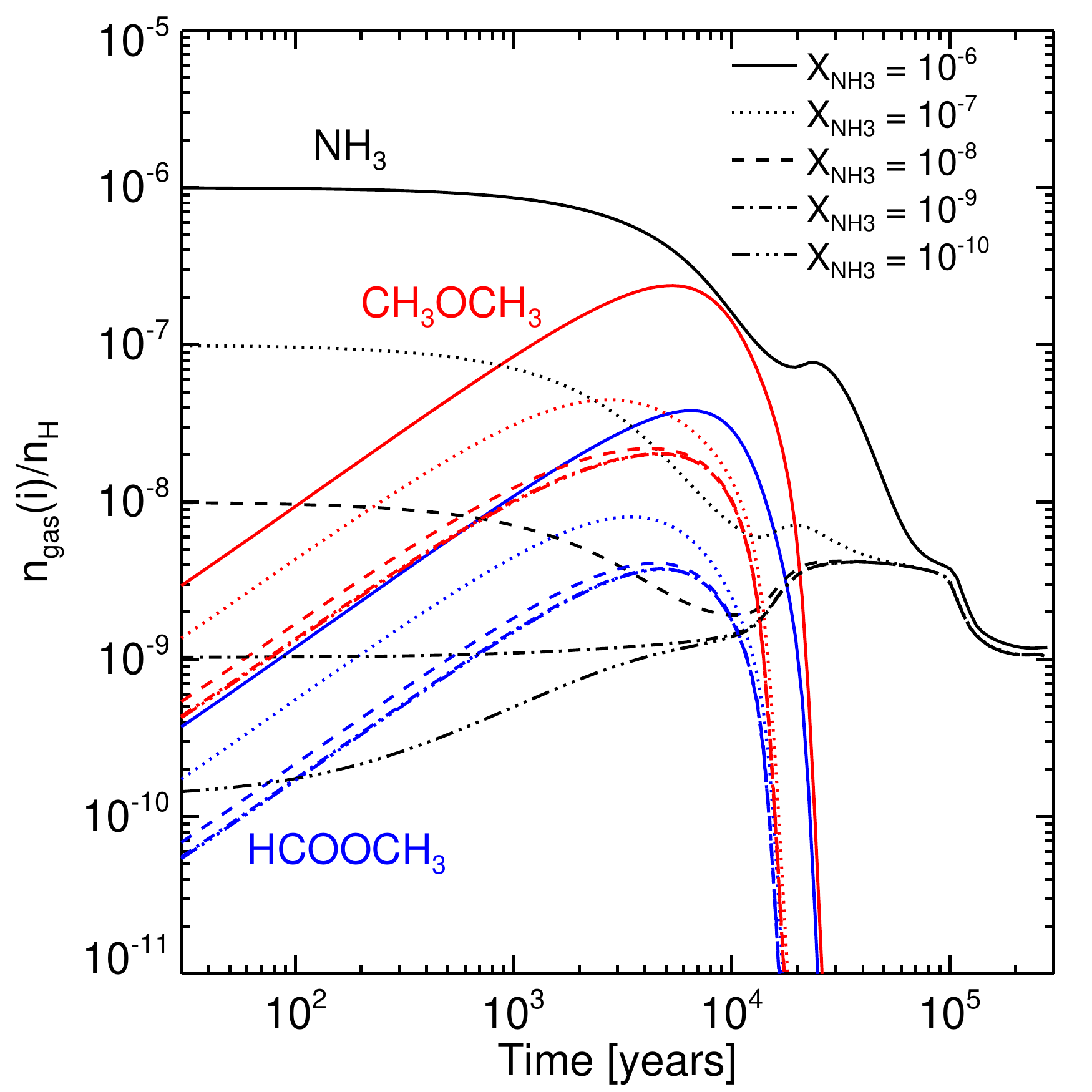}   
\caption{Evolution of CH$_3$OCH$_3$ and HCOOCH$_3$ abundances in the LMC model as a function of the NH$_3$ abundance  sublimated  from ices.}
\label{fig:VTnh3}
\end{center} \end{figure*}

The calculations were performed with the dynamical-chemical model of \cite{taquet2016} but modified according to the simplest version of these models in which the physical conditions are fixed in time and the ices are injected instantaneously into the hot gas at $t=0$  (cf. \cite{charnley1992}), as opposed to computing the physical and chemical  evolution of the core (e.g., gravitational collapse, protostellar luminosity heating, time-dependent molecule desorption). Table \ref{tab:ices} lists the assumed ice mantle compositions.  These were obtained by scaling the ice abundances from  Table 1 of \cite{taquet2016} to the elemental depletions observed in the SMC and LMC clouds as listed in Table 1 of \cite{acharyya2015} ($f_{{\rm LMC}, i}$) and \cite{acharyya2016} ($f_{{\rm SMC}, i}$), respectively.  The (C, O, N) scaling factors ($1/f$) are (0.41, 0.46, 0.21) for the LMC and (0.20, 0.22, 0.064) for the SMC; for each molecule listed in Table~\ref{tab:ices}, we used the averaged $1/f$ of its corresponding elements to determine initial ice abundances in the LMC/SMC from their Galactic values used by \cite{taquet2016}.  Typical hot core physical conditions were assumed:  $n_{\rm H_2} = 5 \times 10^7$ cm$^{-3}$,   $T$=150 K, and a cosmic-ray ionization rate $\zeta = 3\times10^{-17}$ s$^{-1}$.  The chemical network is as discussed in \cite{taquet2016}. Figure \ref{fig:VT} shows COM chemical evolution in hot core models applicable to the LMC and SMC. Generally, one can notice a decrease of all gas phase abundances by a factor of a few depending on the elemental depletion of C, N, and O in the two galaxies. However, the stronger N depletion relative to C and O (by a factor of 2 and 4 for the LMC and SMC, respectively) induces a more pronounced decrease of NH$_3$ relative  to other  ice species. 

As found by \cite{taquet2016}, a decrease of the injected ice mantle  NH$_3$/CH$_3$OH abundance ratio from  $\sim$1 decreases the production efficiency of complex organic molecules;  for  CH$_3$OCH$_3$ and HCOOCH$_3$ the Magellanic Cloud abundances are lower by a factor of 3 for the LMC and of 5 for the SMC. The NH$_3$ abundance is not observationally well-constrained  in the LMC and SMC and is unknown in the N\,113 region.  In the LMC,  \cite{ott2010} measured an ammonia abundance of $\sim$4$\times$10$^{-10}$ in N\,159W and Fig.~\ref{fig:VTnh3}  shows that systematically  reducing the injected NH$_3$ abundance from ices  dramatically lowers the  COM abundances. However, any decline in COMs below an NH$_3$ abundance of $\sim$10$^{-8}$, down to the level measured by \cite{ott2010},  is halted as electron dissociative recombinations come to dominate the loss of molecular ions.  Although models of ice formation in the Magellanic Clouds do predict NH$_3$/H$_2$O mantle fractions sufficiently large  to allow efficient ion--molecule formation in the LMC (see Section~\ref{sec:magmodels}),   NH$_3$ observations of  N113 and other COM-containing regions are clearly required to validate the efficacy of this process.

\begin{figure*} [h!]
 \includegraphics[angle=0,width=0.6\textwidth]{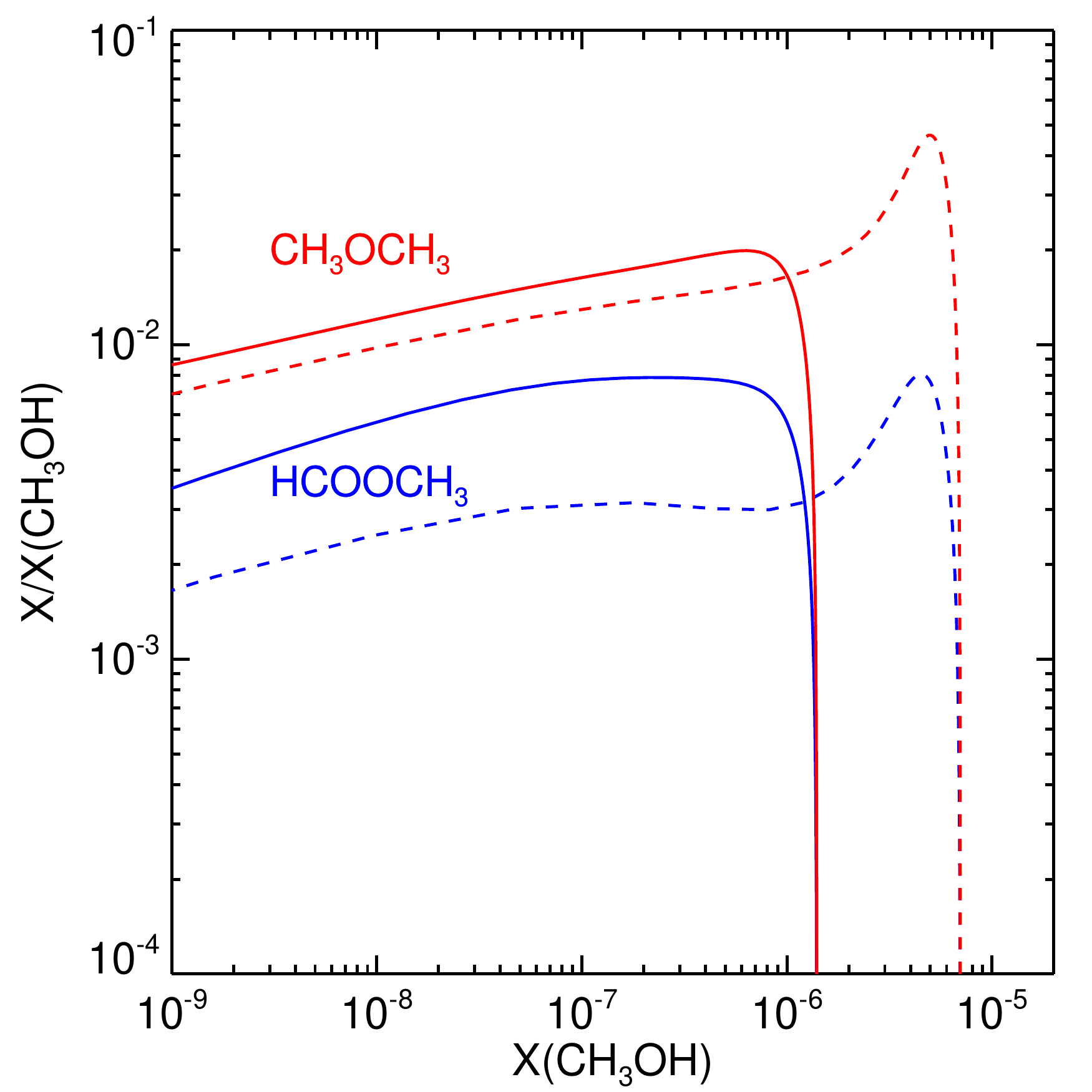}   
 \caption{\small  
 {Evolution of the CH$_3$OCH$_3$/CH$_3$OH and HCOOCH$_3$/CH$_3$OH abundance ratios in the models of Fig.~\ref{fig:VT} for the LMC (solid curves) and SMC (dashed curves).
 }   }
\label{fig:VTratios}
\end{figure*}

 The  observed abundances of CH$_3$OCH$_3$ and HCOOCH$_3$ in the A1 and B3 hot cores of N\,113 (see Table 3) can be reproduced in the models of Fig.~\ref{fig:VT} despite the lower elemental depletions.  However, as in the case of Galactic hot core models these do not occur  at a time when the   CH$_3$OCH$_3$/CH$_3$OH and HCOOCH$_3$/CH$_3$OH  abundance ratios have their observed values  (Fig.~\ref{fig:VTratios}). In the LMC, when the methanol abundance is $\sim$10$^{-8}$, the observed  CH$_3$OCH$_3$/CH$_3$OH ratios in A1 ($\sim$0.1) and B3 ($\sim$0.2) are about an order of magnitude higher than the calculated values ($\sim$0.01); for HCOOCH$_3$/CH$_3$OH the calculated value of ($\sim$0.003) is about a factor 20 too low.

 \cite{acharyya2018} claimed a very good agreement with the abundances of \cite{sewilo2018} to within a factor of 5. However, calculated abundances and the observed abundances were presented as relative to total hydrogen density and molecular hydrogen density, respectively. Correcting the abundance comparison by the factor of 2 means that for \ch{CH3OCH3} in A1 and B3, the models overpredict the abundances by factors of 12--20; only the calculated abundances of \ch{CH3OH} and \ch{HCOOCH3} in A1 agree to within a factor of 5.

When a detailed treatment of the collapse physics  is included in hot core chemistry models,  a monotonic increase in the radial gas and dust   temperatures occurs as a function of the increasing (accretion)  luminosity  of the protostar (e.g., \cite{rodgers2003}).  In the case of Galactic hot cores/corinos, \cite{taquet2016} showed that molecular {\it recondensation} may be just as chemically important as molecular depletion onto dust in cold cores, and  sublimation  from them in hot cores.  This can occur if the accretion onto the protostar is not monotonic but sporadic \cite{audard2014}. Once grains have been raised to a temperature of about 120~K or more,  most of the ice mantle is removed and, as they cool to  50 K or  lower in about 100 years,  selective recondensation can occur  due to the  differing  binding energies for physisorption. Production of organic molecules ensues in the brief post- sublimation  period,  as in the standard model,  except that the products   subsequently condense as ices.  \cite{taquet2016} demonstrated that the observed large CH$_3$OCH$_3$/CH$_3$OH and HCOOCH$_3$/CH$_3$OH ratios could be reproduced due to the relatively higher  binding energy of CH$_3$OH.  This may be a fundamental process for  the organic chemistry of material near protostars  as, between sublimation--recondensation  events, there can be a mantle-driven gaseous  chemistry leading to the dust grains being  coated with the products as they cool.  This scenario remains to be evaluated in the context of Magellanic Cloud hot cores.


\section{Discussion}
\label{s:discussion}

Although single-dish observations provided the first detections of methanol in two bright regions in the LMC (N\,159 and N\,113), it was ALMA with its unprecedented  angular resolution and sensitivity that has started and continue revolutionizing the field of complex organic chemistry in  low-metallicity environments.  To date, the sample of sources with  COMs detection is very small and diverse.  Only two {\it bona fide} hot cores with COMs are known in the Magellanic Clouds, both in the N\,113 star-forming region in the LMC (N\,113 A1 and B3; \cite{sewilo2018}). Another hot core in the LMC (ST\,11) was found by \cite{shimonishi2016b}, but no methanol or other COMs were detected.  The detection of methanol was reported for two other sources, one in the LMC (N\,159W--S, Section~\ref{s:newalma}) and one in the SMC (IRAS\,01042$-$7215, \cite{shimonishi2018}), but not in hot cores (`cold methanol').  Several sources in N\,113 (in addition to two hot cores) are associated with methanol, but due to a very low spectral resolution of the observations, a determination of their physical parameters based on the existing data is currently impossible.

No firm conclusions about complex organic chemistry under reduced metallicity conditions can be drawn based on the current sample of sources with the detection of COMs.  However, some general trends are noticeable. 

None of the sources with a COMs detection is associated with an infrared source: COMs are detected toward mm continuum sources in the vicinity of bright {\it Spitzer} YSOs.  ST\,11 with no COMs detection but with evidence for hot core chemistry is the only source associated with the infrared source. In fact, this is the brightest {\it Spitzer} YSO in the entire N\,144 H\,{\sc ii} region \cite{carlson2012}; higher resolution near-infrared observations reveal a cluster with one dominating source (see Fig.~\ref{f:OtherMosLMCST11}), which likely has the largest contribution to the emission at longer wavelengths.  Multiple mm continuum sources are detected toward all the other fields. The N\,113 B3 field is particularly interesting since it harbors sources at different evolutionary stages including a hot core and ultra-compact H\,{\sc ii} regions with younger sources distributed along a dense gas filament (see Figs.~\ref{f:N113mos} and \ref{f:N113SiO}). Several mm continuum sources including one associated with the brightest \ch{CH3OH} peaks in the region, are also found along the filament in N\,159W--S (see Figs.~\ref{f:N159WSKMOS}--\ref{f:N159WSmeth} and \cite{tokuda2018}).  

The detection of cold methanol in N\,159W--S and IRAS\,01042$-$7215 P2/P3 is puzzling, however, similar results are being obtained in our Galaxy.  In Galactic dark clouds, gas--phase methanol is found to have a clumpy distribution, often not following the density distribution traced by continuum and molecular tracers \cite{buckle2006}, and thus provides evidence of rapid ice desorption processes. This methanol tends to be cold, at excitation temperatures of 5--20 K, similar to what is observed in N\,159W--S and IRAS\,01042$-$7215. In the Barnard~5 cloud in particular, a cold methanol clump with the \ch{CH3OH} abundance with respect to \ch{H2} of $\sim4 \times 10^{-8}$, offset from the IRAS sources, has proven to be a signpost of ice desorption -- first by the detection of cold gas--phase water at abundances similar to methanol \cite{wirstrom2014} and then other COMs like \ch{CH3CHO} and \ch{HCOOCH3}, likely formed by gas--phase chemistry following an event of ice desorption \cite{taquet2017}. However, the low yield of COMs from ice desorption at cold temperatures is at most a few percent of methanol even at Galactic metallicities, thus consistent with non-detections of COMs towards N\,159W--S, IRAS\,01042$-$7215 P2 and P3.

For IRAS 01042$-$7215 where the \ch{CH3OH} emission is detected toward two sources corresponding to the continuum peak, \cite{shimonishi2018} discussed three possibilities for the nature of P2/P3 (their `East core') where cold methanol was detected -- they suggested it can be either a massive starless core, an embedded high-mass YSO (or multiple YSOs) before the emergence of infrared emission, or a cluster of low-mass embedded YSOs.

Observations confirm that COMs can form in extragalactic environments. Formation pathways have been suggested that involve either grain--surface chemistry or gas--phase chemistry, or a combination of both. Theoretical models accounting for the physical conditions and metallicity of  hot molecular cores in the Magellanic Clouds have been able to broadly account for the existing observations. 
 
 As cosmic metallicity is increasing with time, understanding interstellar chemistry in low metallicity environments also gives us insight into the chemistry of the past Universe. The detection of COMs in the Magellanic Clouds has important implications for astrobiology.  The metallicity of the LMC/SMC is similar to  the mean metallicity of the interstellar medium during the epoch of peak star formation in the Universe, thus the presence of COMs in these systems indicates that a similar prebiotic chemistry leading to the emergence of life, as it happened on Earth, is possible in low-metallicity systems in earlier epochs of the Universe.  According to one of the theories on the emergence of life on Earth,  interstellar COMs (after incorporating into comets) might have been delivered to early Earth to provide important ingredients for the origin of life (e.g., \cite{ehrenfreund2000,mumma2011,caselli2012}). 
 
COMs more complex than \ch{CH3OH} are detected in the LMC toward hot cores which are associated with massive stars and therefore are short-lived; a detection of COMs does not have a direct relation to the origin of life at these locations. However, if COMs are detected toward hot cores, we can expect them to be present in hot corinos, which are associated with low-mass stars. The evolution of low-mass stars is long enough for life to emerge if favorable conditions exist.

\small
\begin{table*}
  \caption{COM Transitions Detected in the Magellanic Clouds with ALMA\textsuperscript{\emph{a}}}
  \label{tbl:transitions}
  \begin{tabular}{ccccll}
    \hline
    \hline
    Species  & Transition & Frequency & $E_{\rm U}$ & \multicolumn{2}{l}{Detection\textsuperscript{\emph{b}}} \\
             &            & (GHz)     & (K)         &   & \\ 
    \hline
        \multicolumn{4}{c}{LMC N\,113} & A1 & B3\\
    \hline
CH$_{3}$OH v$_t$=0& 5$_{1,4}$--4$_{2,2}$ E & 216.94560 & 55.87   &       $+$ & $+$   \\
CH$_{3}$OH v$_t$=1 & 6$_{1,5}$--7$_{2,6}$  A$^{-}$ & 217.299205 & 373.93  &    $+$ & $+$ \\
CH$_{3}$OH v$_t$=0 & 20$_{1,19}$-20$_{0,20}$ E & 217.88639 & 508.38 & $+$ & $-$\\
CH$_{3}$OH v$_t$=0 & 10$_{2,9}$-9$_{3,6}$ A$^{-}$ & 231.28110 & 165.35 &       $+$  &  $+$ \\
CH$_{3}$OH v$_t$=0 & 10$_{2,8}$-9$_{3,7}$ A$^{+}$ & 232.41859 & 165.40 &       $+$ &  $+$\\
CH$_{3}$OH v$_t$=0 & 18$_{3,16}$-17$_{4,13}$  A$^{+}$ & 232.78350 &        446.53 &  $+$ & $+$ \\
CH$_{3}$OH v$_t$=0\textsuperscript{\emph{c}}  & 2$_{-1,2}$--1$_{-1,1}$ E & 96.739362 & 12.54& $+$ & $+$ \\
& 2$_{0,2}$--1$_{0,1}$ A$^{+}$ & 96.741375 & 6.97&& \\
& 2$_{0,2}$--1$_{0,1}$  E & 96.744550 & 20.09&& \\
& 2$_{1,1}$--1$_{1,0}$  E & 96.755511 & 28.01 && \\
HCOOCH$_3$ v=0 & 18$_{2,16}$--17$_{2,15}$ E & 216.83020 & 105.68&     $+?$ &  $-$ \\
HCOOCH$_3$ v=0 & 18$_{2,16}$--17$_{2,15}$ A & 216.83889 &  	105.67  & $+?$  & $-$ \\
HCOOCH$_3$ v=0 & 20$_{1,20}$--19$_{1,19}$ E & 216.96476 & 111.50 &     $+$ & $+?$ \\
HCOOCH$_3$ v=0 & 20$_{1,20}$--19$_{1,19}$ A & 216.96590 & 111.48  &    $+$ & $+?$\\
HCOOCH$_3$ v=0 & 20$_{0,20}$--19$_{0,19}$ E & 216.96625 &  	111.50&    $+$ & $+?$ \\
HCOOCH$_3$ v=0 & 20$_{0,20}$--19$_{0,19}$ A & 216.96742 & 111.48&       $+$ & $+?$ \\
CH$_3$OCH$_3$ & 13$_{0,13}$--12$_{1,12}$ EE & 231.98782& 80.92&        $+$ & $+$ \\
    \hline
      \multicolumn{4}{c}{LMC N\,159W--S} & \multicolumn{2}{c}{ MMS--3\textsuperscript{\emph{d}}}  \\

     \hline
CH$_{3}$OH v$_t$=0   & 5$_{0,5}$--4$_{0,4}$ E & 241.700219 & 47.93 & \multicolumn{2}{c}{$+$} \\
CH$_{3}$OH v$_t$=0   & 5$_{-1,5}$--4$_{-1,4}$ E  & 241.767224 & 40.39 & \multicolumn{2}{c}{$+$} \\
CH$_{3}$OH v$_t$=0   & 5$_{0,5}$--4$_{0,4}$ A$^{+}$ & 241.791431 & 34.82 & \multicolumn{2}{c}{$+$} \\  
CH$_{3}$OH v$_t$=0   & 5$_{1,4}$--4$_{1,3}$  E & 241.879073 & 55.87 &  \multicolumn{2}{c}{$+$} \\
CH$_{3}$OH v$_t$=0\textsuperscript{\emph{e}}   & 5$_{-2,4}$--4$_{-2,3}$ E & 241.904152 & 60.72 &  \multicolumn{2}{c}{$+$} \\
                                      & 5$_{2,3}$--4$_{2,2}$ E & 241.904645 & 57.07 & \\
    \hline
    \multicolumn{4}{c}{SMC IRAS\,01042$-$7215} & P2 & P3 \\
    \hline
CH$_{3}$OH v$_t$=0   & 5$_{0,5}$--4$_{0,4}$ A$^{+}$ & 241.791431 & 34.82 & $+$ & $+$ \\ 
CH$_{3}$OH v$_t$=0   & 5$_{-1,5}$--4$_{-1,4}$ E  & 241.767224 & 40.39& $+$ & $+$ \\
CH$_{3}$OH v$_t$=0   & 5$_{0,5}$--4$_{0,4}$ E & 241.700219 & 47.93 & $+?$ & $-$\\
CH$_{3}$OH v$_t$=0\textsuperscript{\emph{e}}   & 5$_{-2,4}$--4$_{-2,3}$ E & 241.904152 & 60.72 & $+$ & $-$\\
                                      & 5$_{2,3}$--4$_{2,2}$ E & 241.904645 & 57.07 & & \\
    \hline
      \end{tabular}
      
      \textsuperscript{\emph{a}} The table uses the CDMS notation from the Splatalogue.  See footnotes to Table~\ref{tbl:transitionssingle} for details; 
      \textsuperscript{\emph{b}} The symbols in these columns indicate a detection (`$+$'), a tentative detection (`$+?$'), or a non-detection (`$-$') of a given molecular line transition;    
       \textsuperscript{\emph{c}} The four transitions at $\sim$96.7 GHz are blended; the spectral resolution of these observations which were dedicated to the continuum detection is $\sim$50 km s$^{-1}$. In addition to hot cores A1 and B3, there are several other $\sim$96.7 GHz \ch{CH3OH} peaks in N\,113 (see Section~\ref{s:N113});
        \textsuperscript{\emph{d}} The two methanol peaks associated with MMS--3 where the largest number of \ch{CH3OH} lines are detected and the temperature fitting is the most reliable (see Section~\ref{s:N159WS});
      \textsuperscript{\emph{e}} The \ch{CH3OH} lines at 241.904152 GHz and 241.904645 GHz are blended.
\end{table*}
\normalsize

\section{Future Prospects}

Systematic studies on YSOs in different environments in the LMC and SMC are needed to build a statistically significant sample of sources with COMs detections in the Magellanic Clouds that would allow us to draw reliable conclusions on the formation and evolution of
COMs in reduced metallicity environments.  Both the LMC and SMC samples will be important to get as wide a range of metallicities as possible. In the SMC, the metallicity is lower than in the LMC (as low as 0.1 $Z_{\odot}$ in the SMC tail where star formation occurs in pockets; e.g., \cite{oliveira2009IAUS}) and the formation and survival of COMs can be tested in an even harsher environment. Enlarging the sample in both galaxies is important to distinguishing between local environmental conditions in individual clouds, and the more global effect of metallicity.

Reliable determination of Magellanic Clouds hot cores' physical and chemical properties and a comparison between the two Clouds and the Galaxy are of paramount importance to understand the impact of metallicity on complex organic chemistry.  High-resolution and sensitivity observations of known hot cores in the Magellanic Clouds as their number starts increasing are a natural next step in reaching this goal.  This direction of research involves chemical modeling of hot cores to understand their chemical evolution.  

ALMA programs are underway targeting tens of YSOs identified by {\it Spitzer} and {\it Herschel}, both specifically designed to search for COMs and those with different science goals, but covering frequency ranges where COMs can be detected (serendipitous detections as in N\,113).  ALMA's broad spectral windows allow to observe many spectral lines simultaneously with the spectral setup covering the most important molecules. 

What new COMs await discovery in the Magellanic Clouds? Chemical models of the LMC have not been successful (to within an order of magnitude) in accounting for the observed presence of dimethyl ether in the LMC (see Section~\ref{sec:icegas}) and so may not be reliable indicators for new molecules. At present, taking the inventory of well-studied Galactic hot cores may provide the best guidelines for future searches. Molecules that could be expected to present include acetaldehyde, ethanol, ethyl cyanide, isocyanic acid, and formamide.

The NASA's  {\it James Webb Space Telescope} ({\it JWST}; scheduled to launch in 2021) near- (0.6--5.3$\mu$m; the Near Infrared Spectrograph -- NIRSpec instrument) to mid-infrared (4.9--28.8 $\mu$m; Mid-Infrared Instrument -- MIRI) observations will enable studies of the solid phase material in different Galactic and extragalactic environments.  The high sensitivity and angular resolution of {\it JWST} will enable detailed studies of hot cores in the Magellanic Clouds on $\sim$0.04 pc scales in the near-infrared, without contamination from the surrounding regions.  It will be possible to do an inventory of simple (e.g., \ch{H2O}, \ch{CO2}, \ch{CO}) and complex (\ch{CH3OH} and beyond, including molecules of prebiotic significance) ices in the low-metallicity environments in the Magellanic Clouds. For example, in the 5--8 $\mu$m range (MIRI) COMs such as ethylene glycol, glycolaldehyde, methyl formate, and dimethyl ether (the latter two already detected in N\,113 A1/B3 in the gas phase with ALMA), and others can be detected \cite{fedoseev2017}.  The laboratory experiments are underway predicting which COMs can be detected at shorter wavelengths covered by NIRSpec. Such studies will provide information on differences in chemical complexity of hot cores in different environments. 

\appendix

\section{Technical Details on Unpublished Observations Reported in this Paper}

The ALMA and {\it VLT}/KMOS data presented in Section~\ref{s:N159WS} for N\,159W--S have not been published in literature yet.  Below we summarize technical details of these observations. 

\subsection{ALMA}
\label{s:newalma}

N159W--S was observed with ALMA (Atacama Large Millimeter/submillimeter Array; ALMA Partnership et al. 2015) during its Cycle 4 in October 2016 as part of  project number 2016.1.00308.S. We used the main array (i.e.,12m antennas) consisting of 40 antennas. The phase center was set to RA (J2000) = 05$^{\rm h}$39$^{\rm m}$41$^{\rm s}$, Dec (J2000) = 69$^{\circ}$46$'$11$''$. The ALMA correlator was configured to cover specific frequency ranges within the Band 6 of ALMA. Four broad spectral windows (with a bandwidth of 1875 MHz each) centered at frequencies 242.4, 244.8, 257.85 and 259.7 GHz were tuned at the frequencies of specific molecular transitions of dense gas, shock/hot core tracers (e.g., CS, \ch{SO2}, \ch{CH3OH} and SiO). The ALMA data were calibrated using the ALMA calibration pipeline available in {\sc CASA} version 5.5.1. Flux calibration was obtained through observations of the bright quasar J0519--4546. The gains were calibrated by interleaved observations of the quasar J0601--7036.  The bandpass response was obtained by observing the bright quasar J0635--7516. After the calibration was applied, the line-free channels of the broad spectral windows were identified and used to create the continuum images. The {\sc CLEAN}ing process was done with the task {\sc TCLEAN}, and setting the robust parameter of Briggs equal to 0.5 as compromise between resolution and sensitivity.  The resulting images were restored with a synthesized beam of 0$\rlap.{''}$22$\times$0$\rlap.{''}$12.  Sensitivity of 1.4 mJy beam$^{-1}$ was achieved in the data cube covering the \ch{CH3OH} lines.

\subsection{{\it VLT}/KMOS}
\label{s:newkmos}

N159W--S was observed with {\it VLT}/KMOS ({\it Very Large Telescope}/$K-$band Multi-Object Spectrograph) under program 0101.C-0856(A) using the $H+K$ grating with a spectral resolving power of 2000. The observations were carried out using a standard nod-to-sky procedure with an integration time of 150\,s, four DITs and three dither positions, yielding a total on-source integration time of 1800\,s. Telluric absorption correction, response curve correction, and absolute flux calibration was carried out using observations of telluric standard stars using three IFUs. The data were reduced with the standard {\it VLT}/KMOS pipeline using the {\sc esoreflex} data reduction package. The $K-$band continuum image is produced by integrating over a third-order polynomial fit to the data for every spatial pixel (spaxel) over the spectral range 2.028--2.290\,$\mu$m. The Br$\gamma$ and H$_{2}$ line emission images are produced by fitting a Gaussian profile to the emission lines at every position in the image.

\begin{acknowledgement}
We thank Dr. Kazuki Tokuda for providing the ALMA 1.3 mm continuum image of N159W--S and Dr. Julia Roman-Duval for providing the \ch{CH3OH} $\sim$96.7 GHz image of N\,113. We also thank Dr. Gerard Testor for providing the {\it VLT}/NACO near-infrared images that were used to improve astrometry of the  {\it VLT}/KMOS images. The work of M.~S. was supported by NASA under award number 80GSFC17M0002.  The work of S.~B.~C. was supported by the NASA Astrobiology Institute through the Goddard Center for Astrobiology.  P.~S. acknowledges support from the University of Cologne, Collaborative Research Centre 956, sub-project C3, funded by the Deutsche Forschungsgemeinschaft (DFG) -- project ID 18401886, and Verbundforschung Astrophysics, Project 05A17PK1, funded by BMBF (German Ministry of Science and Education).  V.~T. acknowledges the financial support from the European Union's Horizon 2020 research and innovation programme under the Marie Sk{\l}odowska-Curie grant agreement No. 664931. E.~S.~W. received funding from the Swedish National Space Agency, grants dnr 98/14 and dnr 246/16. The work of J.~W. was supported by  Sonderforschungsbereich SFB 881 "The Milky Way System" (subproject B2) of the German Research Foundation (DFG).  
The work of S.~Z. and T.~O. were supported by NAOJ ALMA Scientific Research Grant Number 2016-03B. 
The work of T.~O. was also supported by JSPS KAKENHI (Grant Nos. 22244014, 26247026, and 18H05440).
The work of A.~K. was supported by JSPS KAKENHI (Grant No. 23403001). 
The authors acknowledge financial support from their home institutions. 
\end{acknowledgement}

\bibliography{refs.bib}

\end{document}